 \documentclass[aps,12pt,floatfix,tightenlines,showkeys,showpacs]{revtex4}
 \usepackage{amsmath, amsthm, amssymb}
 \usepackage{graphicx} 
\usepackage{subfigure}
\newcommand{\be}{\begin{equation}}
\newcommand{\ee}{\end{equation}}
\newcommand{\bea}{\begin{eqnarray}}
\newcommand{\eea}{\end{eqnarray}}
\newcommand{\eq}[1]{Eq.~(\ref{#1})}

\newcommand{\ben}{\begin{displaymath}}
\newcommand{\een}{\end{displaymath}}

\begin {document}

\title{\bf \hskip10cm NT@UW-13-10\\
{Incoherent $J/\psi$ electroproduction from the deuteron at JLab energies and the elastic $J/\psi$-nucleon scattering amplitude 
  	}}
                                                                           
\author{Gary T.~Howell, Gerald A.~Miller}
\affiliation{Department of Physics, Univ. of Washington\\
Seattle, WA 98195-1560}
\date{\today}

\begin{abstract}
Calculations are presented for incoherent $J/\psi$ electroproduction from the deuteron at JLab energies, including the effects of $J/\psi$-nucleon rescattering in the final state, in order to determine the feasibility of measuring the $J/\psi$-nucleon scattering length, or the $J/\psi$-nucleon scattering amplitude at lower relative energies than in previous measurements.  It is shown that for a scattering length of the size predicted by existing theoretical calculations, it would not be possible to determine the scattering length.  However, it may be possible to determine the scattering amplitude at significantly lower relative energies than the only previous measurements.
\end{abstract}    


\maketitle

\section{Introduction}
\label{sec:intro}

With the upcoming $12\;GeV$ upgrade at JLab, electroproduction of the $J/\psi$ at JLab on a proton or deuteron will be possible.  With the mass of the $J/\psi$ being $3.097\;GeV$, the threshold virtual photon energy for electroproduction (at small $Q^2$) on a single nucleon is $\nu_{thresh}\simeq8.2\;GeV$ in the LAB frame, and is thus accessible with a $12\;GeV$ electron beam.  Most of the existing data on $J/\psi$ photo- and electroproduction is at much higher energy.  The $12\;GeV$ upgrade provides the opportunity to measure $J/\psi$ production near threshold~\cite{jlab12}.  In addition, measuring electroproduction on the deuteron provides the opportunity to measure the $J/\psi$-nucleon elastic scattering amplitude at lower energies than previous measurements, if the rescattering of the produced $J/\psi$ on the spectator nucleon in the deuteron is non-negligible. This paper presents calculations of the incoherent $J/\psi$ production amplitude from the deuteron near threshold (in order to determine if the $J/\psi$-nucleon scattering length can be determined) and at somewhat higher energy where the $J/\psi$-nucleon scattering amplitude may be determined.  We use a model which takes into account quasi-elastic production from the deuteron (with production on one nucleon while the spectator nucleon recoils freely) as well as rescattering effects in the final state (proton-neutron rescattering and $J/\psi$-nucleon rescattering).

The motivation for the work in the first part of this paper (Secs. \ref{sec:electroproduction} - \ref{sec:invaramps}) was a proposal at JLab~\cite{jlab10} to measure the $J/\psi $-nucleon scattering length by the reaction $\gamma^*+d\to J/\psi +p+n$.  The proposed experiment would detect the outgoing proton and the decay products ($e^+e^-$) of the $J/\psi$, for a kinematically complete measurement.  The reason the $J/\psi $-nucleon scattering length is of interest is that several authors have argued that a nuclear bound state of the $J/\psi$ may exist~\cite{savage92,brodsky97}.  They propose that the force between a $J/\psi$ and a nucleon is purely gluonic in nature, and therefore is the analogue in QCD of the van der Waals force in electrodynamics, since the hadrons are color neutral objects.  There is very little experimental data on elastic $J/\psi$-nucleon scattering.  There has only been one experimental measurement of it, at SLAC in 1977, where the $J/\psi$-nucleon total cross-section was extracted by measuring production of $J/\psi$'s on nuclei and using an optical model for the re-scattering of the $J/\psi$ on the spectator nucleons~\cite{psidata77}.  

Measurement of the scattering length provides information on the bound states of the two particles involved in the scattering.  In particular, for an attractive potential, if the scattering length is positive then there exists a bound state.  Theoretical calculations of the $J/\psi$-nucleon scattering length~\cite{savage92,brodsky97,kawanai2010,Hayashigaki:1998ey} predict a scattering length too small to produce a $J/\psi$-nucleon bound state.  However, it would be large enough that there could exist a nucleus-$J/\psi$ bound state in a large enough nucleus.  Since the scattering length is the (negative of) the zero-energy scattering amplitude, in order to measure this it is necessary for the two particles to scatter with small relative-momentum.  In the case of $\gamma^*+d\to J/\psi +p+n$ at the energies which are kinematically allowed in the proposed JLab experiment, it isn't possible to have an on-mass-shell nucleon and $J/\psi$ scatter at small relative momentum.  For an incident virtual photon of energy $\nu=9\;GeV$, and an outgoing $J/\psi$-neutron pair with zero relative momentum, the minimum possible momentum of the neutron in the LAB frame (deuteron at rest) is $\simeq 0.85\;GeV$; for $\nu=6.5\;GeV$ and zero relative momentum of the $J/\psi$-neutron pair, the minimum LAB momentum of the neutron is $\simeq 1\;GeV$.  For zero relative momentum of the outgoing pair, the initial LAB momentum of an on-shell neutron in the deuteron (before the collision with the $J/\psi$) must equal the final LAB momentum of the neutron.  Therefore, the momentum of the neutron inside the deuteron would have to be $0.85\;GeV$ (for $\nu=9\;GeV$). However,  the deuteron wavefunction at that momentum is very small (essentially zero).  

So although the proposed experiment~\cite{jlab10} may not be able to measure the $J/\psi$-nucleon scattering length, it might still be possible to measure the on-mass-shell $J/\psi$-nucleon scattering amplitude, but at higher relative energies.  The relative energy of the $J/\psi$-neutron pair would still be significantly smaller than in the only existing data (from the 1977 experiment at SLAC).  Under certain kinematic conditions, the dominant contributions to the amplitude will come from p-n rescattering and/or $J/\psi-n$ rescattering after the $J/\psi$ is produced.  If we fix the magnitude of the outgoing neutron's momentum at a moderately large value (here taken to be 0.5 GeV) the contribution of the impulse diagram (where the $J/\psi$ is produced on the proton and the neutron recoils freely) will be negligible, since the impulse diagram is proportional to the value of the deuteron wavefunction at that momentum.  This higher-energy rescattering is the subject of the second part of this paper (Sec.  \ref{sec:intermedenergy}).

This paper is organized as follows.  Sec. \ref{sec:electroproduction} reviews the general expression for the cross-section of electroproduction from a nucleus.  In Sec. \ref{sec:kinematics} the kinematics for the case of zero and small relative momentum of the outgoing $J/\psi$-neutron pair is discussed.  In Sec. \ref{sec:invaramps} the calculation of the invariant amplitudes for $\gamma^*+D\to J/\psi +p+n$ are presented, including the one-loop diagrams corresponding to the $p-n$ and $J/\psi$-nucleon rescattering processes.  In order to calculate the amplitude corresponding to the low-energy $J/\psi$-neutron scattering, which involves the scattering length, model $J/\psi$-neutron scattering wavefunctions and potentials are used, and it is shown that the resulting amplitude is insensitive to the model used.  In addition, it is shown that for the kinematic conditions of the JLab experiment, the dominant amplitude is the impulse diagram, corresponding to $J/\psi$ production on the neutron with the proton recoiling freely, with no rescattering of any particles.  This demonstrates that the measurement of the $J/\psi$-nucleon scattering length is not feasible for the JLab experiment.  Finally, in Sec. \ref{sec:intermedenergy} calculations of the amplitude for $\gamma^*+D\to J/\psi +p+n$ are presented under different kinematic conditions (not restricting the outgoing $J/\psi$-neutron pair to small relative momentum).  There it is shown that if the $J/\psi$-neutron elastic scattering amplitude is somewhat larger than the value measured at SLAC at higher energy, it may be possible to extract this amplitude from the JLab experiment.

\section{Electroproduction from a Nucleus}
\label{sec:electroproduction}

We consider here electron scattering from the deuteron with production of a vector meson, with the final state of the proton-neutron system being a continuum state.  The formalism for the cross-section for electroproduction from a nucleus can be found in~\cite{Raskin:1988kc}.  Here we summarize the relevant facts.   We consider here the completely unpolarized electron (initial and final) cross-section.  Then the cross-section can be written in terms of the amplitudes for $\gamma^*+d\to p+n+V$, i.e. vector meson production from virtual photons.  In the LAB frame, with $\epsilon^\prime$ the final electron energy , $\Omega^\prime$ the final electron solid angle, $\Omega_V$ the vector meson solid angle , and $\mathbf{p}_{pn}^*$ the final proton-neutron relative momentum in the $p-n$ center-of-mass frame,    the 8-fold differential cross-section has the form
\be
\frac{d^8\sigma}{d\epsilon^\prime d\Omega^\prime d\Omega_V d^3p_{pn}^*}=(kinematic\;factors)\times (v_T R^T_{fi}+v_{TT} R^{TT}_{fi}+v_L R^L_{fi}+v_{TL} R^{TL}_{fi}),
\ee
where the factors
\be
 R^T_{fi}=\vert \langle f\vert J_{+1}(\mathbf{q})\vert i \rangle\vert^2+\vert \langle f\vert J_{-1}(\mathbf{q})\vert i \rangle\vert^2
\ee 
\be
 R^{TT}_{fi}=2 \operatorname{Re}\langle f\vert J^*_{+1}(\mathbf{q})\vert i \rangle \langle f\vert J_{-1}(\mathbf{q})\vert i \rangle
\ee 
\be
 R^{TL}_{fi}=-2 \operatorname{Re}\langle f\vert \rho^*(\mathbf{q})\vert i \rangle (\langle f\vert J_{+1}(\mathbf{q})\vert i \rangle - \langle f\vert J_{-1}(\mathbf{q})\vert i \rangle)
\ee 
\be
 R^{L}_{fi}=\vert\langle f\vert \rho(\mathbf{q})\vert i \rangle\vert^2, 
\ee 
are in terms of the matrix elements of the spherical vector components of the electromagnetic current operator $J$ between the initial deuteron state $\vert i\rangle$ and final hadron ($p+n+J/\psi$) state $\vert f\rangle$.  $v_T$, $v_{TT}$, etc., are kinematic factors that only depend on the electron momenta.  $ R^T_{fi}$ is the sum of the squares of the amplitudes for $\gamma^*+d\to p+n+V$ for transversely polarized virtual photons, whereas $R^{TT}_{fi}$ is an interference term between these two amplitudes.  $R^{L}_{fi}$ is the square of the amplitude for a longitudinally polarized virtual photon, while $ R^{TL}_{fi}$ is an interference term between the amplitudes for production from transverse and longitudinally polarized photons. The matrix element $R^{TL}_{fi}$ is proportional to $\cos{\phi}$, while $R^{TT}_{fi}$ is proportional to $\cos{2\phi}$, where $\phi$ is the angle between the plane including the initial and final electron momenta, and the plane including the 3-momentum transfer $\mathbf{q}$ and the $J/\psi$ momentum $\mathbf{p}_V$.  Thus if we integrate the cross-section over $\phi$, the terms $R^{TT}$ and $R^{TL}$ drop out.  Or, if we assume helicity conservation (i.e. the helicity of the outgoing $J/\psi$ is equal to the helicity of the photon) then $R^{TT}=R^{TL}=0$.  Moreover, several theoretical models~\cite{Dosch:2003dh,Fiore:2009xk} indicate that for small $Q^2$, the amplitude for $J/\psi$ electroproduction from transverse virtual photons is much larger than the amplitude for production from longitudinally polarized virtual photons; for $Q^2=0$ (photoproduction) the production amplitude for longitudinal photon polarization is of course exactly zero.  Therefore in what follows we will neglect  $R^{L}_{fi}$, and so the differential cross-section is simply given by $ R^T_{fi}$ multiplied by  kinematic factors.  Thus our task is to calculate
\be
 R^T_{fi}=\vert \langle f\vert J_{+1}(\mathbf{q})\vert i \rangle\vert^2+\vert \langle f\vert J_{-1}(\mathbf{q})\vert i \rangle\vert^2\equiv \vert F_+\vert^2 +\vert F_-\vert^2
\ee
where $F_\pm$ are the amplitudes for $J/\psi$ production from positive and negative helicity virtual photons.  In the following we will calculate the amplitude for $\gamma^*+d\to p+n+V$ by evaluating Feynman diagrams corresponding to the various processes contributing to it.

\section{Kinematics for small $J/\psi$-neutron relative momentum}
\label{sec:kinematics}

We will assume kinematics where the outgoing $J/\psi$ and neutron have a small relative momentum.   Since the scattering length is the zero-energy limit of the scattering amplitude, in order to measure it the relative momentum of the $J/\psi-n$ system must be small.  An estimate of how small can be obtained by requiring only S-wave scattering, meaning the contribution of higher partial waves should be negligible.  The classical relation between impact parameter and angular momentum yields an estimate for the maximum $l$ that contributes.  If the relative momentum of the $J/\psi-n$ pair is $p^*$ and the impact parameter is $b$, then the orbital angular momentum in the $J/\psi-n$ c.m. frame is $L\simeq p^*b=b\sqrt{2\mu T^*}$ where $T^*$ is the total kinetic energy in the 2-body c.m. frame and $\mu$ is the reduced mass.  The largest angular momentum partial wave which will be scattered is obtained by setting $b$ equal to the range of the potential.  With $L^2=l(l+1)$ (we take $\hbar=1$), the condition for only S-wave scattering is that $l\ll1$, which implies $L^2=b^2 2\mu T^*\ll 1$.   Taking the range of the interaction to be $\simeq 1\;fm$ yields $T^*\ll\;30 MeV$.

Experimentally, perhaps the simplest quantity to measure is the total production cross-section, integrated over all available phase space, for a given incident photon energy.  However, if we restrict the photon energy such that the maximum $J/\psi$-neutron c.m. kinetic energy $T^*_{max}$ is small in of all the available phase space, then that means that the maximum proton-neutron relative energy will also be small everywhere in the available phase space: for a given value of the Mandlestam variable $s$ for a system consisting of 3 particles, the total kinetic energy of any two of the particles (say 1 and 2) in their c.m. frame satisfies
\be
T^*_{12}\le \sqrt{s} - m_1-m_2-m_3.
\ee 
The low-energy $J/\psi-n$ scattering amplitude is expected to be much smaller than the low-energy $p-n$ scattering amplitude, and therefore  the $p-n$ rescattering would dominate over the $J/\psi-n$ rescattering, as contributions to the total production cross-section.  Thus we need to restrict our considerations to a kinematic range where the $p-n$ rescattering is at relatively high energy, while the $J/\psi-n$ rescattering is at very low energy, in order to have the possibility that the $J/\psi-n$ rescattering makes a noticeable contribution to the differential cross-section.

The ideal situation would be to have the final $J/\psi$ and neutron sitting at rest in the LAB, with the proton moving off at high velocity.  Such a final state is kinematically allowed for other reactions, e.g. $\pi^+ d\to\eta pp$, $\gamma^* d\to\eta p n$, but it is not possible for the reaction $\gamma^* d\to J/\psi\; p n$, for any real or virtual photon 4-momentum.

\begin{figure}[tbp]
     \begin{center}

            \includegraphics[width=4.5in,height=2.5in]{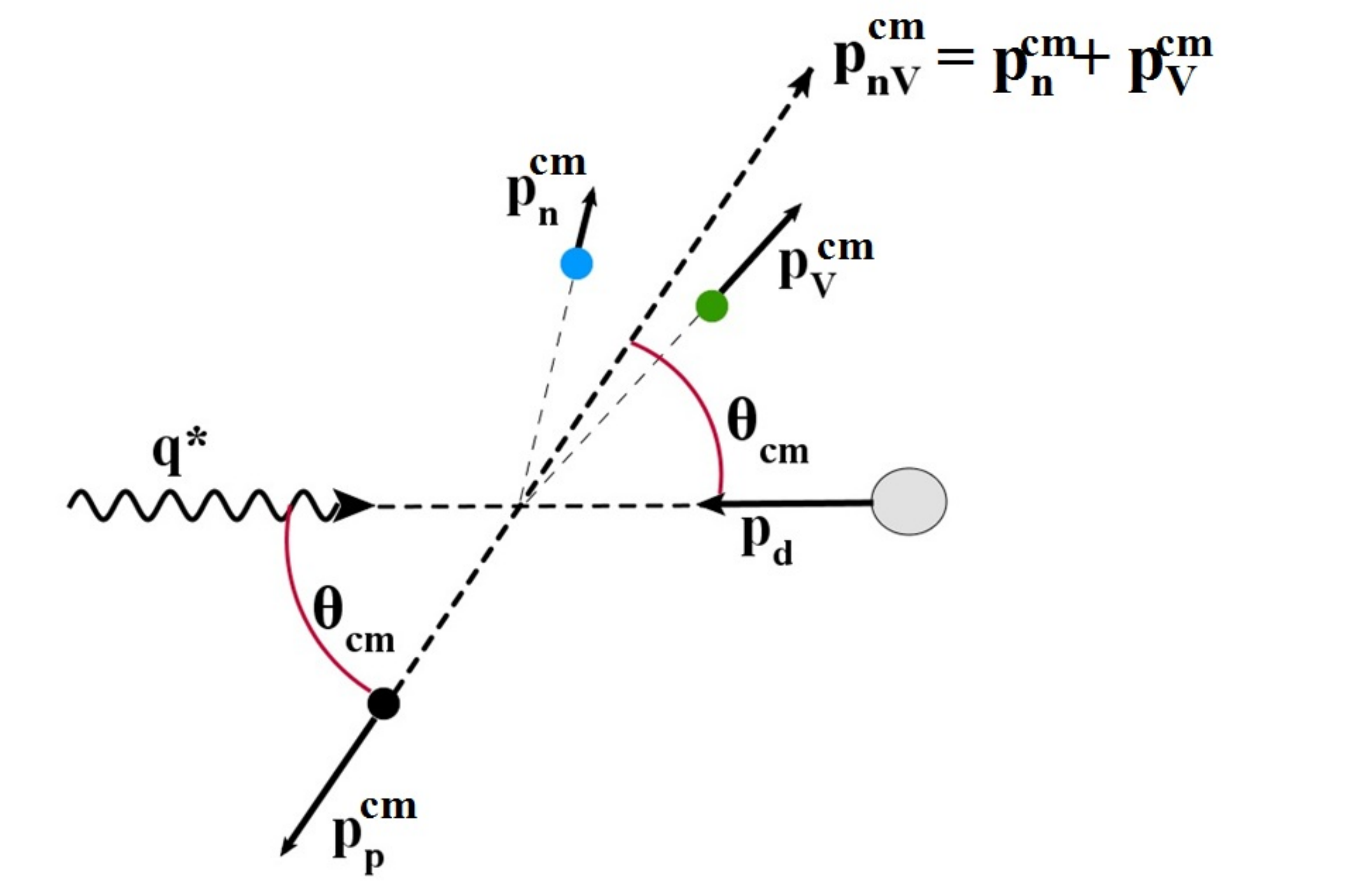}

    \end{center}
    \caption{%
        Momenta and angles in the overall c.m. frame, for coplanar kinematics.  
     }%
   \label{fig:cmkin}
\end{figure}

\begin{figure}[bth]
     \begin{center}
        \subfigure[  $\nu=9$ GeV]{%
            \label{fig:lab}
            \includegraphics[width=0.4\textwidth]{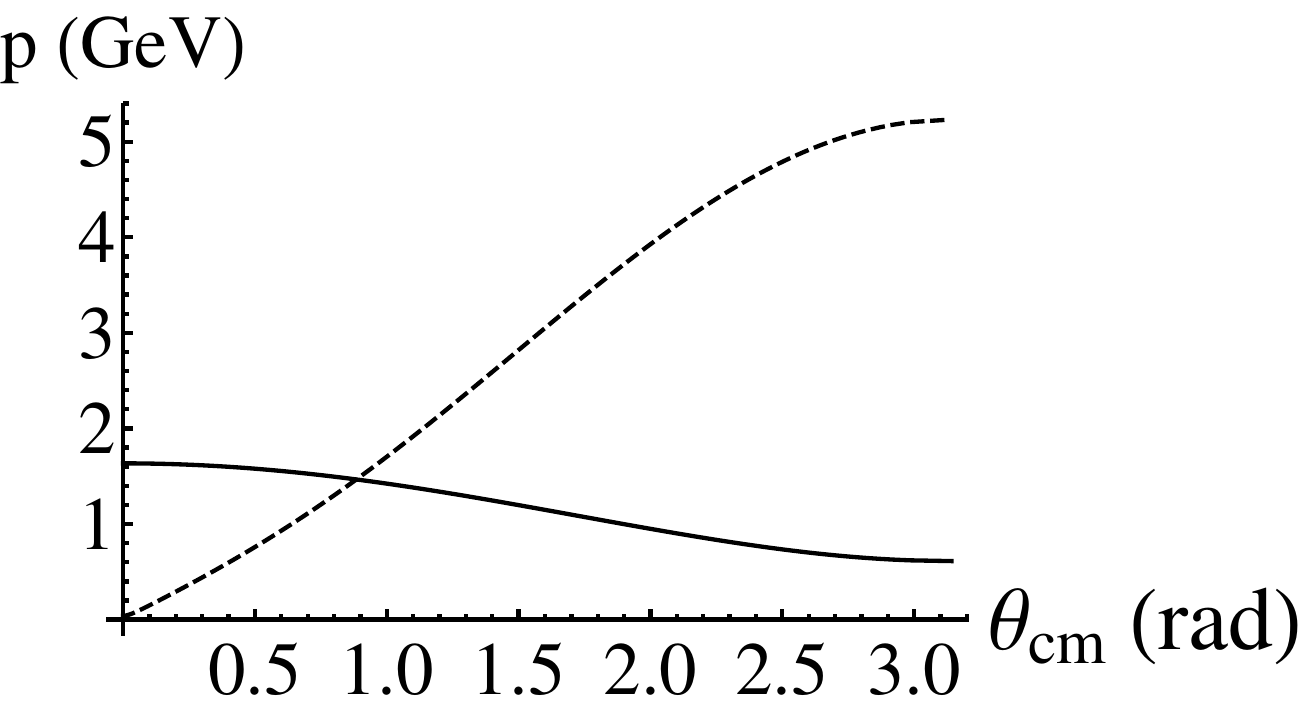}
        }%
        \subfigure[ $\nu=6.5$ GeV]{%
           \label{fig:cm}
           \includegraphics[width=0.4\textwidth]{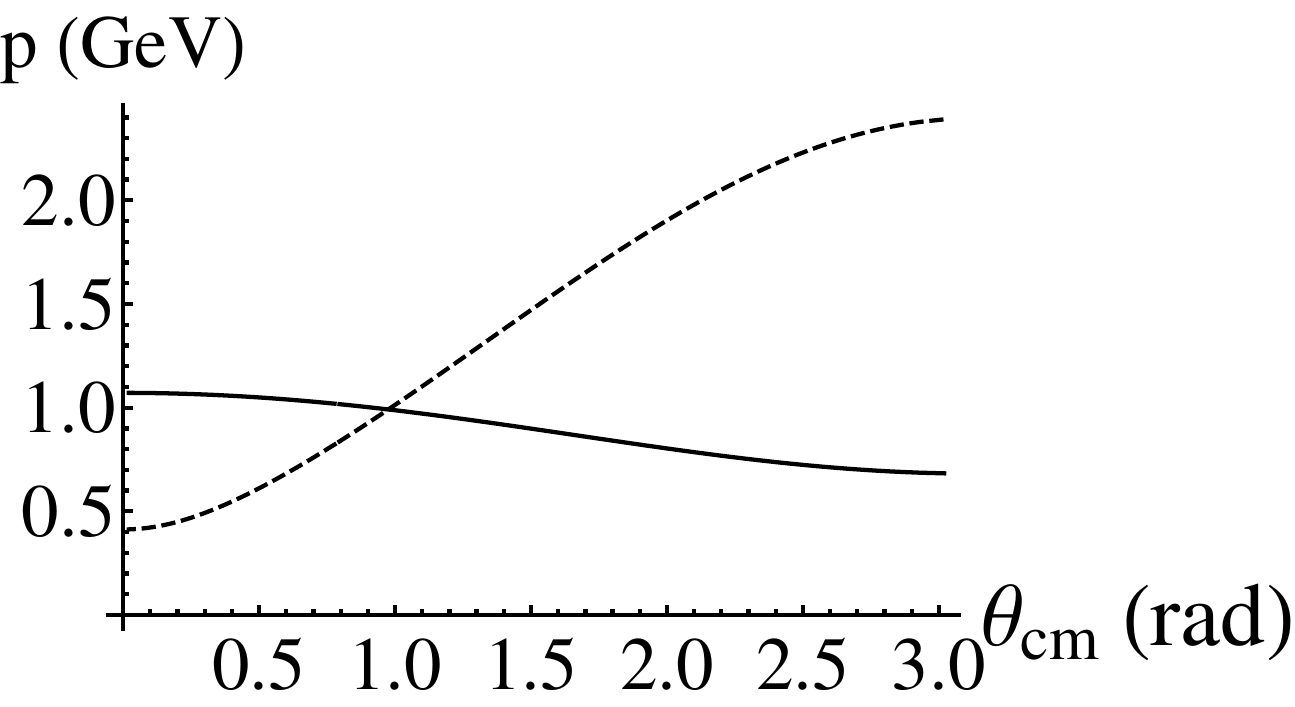}
        }\\ 

%
    \end{center}
    \caption{%
       Minimum possible neutron LAB momentum (solid curve), and the corresponding proton LAB momentum (dashed curve), vs. $\theta_{cm}$, for $T^*_{Vn}=30$ MeV and two values of photon LAB energy $\nu$.
     }%
   \label{fig:nandpmomentum}
\end{figure}

Fig. \ref{fig:nandpmomentum} shows the minimum possible outgoing neutron LAB momentum and the corresponding proton LAB momentum vs. $\theta_{cm}$ for fixed $T^*_{Vn}=30$ MeV (see Fig. \ref{fig:cmkin} for the definition of $\theta_{cm}$); graphs are shown for photon LAB energy $\nu=9$ GeV and for $\nu=6.5$ GeV.   One can see that the neutron's momentum is always greater than at least $0.6$ GeV.  Since the maximum nucleon momentum in the deuteron is around $0.3$ GeV (the deuteron momentum-space wavefunction is negligible for momenta larger than that) that means that in order for these final-state kinematics to occur,  the neutron must have acquired its large momentum through a scattering event.  In fact it will turn out that the dominant process corresponds to the impulse approximation wherein the $J/\psi$ is produced on the neutron itself, and the proton simply recoils freely.  For the kinematics of interest here, rescattering processes (e.g. $J/\psi$-neutron rescattering, $J/\psi$-proton rescattering, proton-neutron rescattering)  make  very small contributions to the total amplitude.              

We use the following notation throughout this paper:  $q=(\nu,\mathbf{q})$ is the virtual photon 4-momentum in the LAB, with $q^2=-Q^2<0$; $\mathbf{p}_p$ is the outgoing proton LAB 3-momentum; $\mathbf{p}_n$ is the outgoing neutron LAB 3-momentum; $\mathbf{p}_V$ is the $J/\psi$ LAB 3-momentum; and the same variables with $cm$ superscripts denote their values in the overall (3-body) center-of-mass frame.  $\theta_p$, $\theta_n$, and $\theta_V$ denote the angle that the outgoing proton, neutron, and $J/\psi$ momenta, respectively, make with $\mathbf{q}$, in the LAB frame.

\section{Invariant Scattering Amplitudes}
\label{sec:invaramps}

\begin{figure}[tbp]
     \begin{center}
        \subfigure[$F_{1a}$:  Impulse diagram]{%
            \label{fig:F1a}
            \includegraphics[width=0.4\textwidth]{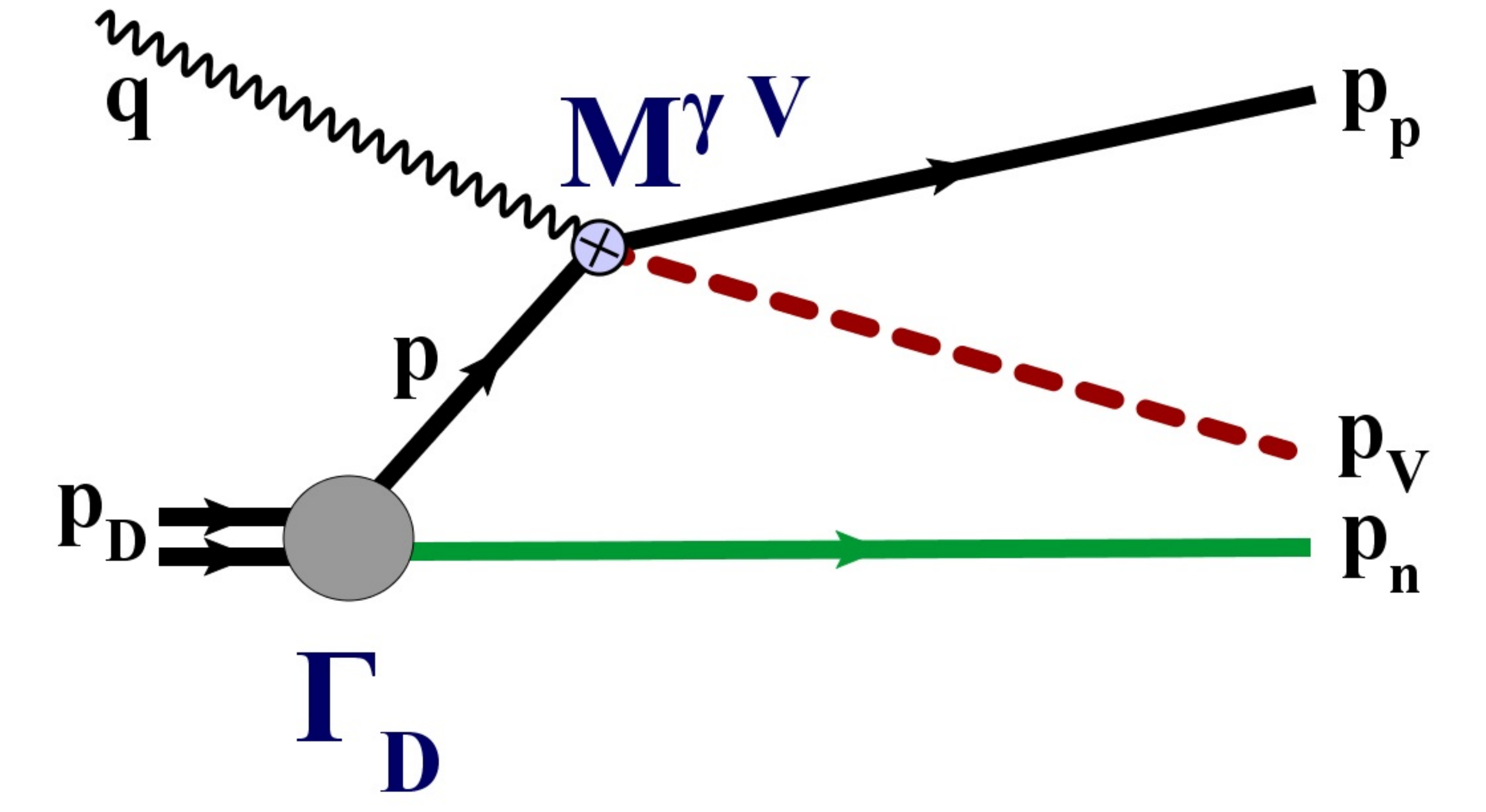}
        }%
        \hspace{0.5in}
         \subfigure[$F_{2a}$:  p-n rescattering diagram]{%
           \label{fig:F2a}
           \includegraphics[width=0.4\textwidth]{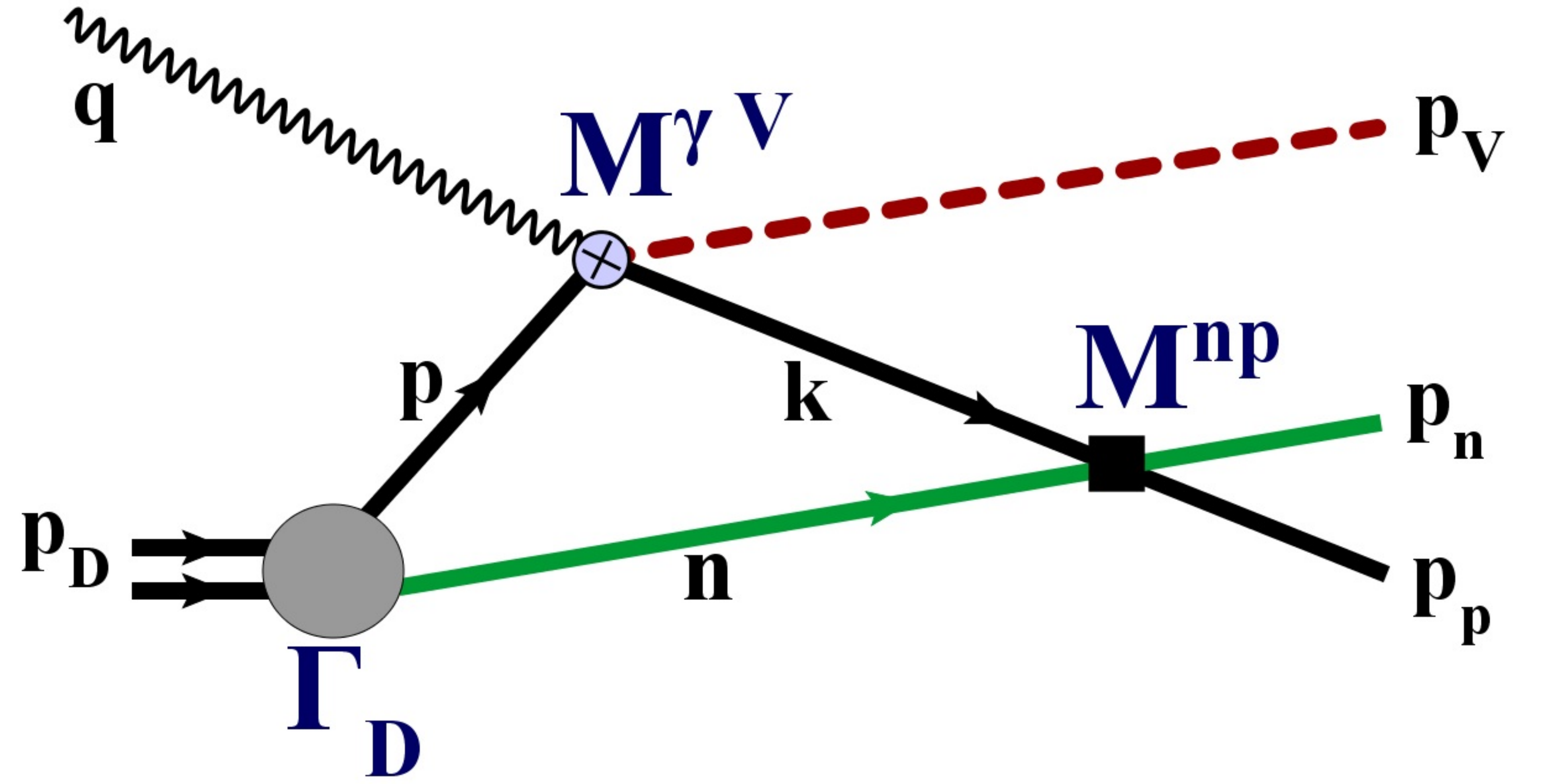}
        }\\ 

        \subfigure[ $F_{3a}$:  $J/\psi$-n rescattering diagram]{%
            \label{fig:F3a}
            \includegraphics[width=0.4\textwidth]{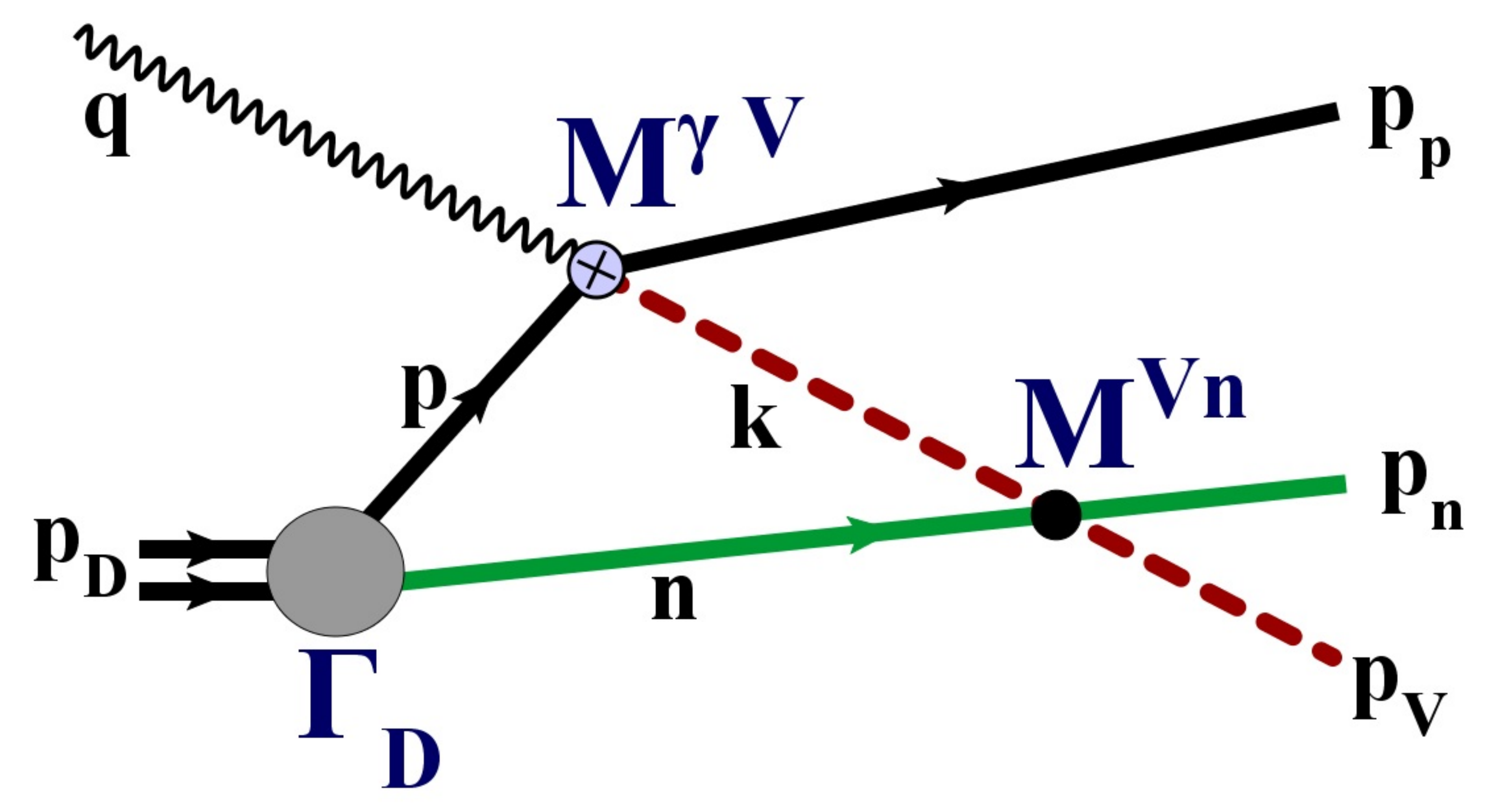}
        }%

    \end{center}
    \caption{%
       Feynman diagrams for $\gamma^*+d\to J/\psi + p+n$, for production on the proton.
     }%
   \label{fig:diagrams}
\end{figure}

\begin{figure}[tbp]
     \begin{center}
        \subfigure[$F_{1b}$:  Impulse diagram]{%
            \label{fig:F1b}
            \includegraphics[width=0.4\textwidth]{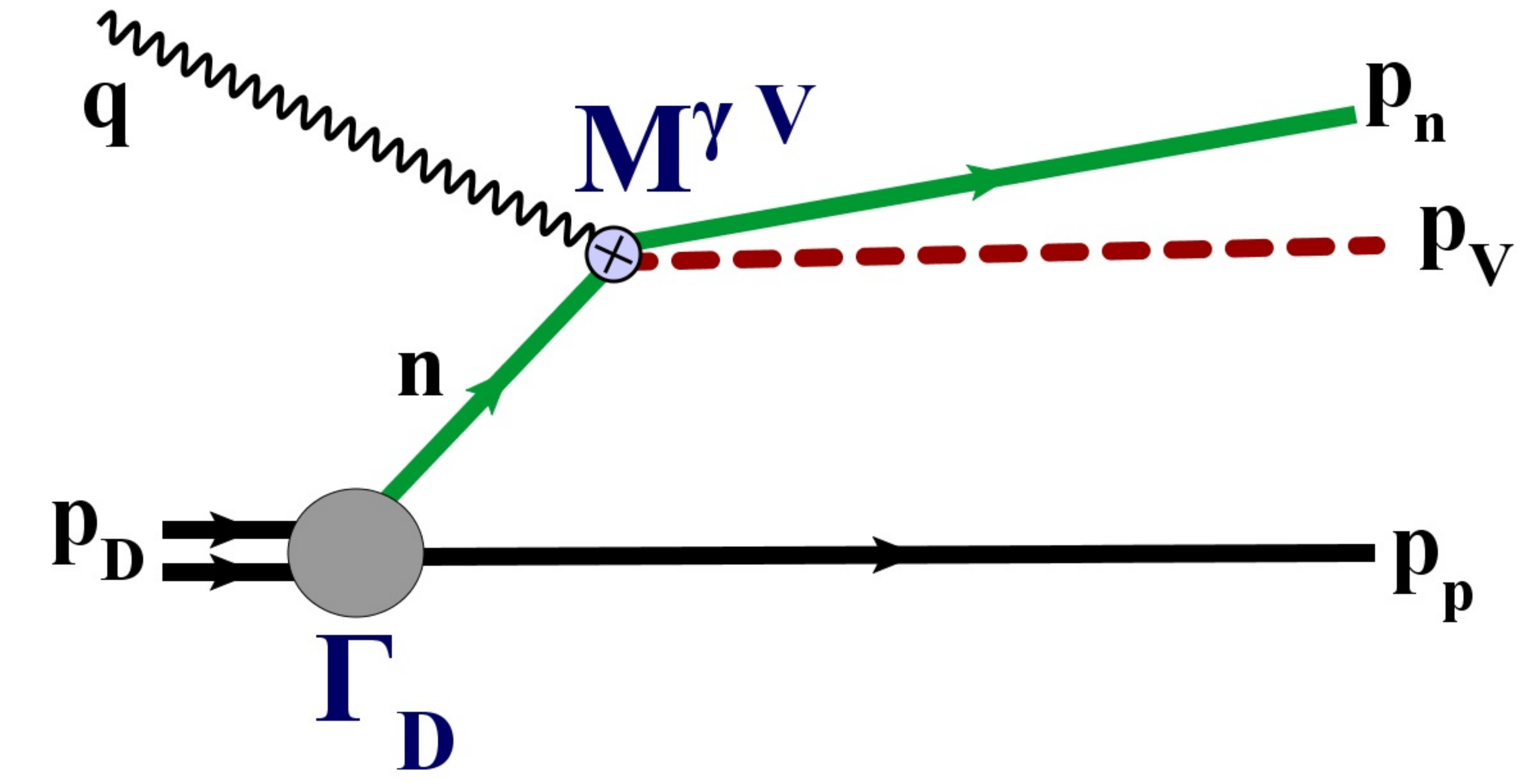}
        }%
        \hspace{0.5in}
         \subfigure[$F_{2b}$:  p-n rescattering diagram]{%
           \label{fig:F2b}
           \includegraphics[width=0.4\textwidth]{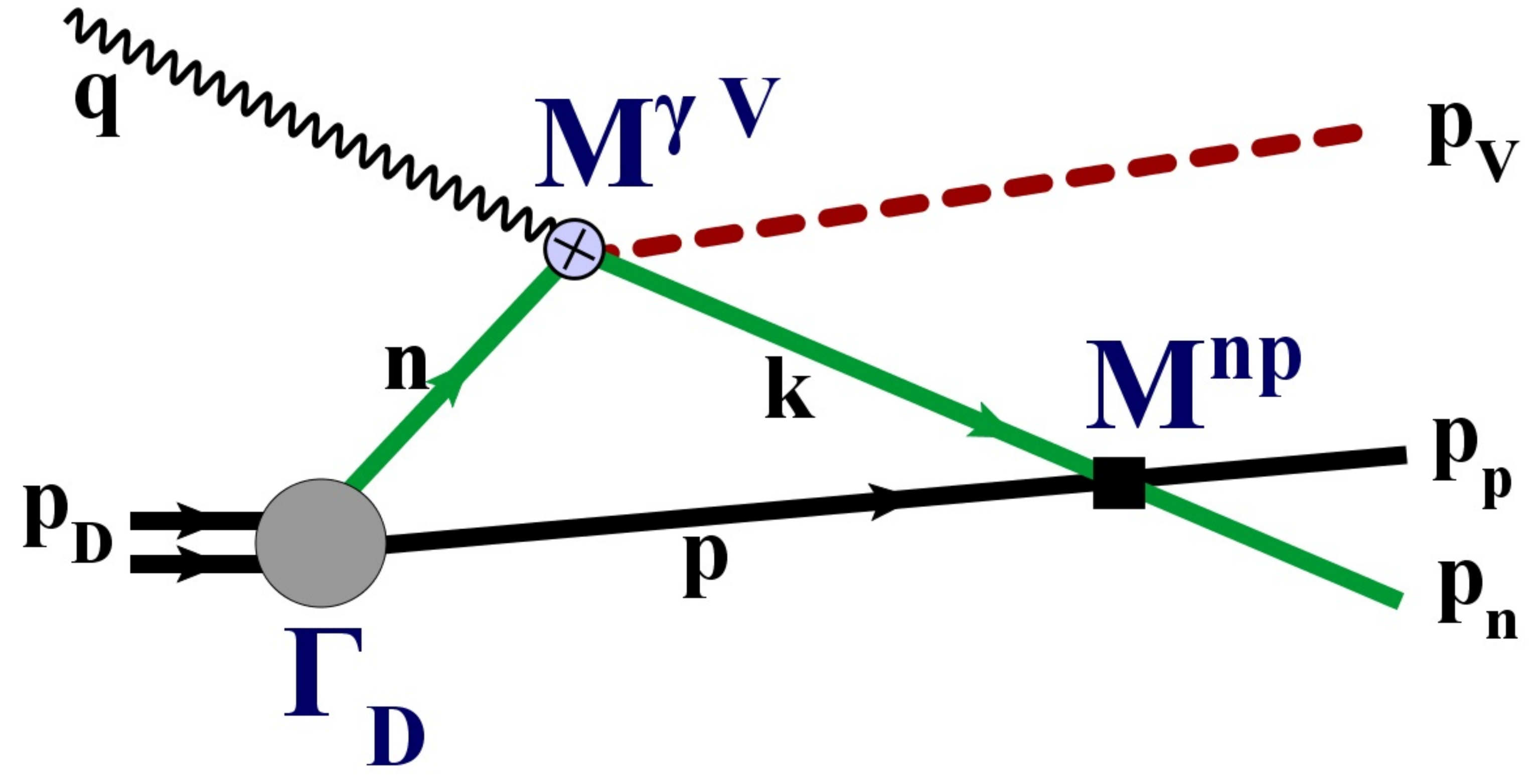}
        }\\ 

        \subfigure[ $F_{3b}$:  $J/\psi$-p rescattering diagram]{%
            \label{fig:F3b}
            \includegraphics[width=0.4\textwidth]{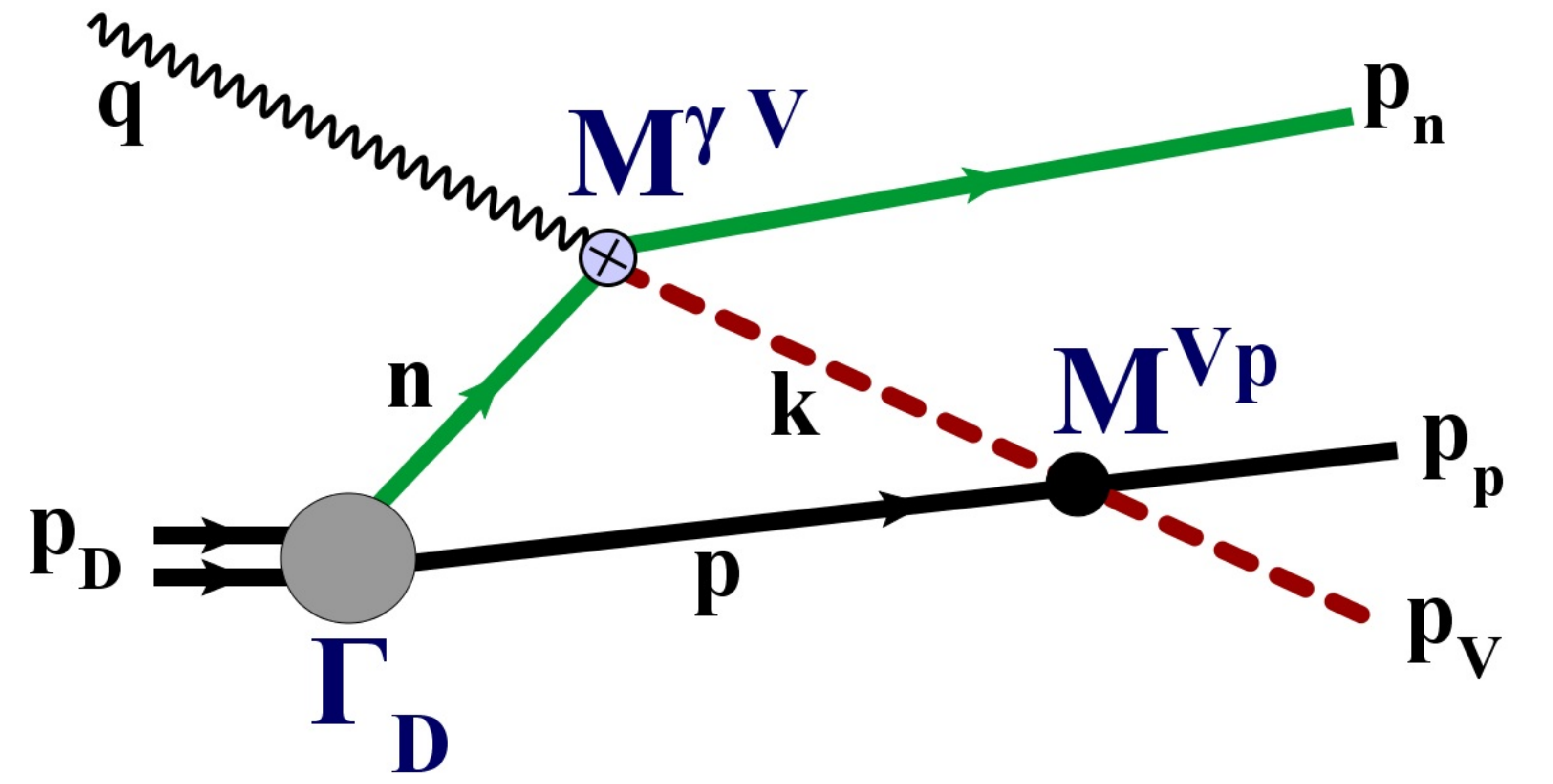}
        }%

    \end{center}
    \caption{%
       Feynman diagrams for $\gamma^*+d\to J/\psi + p+n$, for production on the neutron.
     }%
   \label{fig:diagrams2}
\end{figure}

The Feynman diagrams considered here are shown in Figs. \ref{fig:diagrams} and \ref{fig:diagrams2}.  There are 3 diagrams for production on the proton, and 3 similar diagrams where the $J/\psi$ is produced on the neutron.  In all cases we are interested in kinematics where the $J/\psi$ and neutron have small relative momentum.  The diagrams are covariant, and hence give Lorentz invariant amplitudes.  In the diagrams,   ${\cal M}^{\gamma V}$ is the Lorentz invariant amplitude for the quasi-2-body process $\gamma^*+N\to V+N$ (where $N$ is a nucleon, and $V$ stands for the $J/\psi$), while ${\cal M}^{V n}$ is the  Lorentz invariant amplitude for the elastic scattering process $V+n\to V+n$ (with $n$ meaning neutron), and ${\cal M}^{V p}$ and ${\cal M}^{np}$ are the same for elastic $J/\psi$-proton scattering and neutron-proton scattering, respectively.

\subsection{Impulse Diagrams}

Amplitudes $F_{1a}$ and $F_{1b}$ are the impulse diagrams, where the $J/\psi$ is produced on one of the nucleons and the other nucleon (the ``spectator") recoils freely without interacting with the other particles.  In $F_{1a}$ the vector meson is produced on the proton and the neutron is the spectator, while in $F_{1b}$ the production occurs on the neutron and the proton is the spectator.  The invariant amplitudes in this case are
\begin{equation}
F_{1a}={\cal M}^{\gamma V}(s_{1a},t_{1a})\;\frac{\Gamma_D(p)}{D(p)}\end{equation} 
\begin{equation}
F_{1b}={\cal M}^{\gamma V}(s_{1b},t_{1b})\;\frac{\Gamma_D(n)}{D(n)}.\end{equation} 
Here ${\cal M}^{\gamma V}$ is the Lorentz invariant amplitude for the quasi-2-body process $\gamma^*+N\to V+N$ (where $N$ is a nucleon, and $V$ stands for the $J/\psi$), $\Gamma_D$ is the covariant vertex function for the virtual dissociation $D\to p+n$, and $D(p)$ is the propagator denominator for the intermediate-state nucleon, $D(p)\equiv-p^2+m^2-i\epsilon$.  $s_{1a},t_{1a},s_{1b}$, and $t_{1b}$ are the Mandlestam variables for the 2-body production process $\gamma^*+N\to V+N$.        Evaluated in the LAB frame, and neglecting any contributions to the deuteron vertex from antinucleons, the deuteron vertex function is related to the nonrelativistic deuteron wavefunction by~\cite{gross80}
\begin{equation}
\label{wfandvertex}
\psi_D(\mathbf{k}_{rel})=\frac{-\Gamma_D(p)}{\sqrt{2p^0 (2\pi)^3}\;D(p)},
\end{equation}
where in the LAB frame, $\mathbf{k}_{rel}=\mathbf{p}=-\mathbf{p}_n$ for $F_{1a}$, and $\mathbf{k}_{rel}=\mathbf{p}_p=-\mathbf{n}$ for $F_{1b}$ ($\mathbf{k}_{rel}$ is the proton's momentum inside the deuteron, in the LAB frame, for both).
In terms of the deuteron wavefunction, the amplitudes are thus:
\begin{equation}
F_{1a}=-{\cal M}^{\gamma V}(s_{1a},t_{1a})\;\psi_D(-\mathbf{p}_{n})\sqrt{2m (2\pi)^3}\end{equation}
\begin{equation}
F_{1b}=-{\cal M}^{\gamma V}(s_{1b},t_{1b})\;\psi_D(\mathbf{p}_{p})\sqrt{2m (2\pi)^3}.\end{equation}
The amplitudes ${\cal M}^{\gamma V}$ used in calculations are to be taken from experimental data on $J/\psi$ production on a single nucleon; we assume here that the amplitudes for production from a neutron is the same as for production from a proton.

In the above expressions for the amplitudes $F_{1a}$ and $F_{1b}$, spin labels have been suppressed.  The initial virtual photon and the deuteron are in specific spin states, the final hadrons are in specific spin states, and there is a sum over the spin states of the intermediate-state virtual nucleon.  For example, the amplitude $F_{1b}$, including spin state specification, is explicitly:
\begin{equation}
F_{1b}=-\sqrt{2m (2\pi)^3}    \sum_{m_1}{\cal M}^{\gamma V}(m_1,\lambda,m_n,\lambda_V)\;\psi_D^M(\mathbf{p}_{p},m_1,m_p),
\end{equation}
where $m_1$ is the spin state of the intermediate-state neutron (i.e. the line with momentum $n$ in the Feynman diagram), $m_n$ and $m_p$ are the spin states of the final neutron and proton, $\lambda$ is the photon polarization, $\lambda_V$ is the $J/\psi$ polarization, and $M$ is the deuteron spin state.  In what follows we will assume that the 2-body amplitudes for spin flip are negligible compared to the non-spin flip amplitudes, and so the amplitudes are diagonal in the nucleon spin, and also in the photon and $J/\psi$ spin.    In that case we are able to calculate the spin-averaged squares of the various amplitudes $F_{1a}$, $F_{2a}$, etc., and the spin-averaged square of the total amplitude.  We have included the contribution from the $D$-state in the deuteron wavefunction. For $\nu=9$ GeV the $D$-state was found to not make a significant contribution to the amplitudes, but for $\nu=6.5$ GeV the $D$-state did contribute significantly, especially for the impulse diagram $F_{1b}$.  The deuteron wavefunction used was the Argonne $v18$ wavefunction.

All 2-body amplitudes ${\cal M}^{\gamma V}$, ${\cal M}$ are related to the corresponding 2-body differential cross-section by 
\begin{equation}
\label{crossandM}
\frac{d\sigma}{dt}=\frac{1}{16\pi\lambda(s,m_1^2,m_2^2)}\vert{\cal M}\vert^2,
 \end{equation}
where the flux factor $\lambda$ is given in terms of the incident particle masses $m_1$ and $m_2$ by
\begin{equation}
\lambda(s,m_1^2,m_2^2)=(s-m_1^2-m_2^2)^2-4m_1^2m_2^2,\end{equation}
and $s$ and $t$ are the Mandelstam variables for the 2-body process.

\subsubsection{Parameterization of the  amplitudes ${\cal M}^{\gamma V}$}

If the cross-section for $J/\psi$ production on a single nucleon is parametrized as
\begin{equation}
\frac{d\sigma}{dt}=A_1e^{B_1t},  \end{equation}
with the parameters $A_1$, $B_1$ dependent on energy (in principle), then the elementary production amplitude ${\cal M}^{\gamma V}$ is given by 
\begin{equation}
{\cal M}^{\gamma V}=-i\;\sqrt{16\pi A_1 \lambda(s,-Q^2,m^2)}e^{\frac{1}{2}B_1t},
 \end{equation}
where $s$ and $t$ are either $s_{1a}$, $t_{1a}$ or $s_{1b}$, $t_{1b}$.

The parameters $A_{1}$ and $B_{1}$ that are needed for the elementary $J/\psi$ production amplitude ${\cal M}^{\gamma V}$ were estimated from the (scant) existing data on exclusive $J/\psi$ production on a nucleon.  The only available data for the incident photon energy $\nu\simeq 10\;GeV$ is from a photoproduction experiment at Cornell in 1975~\cite{psidata75}.  For $\nu$ in the range 9.3 to 10.4 GeV, they determined $A_{1}=1.1\pm0.17\;nb/GeV^2=(2.8\pm0.43)\times 10^{-6}\;GeV^{-4}$ and $B_{1}=1.31\pm0.19\;GeV^{-2}$.  Those are the values used in this analysis.

\subsubsection{$F_{1a}$ and $F_{1b}$}

Since $F_{1a}$ is proportional to $\Psi(\mathbf{p}_n)$, and as seen in Fig. \ref{fig:nandpmomentum}  the outgoing neutron's momentum is always greater than $\simeq 0.6\;GeV$, the amplitude $F_{1a}$ will therefore be very small, since the deuteron wavefunction is negligible for those values of momentum.  The amplitude $F_{1b}$, on the other hand, is proportional to $\Psi(\mathbf{p}_p)$; thus as seen in Fig. \ref{fig:nandpmomentum} for $\theta_{cm}<0.3$ rad $F_{1b}$ should be non-negligible since the proton momentum is less than $0.4$ GeV over that range of $\theta_{cm}$.

\subsection{One-loop diagrams}
\begin{figure}[tbp]
     \begin{center}

            \includegraphics[width=4.5in,height=2.5in]{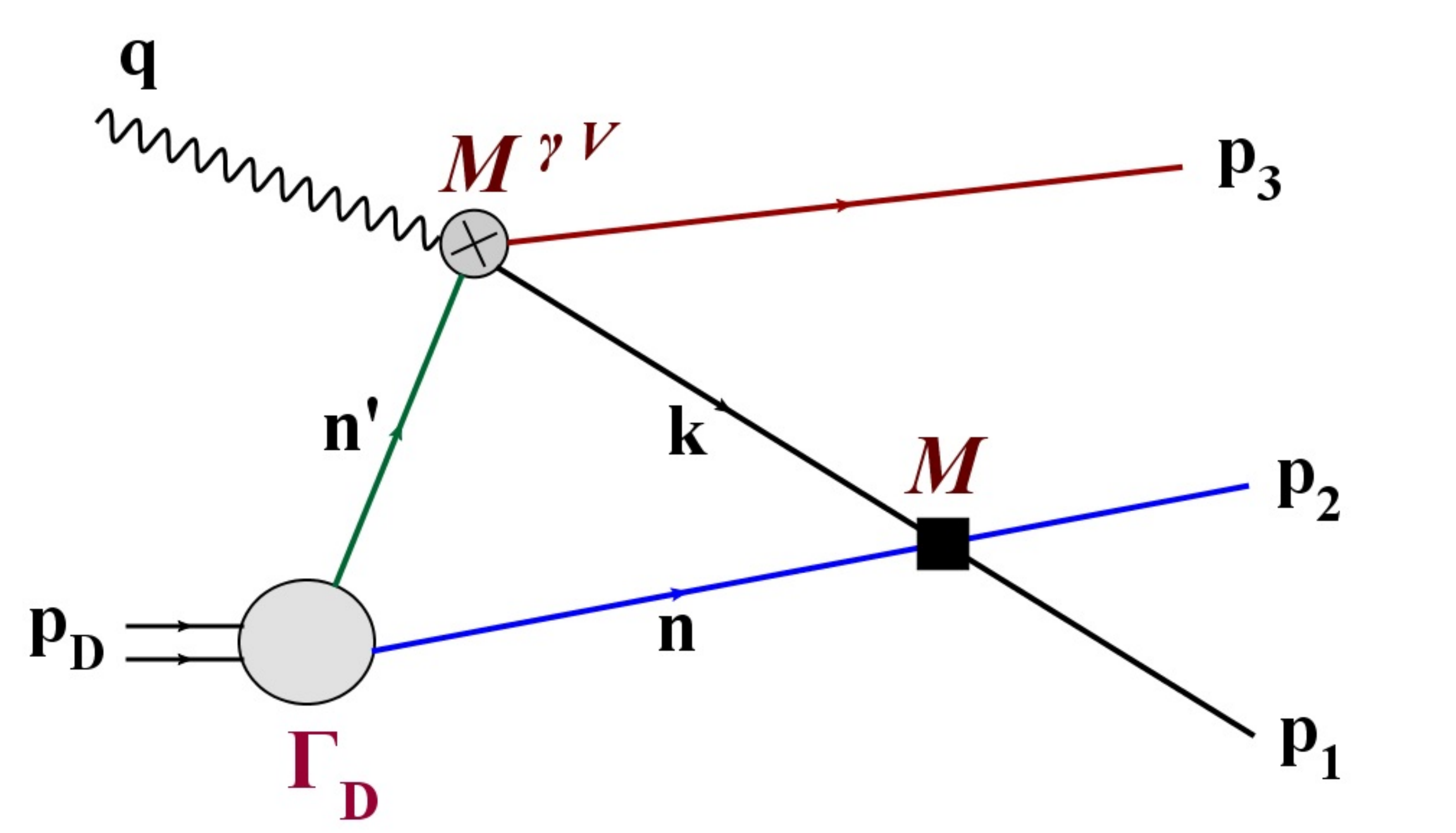}

    \end{center}
    \caption{%
        General one-loop diagram. $n$ and $p_2$ are the same particle (either neutron or proton).  The line $n$ can be either a proton or a neutron.
     }%
   \label{fig:oneloopdiag}
\end{figure}

The covariant expression for a general one-loop diagram (see Fig. \ref{fig:oneloopdiag} ) is
\be
\label{eq:generaloneloop}
F=-\int \frac{d^4n}{i(2\pi)^4} \frac{\Gamma(n^\prime)}{D(n^\prime)}\frac{{\cal M}^{\gamma V}{\cal M}}{D(n)D(k)}
\ee
where $n^\prime,\;n,\;k$ are the internal 4-momenta indicated in the figure, ${\cal M}$ stands for either ${\cal M}^{pn}$,  ${\cal M}^{Vn}$, or ${\cal M}^{Vp}$ (elastic scattering amplitude for proton-neutron, V-neutron, or V-proton scattering, respectively), and $D(p)=p^2-m_p^2+i\epsilon$, etc., are propagator denominators.  Spin labels have been suppressed in \eq{eq:generaloneloop}; in particular, there is an implicit sum over the spin states of the intermediate-state particles (the lines labelled $n$, $k$, and $n^\prime$).  There are 4 diagrams total for a given set of outgoing proton, neutron and $J/\psi$ momenta.  Taking $\mathbf{p}_2=\mathbf{p}_n$, $\mathbf{p}_1=\mathbf{p}_p$ (so that the internal line $n$ is the neutron, and $n^\prime$ and $k$ are the proton) gives one diagram (p-n rescattering diagram).  
The other 3 are:   $\mathbf{p}_2=\mathbf{p}_n$, $\mathbf{p}_1=\mathbf{p}_V$, where the internal line $n$ is the neutron, $k$ is the $J/\psi$, and $n^\prime$ and $p_3$ are the proton (V-n rescattering diagram);  $\mathbf{p}_2=\mathbf{p}_p$, $\mathbf{p}_1=\mathbf{p}_n$, where the internal line $n$ is the proton, and $n^\prime$ and $k$ are the neutron (another p-n rescattering diagram); and  $\mathbf{p}_2=\mathbf{p}_p$, $\mathbf{p}_1=\mathbf{p}_V$, where the internal line $n$ is the proton, $k$ is the $J/\psi$, and $n^\prime$ and $p_3$ are the neutron (V-p rescattering diagram).

All of the one-loop diagrams can be evaluated in the same manner.  We follow closely the method of~\cite{laget81}.  We first integrate over $n^0$ by taking the positive-energy pole at $n^0=\omega_{n}-i\epsilon$, where $\omega_{n}=\sqrt{m^2+\bf{n}^2}$, coming from the propagator denominator $D(n)$.  This corresponds to neglecting antinucleon components of the deuteron wavefunction.  We also use the relation  \eq{wfandvertex} (and we work in the rest frame of the deuteron).   This gives
\be
\label{fullamp1}
F=\sqrt{\frac{2m}{(2\pi)^3}}\int \frac{d^3n}{2\omega_{n}}\Psi(\mathbf{n})\frac{{\cal M}^{\gamma V}{\cal M}}{D(k)}.
\ee

Note that in this expression, the internal nucleon line $n$ is now on-mass-shell, since $n^0=\omega_{n}$.  Thus in the time-ordered diagram of Fig. \ref{fig:oneloopdiag}, only the lines $n^\prime$ and $k$ can be off-shell; all the rest are on-shell.

\begin{figure}[tbp]
     \begin{center}

            \includegraphics[width=3.5in,height=2in]{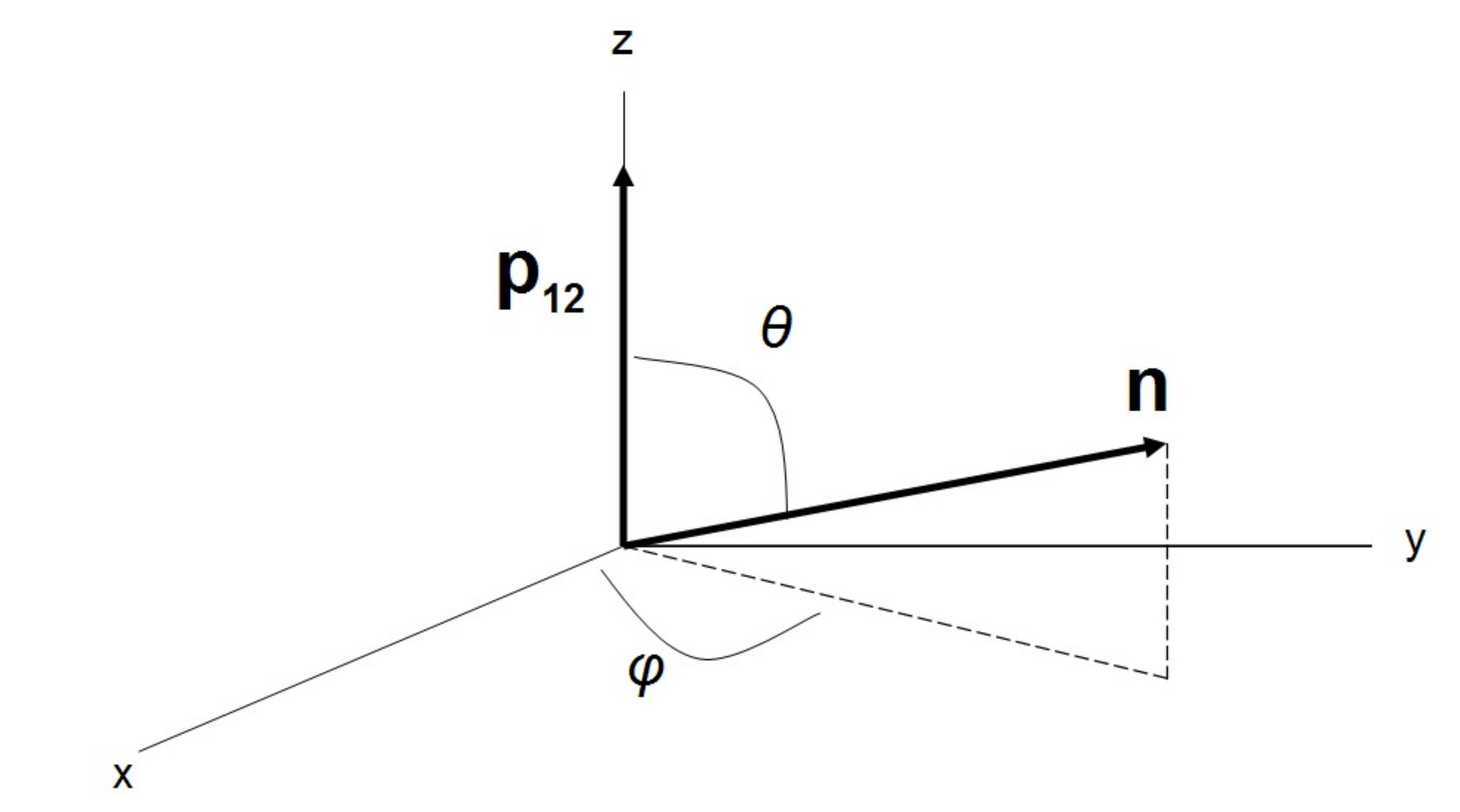}

    \end{center}
    \caption{%
        Coordinate system used.  For the on-shell amplitude, $\theta$ is fixed for a given $n\equiv\vert\mathbf{n}\vert$ and $\mathbf{p}_1$, $\mathbf{p}_2$.
     }%
   \label{fig:coordsystem}
\end{figure}

The above expression for the amplitude $F$ can be separated into two terms, one term in which the line $k$ is on-mass-shell and one in which $k$ is off-mass-shell, by using the relation
\be
\frac{1}{x+i\epsilon}=-i\pi\delta(x)+{\cal P }\frac{1}{x}
\ee
for the remaining propagator in \eq{fullamp1} (with ${\cal P}$ representing the principal value) and by choosing the coordinate system shown in Fig. \ref{fig:coordsystem} where the $n_{z}$ axis is along the direction of the vector $\mathbf{p}_{12}\equiv\mathbf{p}_1+\mathbf{p}_2$.  The delta function term gives the on-mass-shell part of the amplitude, and the principal value term gives the off-mass-shell part.
This separation is useful since the elementary amplitudes ${\cal M}^{\gamma V}$, ${\cal M}$ can be determined (at least in principle) directly from experimental data on the relevant 2-body scattering processes only when all 4 particles involved (2 initial and 2 final) are on-mass-shell.  This is not the case if one of the particles involved is off-mass-shell.  But if for reasonable choices of the off-mass-shell amplitudes the off-mass-shell part is small compared to the on-shell part, then the off-mass-shell part will not play an important role.  The result for the on-shell and off-shell parts of the one-loop amplitude is~\cite{laget81}:
\be
F=F^{on}+F^{off}
\ee
where
\be
\label{Fon}
F^{on}=-i\pi \frac{1}{\sqrt{2m(2\pi)^3}}\frac{1}{2\vert\mathbf{p}_{12}\vert}\int_0^{2\pi}d\phi\int_{\vert n_-\vert}^{n_+}dn\; n\;\Psi(n) {\cal M}^{\gamma V} {\cal M}
\ee
and
\be
\label{Foff}
F^{off}=\frac{1}{\sqrt{2m(2\pi)^3}}\frac{1}{2\vert\mathbf{p}_{12}\vert}\int_0^{2\pi}d\phi\int_0^{\infty}dn\; n\;\Psi(n)\;\;{\cal P}\int d\cos\theta \frac{  {\cal M}^{\gamma V} {\cal M} }{f_{12}(n)+\cos{\theta}}.
\ee
Here 
\be
\label{eq:f12}
f_{12}(n)\equiv\frac{        s_{12}+m^2-m_k^2-2E_{12}\omega_{n}}{    2n\;\vert\mathbf{p}_{12}\vert },
\ee
with $n\equiv \vert \mathbf{n}\vert$.
$s_{12}\equiv(p_1+p_2)^2$ is the Mandelstam $s$-variable for the elastic scattering of particles 1 and 2, $m_k$ is the mass of the real particle which the line $k$ represents, $E_{12}\equiv E_1+E_2$, and $\theta$ is as shown in Fig. \ref{fig:coordsystem}.  Note that $f_{12}$ is independent of $\theta$ and $\phi$.

In the amplitude $F^{on}$, the limits of integration $\vert n_-\vert$ and $n_+$ are the solutions of
\be
f_{12}(n_{\pm})^2=1,
\ee
which are
\be
\label{nplusminus}
n_{\pm}=\frac{E_2^*}{\sqrt{s_{12}}}\vert\mathbf{p}_{12}\vert \pm \frac{p_2^*}{\sqrt{s_{12}}}E_{12}
\ee
where $p_2^*$, $E_2^*$ are the momentum and energy of outgoing particle 2 in the c.m. frame of particles 1 and 2, and particle 2 is the \underline{same} particle as the internal line with momentum $\mathbf{n}$.  
The range of $n$ given by  $\vert n_-\vert\le n \le n_+$ is the range of $n$ for which it is kinematically possible for the line $k$ to be on-mass-shell (given that $n$, $p_1$, and $p_2$ are on-shell).  In $F^{on}$ the value of $\cos\theta$ is fixed at
\be
\label{costheta}
\cos\theta=-f_{12}(n)=-\frac{        s_{12}+m^2-m_k^2-2E_{12}\omega_n}{    2\vert\mathbf{p}_{12}\vert n}.
\ee
The amplitudes $ {\cal M}^{\gamma V},\; {\cal M}$ in $F^{on}$ are evaluated, for a given $n$ and $\phi$, at this value of $\cos\theta$.  The amplitude ${\cal M}$ in $F^{on}$ is  fully on-shell, i.e. all 4 particle lines $n$, $k$, $p_1$, and $p_2$ are on-mass-shell.  The amplitude ${\cal M}^{\gamma V} $ has only one particle off-shell ($n^\prime$), but given that the magnitude of $\mathbf{n}$ is small (due to the deuteron wavefunction), $n^\prime$ is almost on-shell:  $n^{\prime 0}=M_d-\omega_n\simeq M_d-m\simeq m$, and so $(n^\prime)^2=m^2+{\cal O}(\frac{n^2}{m^2})$.

In the amplitude $F^{off}$, $k$ is never on-mass-shell; the principal value imposes this, since for $k$ to be on-mass-shell, $\cos\theta$ must equal $-f_{12}(n)$, which never occurs in the principal value.  Thus the amplitudes  $ {\cal M}^{\gamma V},\; {\cal M}$ that enter into $F^{off}$ have either one particle off-mass-shell (for ${\cal M}$) or two particles off-shell (for  $ {\cal M}^{\gamma V}$).  In our calculations the amplitude $F^{off}$ was much smaller than $F^{on}$, for any of the Feynman diagrams in Fig. \ref{fig:diagrams} and \ref{fig:diagrams2} (except for $F_{3a}$),  for the kinematics considered here, and for reasonable expressions for the off-shell amplitudes $ {\cal M}^{\gamma V},\; {\cal M}$.

\subsection{General features of the one-loop diagrams}

Inspection of \eq{Fon} shows  that a necessary condition for the on-shell part of a given one-loop diagram to be non-negligible is that the corresponding $\vert n_-\vert$ must be small enough so that the range of integration in \eq{Fon} includes the momenta where the deuteron wavefunction is significant.   Since $n_-$ depends on the final-state kinematics (\eq{nplusminus}), graphs of $n_-$ as a function of one of the final state variables reveal regions where the on-shell amplitude can be significant, and regions where it will be negligible.

\begin{figure}[tbp]
     \begin{center}
        \subfigure[$\nu=9$ GeV]{%
            \label{fig:nminusplus3a1a}
            \includegraphics[width=0.4\textwidth]{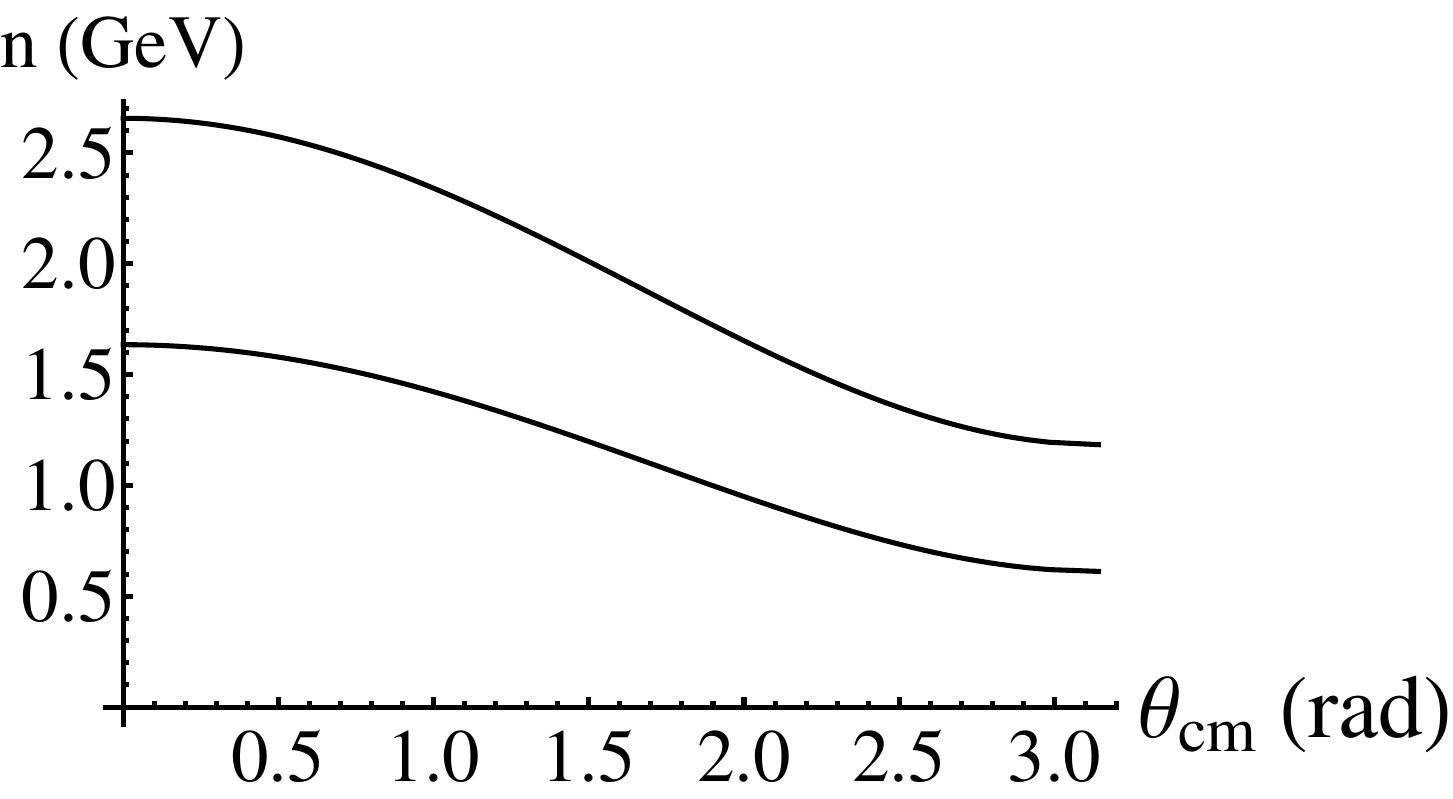}
        }%
        \hspace{0.5in}
         \subfigure[$\nu=6.5$ GeV]{%
           \label{fig:nminusplus3a1b}
           \includegraphics[width=0.4\textwidth]{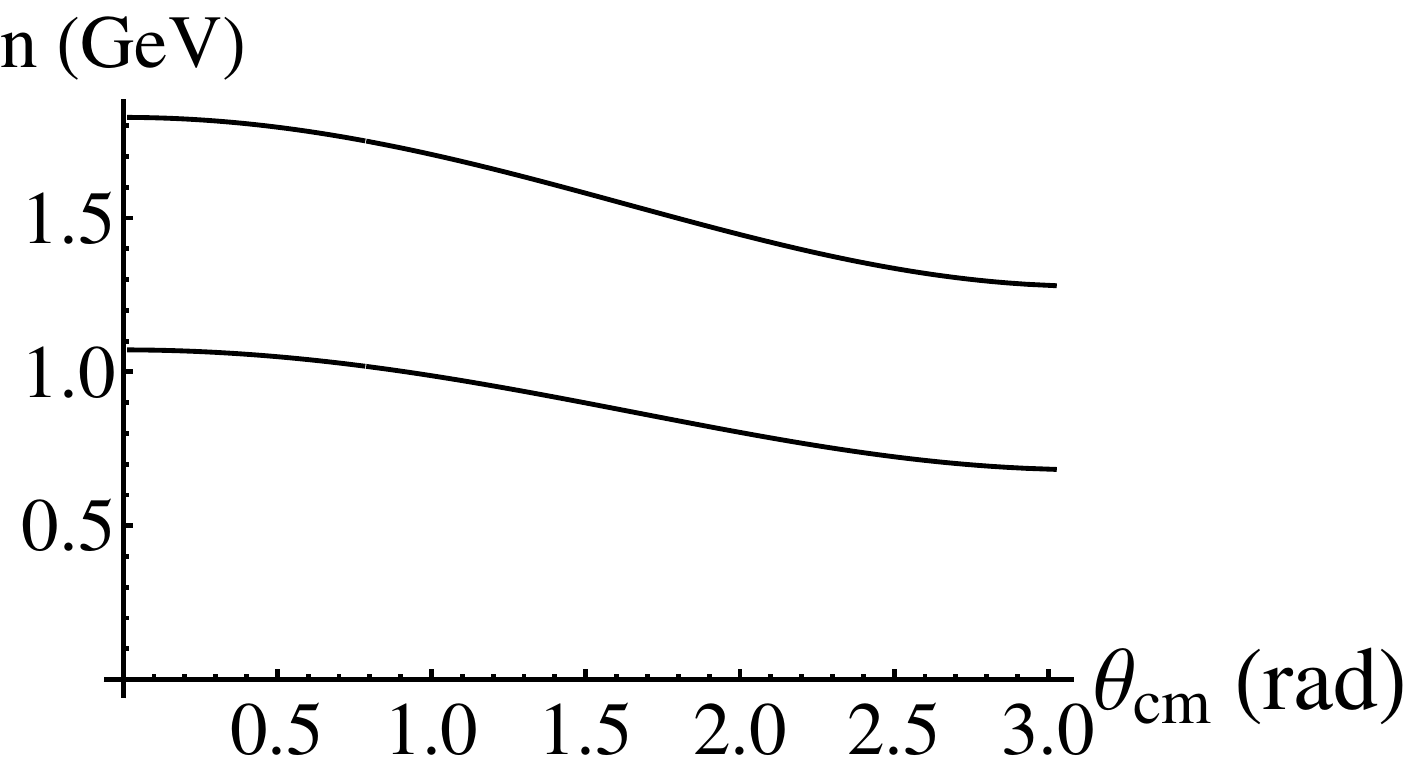}
        }\\ 


    \end{center}
    \caption{%
         $n_{\pm}$ vs. $\theta_{cm}$ for the amplitude $F_{3a}$ ($J/\psi$-neutron rescattering), for $T^*_{Vn}=30\;MeV$, and $\nu=9.0\;GeV$ and $\nu=6.5\;GeV$.  Upper curve is $n_+$, lower curve is $n_-$.
     }%
   \label{fig:nminusplus3a1}
\end{figure}

For the diagram of most interest, $F_{3a}$, particles $1$ and $2$ are the $J/\psi$ and neutron, respectively.    Fig. \ref{fig:nminusplus3a1} shows $n_{\pm}$ vs. $\theta_{cm}$ for $T^*_{Vn}=30\;MeV$, for $\nu=9\;GeV$, and Fig. \ref{fig:nminusplus3a1b} shows the same for $\nu=6.5\;GeV$, for the diagram $F_{3a}$.   At both of these photon energies, $n_-$ is greater than $\sim0.6\;GeV$ for all $\theta_{cm}$, and so the on-shell amplitude will be negligible since the deuteron wavefunction is negligible for that momentum.  

\begin{figure}[tbp]
     \begin{center}

            \includegraphics[width=3.5in,height=2in]{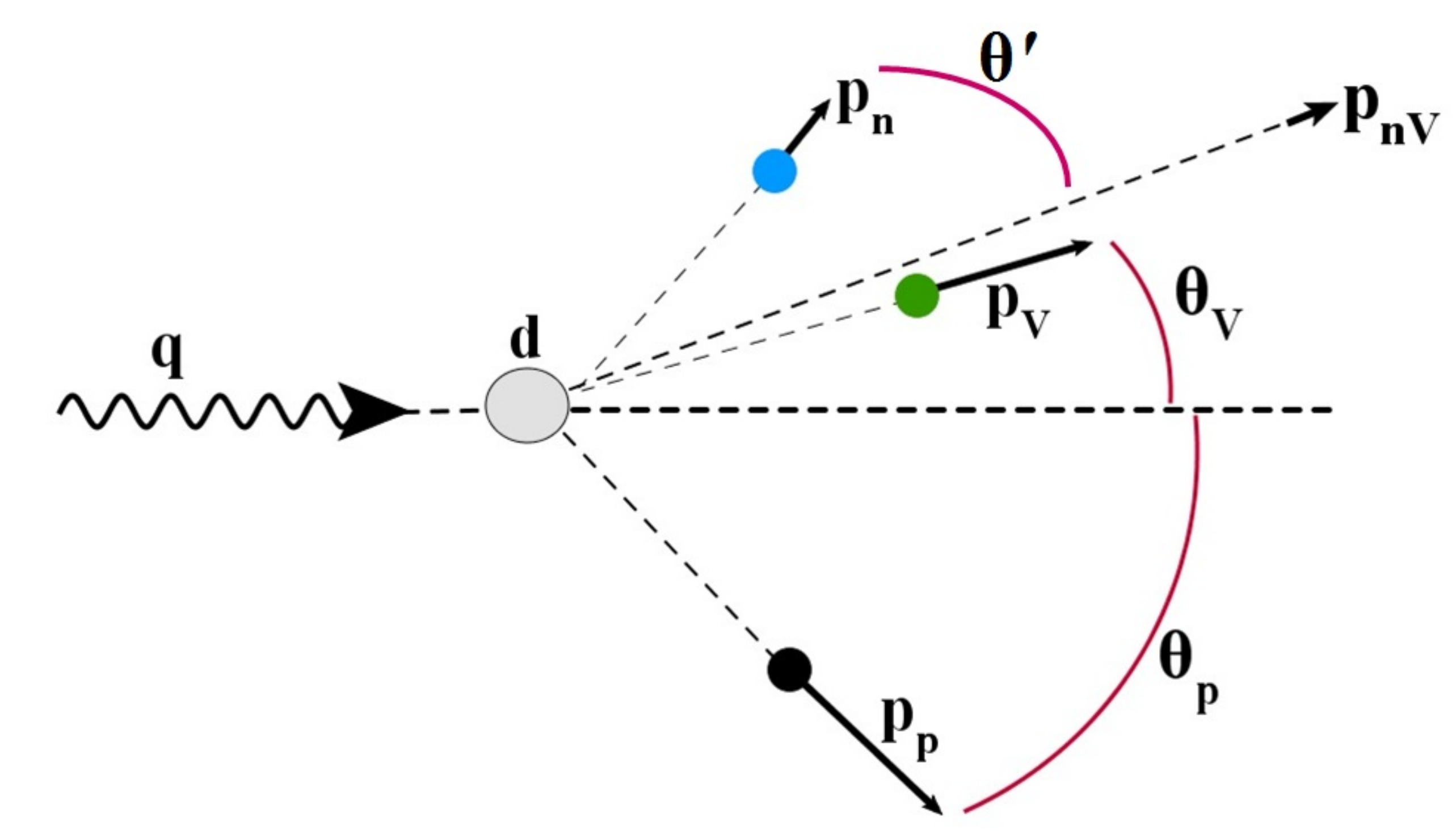}

    \end{center}
    \caption{%
        LAB frame momenta.  For a given $\theta_{cm}$ and $T^*_{Vn}$, the neutron LAB momentum can range between a minimum and maximum value, both of which correspond to $\theta^{\prime}=0$ in the figure.  $\mathbf{p}_{nV}=\mathbf{p}_V+\mathbf{p}_n$.
     }%
   \label{fig:labkinematics}
\end{figure}

\begin{figure}[tbp]
     \begin{center}
        \subfigure [$\nu=9\;GeV$ ]{%
            \label{fig:nminus2bnu9}
            \includegraphics[width=0.4\textwidth]{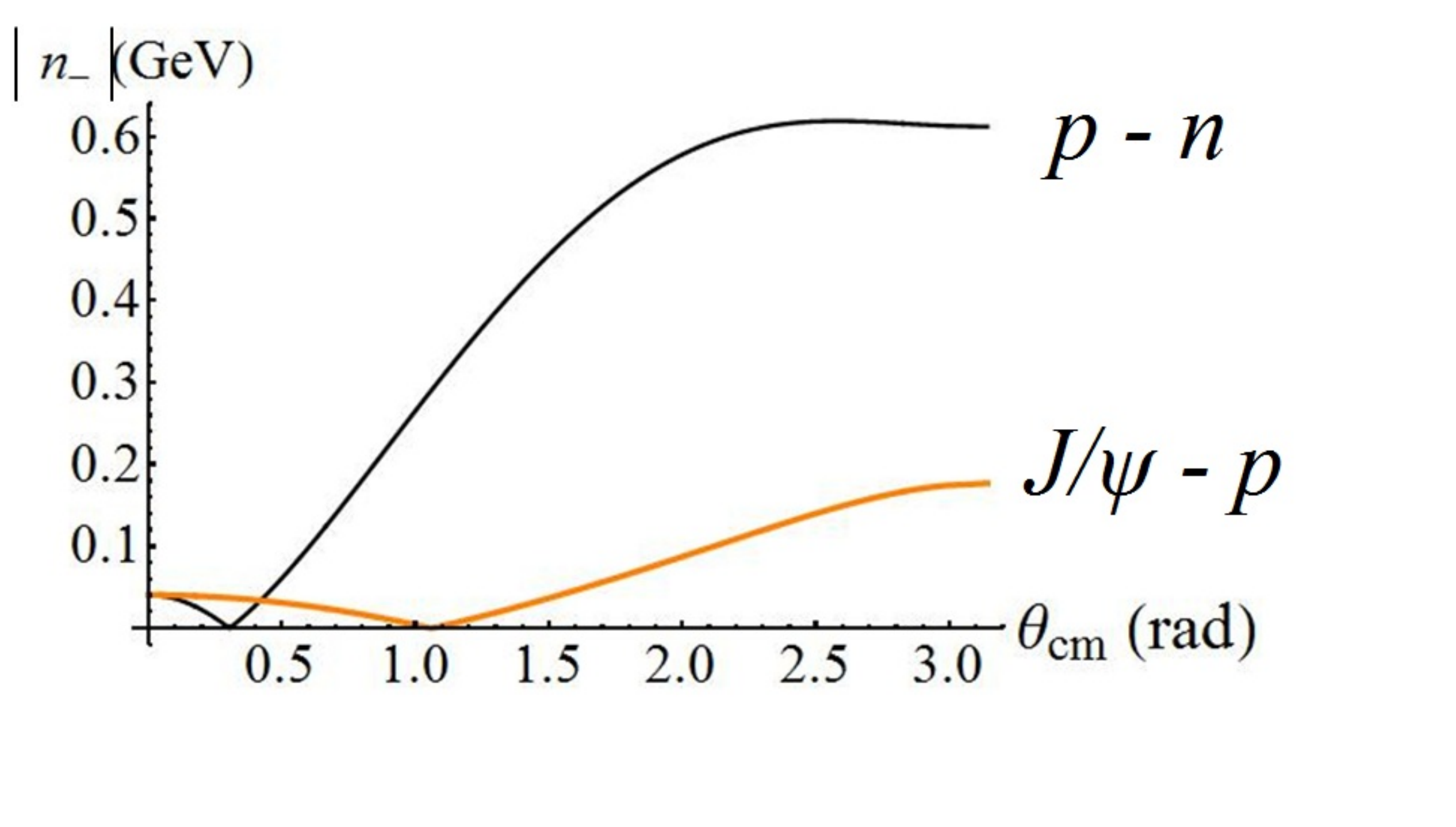}
        }%
        \hspace{0.5in}
         \subfigure[ $\nu=6.5\;GeV$]{%
           \label{fig:nminus3bnu9}
           \includegraphics[width=0.4\textwidth]{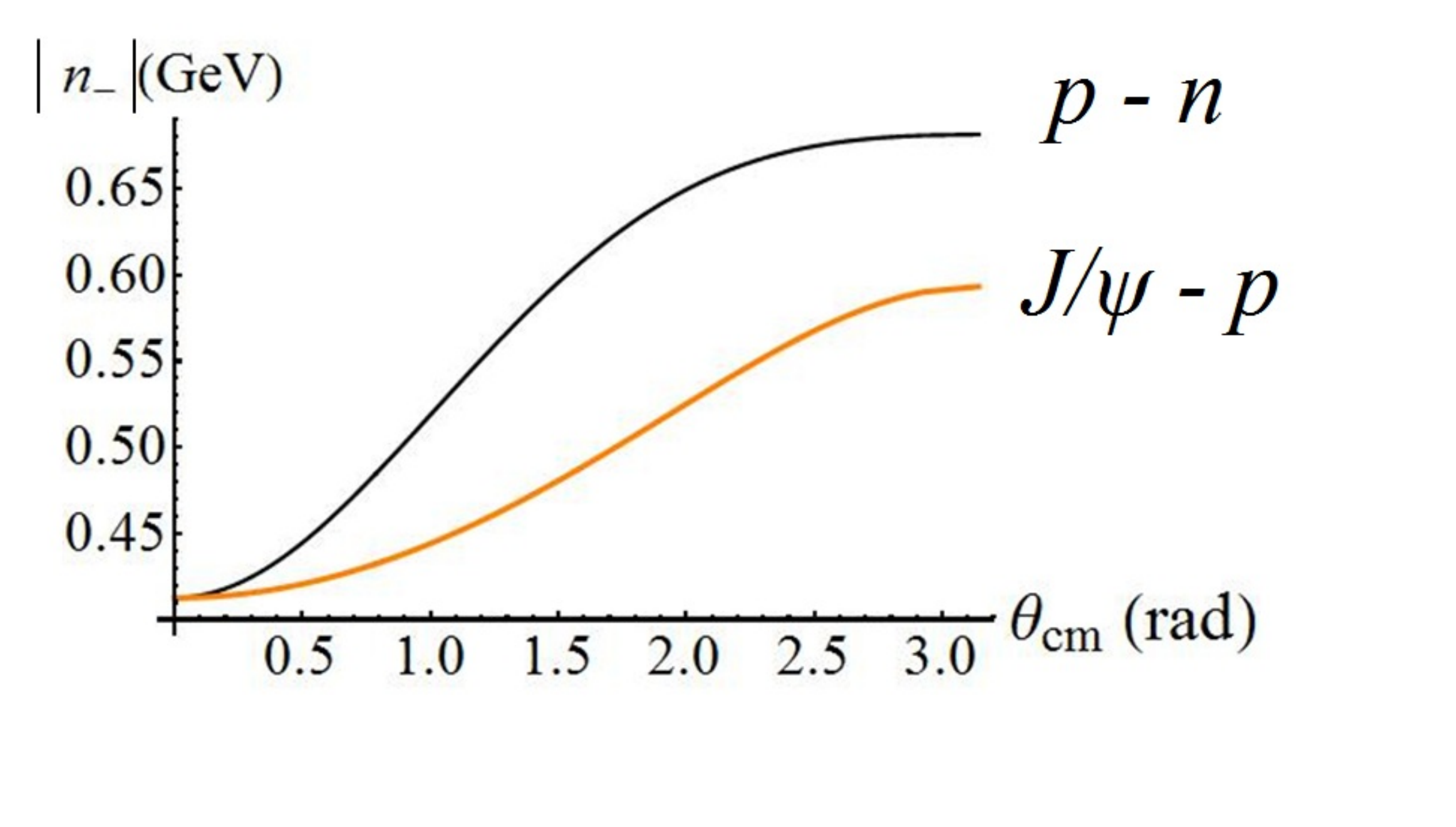}
        }\\ 


    \end{center}
    \caption{%
       $n_-$ vs. $\theta_{cm}$ for proton-neutron rescattering and $J/\psi$-proton rescattering amplitudes, for $T^*_{Vn}=30$ MeV, $p_n=p_{n,min}$.
     }%
   \label{fig:nminus2b3b}
\end{figure}

For the amplitudes $F_{2a}$, $F_{2b}$ and $F_{3b}$, the corresponding $n_-$ graphs are shown in Fig. \ref{fig:nminus2b3b} for $\nu=9\;GeV$ and $\nu=6.5\;GeV$, for $T^*_{Vn}=30$ MeV.  Note that at a given value of $\theta_{cm}$ and $T^*_{Vn}$, the neutron LAB momentum can range from a minimum to a maximum allowed value (with these two values both corresponding to $\mathbf{p}_n$ and $\mathbf{p}_V$ pointing in the same direction in the LAB, with $\theta^{\prime}=0$ in Fig. \ref{fig:labkinematics}), and the values of $n_-$ for $F_{2a}$, $F_{2b}$ and $F_{3b}$ depend on the neutron LAB momentum in addition to  $\theta_{cm}$ and $T^*_{Vn}$.   For our calculations, we have fixed the neutron LAB momentum for a given $\theta_{cm}$ (and $T^*_{Vn}=30$ MeV) at its minimum value; we denote this value by $p_{n,min}$.   One can see from these graphs that for $\nu=9\;GeV$, there are intervals of the variable $\theta_{cm}$ for which the on-shell parts of $F_{2a}$, $F_{2b}$, and $F_{3b}$ should be non-negligible, since $n_- <0.05\;GeV$ there  (note that for the diagram $F_{3b}$, the $J/\psi$-proton rescattering occurs at relatively high energy, if the $J/\psi$-neutron relative energy is small; so $F_{3b}$ is not directly related to the $J/\psi$-nucleon scattering length).  For $\nu=6.5\;GeV$, $n_-$ is larger than $\simeq 0.4$ GeV, and so these on-shell amplitudes should be small.  This is born out by the exact calculations, where the one-loop on-shell  amplitudes for $\nu=6.5\;GeV$ are in general much smaller than those for $\nu=9\;GeV$.

\subsection{Parameters used in the elementary 2-body amplitudes}
\label{sec:parameterssubsec}

For the calculation of the amplitudes, using \eq{Fon} and \eq{Foff}, the elementary amplitudes ${\cal M}^{\gamma V}$ and ${\cal M}$ are phenomenological amplitudes obtained from existing experimental data, related to the 2-body cross-section by \eq{crossandM}.  We take the individual 2-body differential cross-sections to be of the form
\begin{equation}
\frac{d\sigma}{dt}=Ae^{Bt},  
\end{equation}
where the $A$'s and $B$'s can depend on energy.  Thus we have
\be
{\cal M}=-i\sqrt{16\pi\lambda(s,m_1^2,m_2^2) A}\;e^{\frac{1}{2}Bt},
\ee
relating ${\cal M}$ to $A$ and $B$.

The values of $A$ and $B$ depend on the relative momentum (or energy) of the rescattering pair.  Table \ref{table:parameters} lists the values of the momentum $p$ of the neutron in the proton's rest frame (for the $p-n$ subsystem) and the momentum $p$ of the $J/\psi$ in the proton's rest frame (for the $V-p$ subsystem), for the case of $T^*_{Vn}=0$.  Note that these quantities are independent of $\theta_{cm}$ (easily shown in the overall c.m. frame).  Also given in Table \ref{table:parameters} is the total kinetic energy of the pair in the c.m. frame of that pair, $T^*_{12}$.  Given the values of the momentum $p$, we can determine the parameters that enter into the elementary amplitudes ${\cal M}^{pn}$, ${\cal M}^{Vp}$.

\begin{table}[ht]
\caption{Parameters used in elementary scattering amplitudes} 
\centering 
\begin{tabular}{c c c c c c c c} 
\hline\hline 
$\nu$ (GeV) & Subsystem  & $p$ (GeV) & $T^*_{12}$ (GeV) & $B$ (GeV$^{-2}$) & $\sigma_{tot}$ (mb) & $A$ (GeV$^{-4}$) \\ [0.5ex] 
\hline 
9 & $n-p$ & 2.25 & 0.64 & 5.7 - 6.2 & 43 - 46 & 260  \\ 
9 & $V-p$ & 7.4 & 1.02 & 1.31 & 3.5 & 1.61  \\
6.5 & $n-p$ & 0.86 & 0.16 & 6.9 & 35 & 160  \\
6.5 & $V-p$ & 2.84 & 0.247 & 1.31 & 3.5 & 1.61  \\ [1ex] 
\hline 
\end{tabular}
\label{table:parameters} 
\end{table}

\subsubsection{$p-n$ scattering parameters}

For $\nu=9\;GeV$, the existing data~\cite{Perl69} for $p-n$ scattering at incident momentum $p=2.25$ GeV gives $B_{pn}=5.7\;to\;6.2\;GeV^{-2}$.  The value of $A_{pn}$ can be obtained from the total $p-n$ cross-section by using the optical theorem and neglecting the real part of the scattering amplitude:
\begin{equation}
\label{Aandsigma}
\frac{d\sigma}{dt}\vert_{t=0}=\frac{1}{16\pi}\sigma_{tot}^2=A_{pn}  
\end{equation}
The measured value of $\sigma_{tot}$ given in the table is then used to calculate $A_{pn}$.  For $\nu=6.5$ GeV, the existing data for $p-n$ scattering at momentum $p=0.86$ GeV gives $B_{pn}=6.9\;GeV^{-2}$~\cite{Perl69} and $A_{pn}=160\;GeV^{-4}$~\cite{bugg66}.

\subsubsection{$J/\psi-p$ scattering parameters}

There is very little data on elastic $J/\psi$-proton scattering from which to determine the parameters $A_{Vp}$ and $B_{Vp}$ that are needed for the $J/\psi$-proton rescattering amplitude $M^{Vp}$.  For the present analysis, we have assumed that the $t$-slope for elastic $J/\psi$-nucleon scattering is equal to the $t$-slope for the process $\gamma^*+N\to J/\psi+N$, and so we've taken $B_{Vp}=B_{\gamma V}=1.31\pm0.19\;GeV^{-2}$.  We can obtain $A_{Vp}$ from the total $J/\psi$-nucleon cross-section, using the optical theorem; however, there has only been one measurement of $\sigma_{tot}^{J/\psi\; N}$~\cite{brambilla2011}.  In an experiment in 1977 at SLAC~\cite{psidata77} $J/\psi$ photoproduction was measured on beryllium and tantalum targets, and the total $J/\psi$-nucleon cross-section was extracted by using an optical model for the rescattering of the produced $J/\psi$ from the other nucleons in the nucleus.  The value they obtained was $\sigma_{tot}^{J/\psi\; N}=3.5\pm0.8\;mb$, which gives via the optical theorem $A_{Vp}=1.61\pm0.4\;GeV^{-4}$.  In that paper, however, they also note that the measured $J/\psi$-photoproduction cross-section together with vector meson dominance arguments would give a $J/\psi$-nucleon total cross-section of $\simeq 1\;mb$.   So we can assume the value of the $J/\psi$-nucleon total cross-section to be not very well known.  In addition, the photon energy in the SLAC experiment was $20\;GeV$, and so assuming forward production of the $J/\psi$, then the energy of the $J/\psi$ in the LAB frame would also be $\simeq 20\;GeV$, giving a kinetic energy in the LAB of $\simeq 17\;GeV$.  This is significantly larger than the kinetic energy of the $J/\psi$ in the proton rest frame considered here, where for $\nu=9\;GeV$ it is $4.94\;GeV$ and for $\nu=6.5\;GeV$ it is $1.1\;GeV$.  This introduces more uncertainty in the value of $A_{Vp}$ to be used.  In~\cite{brodsky97}, a theoretical calculation of the $J/\psi$-nucleon scattering length yields a value for the total $J/\psi$-nucleon cross-section at threshold of $7\;mb$, and it is argued that the total cross-section should decrease as the energy is increased from threshold.  Thus at the energy of the $J/\psi$-proton rescattering here, the value of $A_{Vp}$ may be larger than the value measured in the experiment at SLAC.  For the purpose of calculating the amplitude $F_{3b}$, however, we will use the value measured at SLAC.  

\subsubsection{Subthreshold $J/\psi$ production}

The threshold photon energy for production on a single nucleon at rest is 
\be
\nu_{thresh}=m_V+\frac{m_V^2+Q^2}{2m}
\ee
while for production on the deuteron it is
\be
\nu_{thresh}=m_V+\frac{m_V^2+Q^2}{2M_d}
\ee
For $Q^2=0.5 \;GeV$, these are $8.47 \;GeV$ and $5.78 \;GeV$, respectively.

For $\nu=6.5\;GeV$, which is below threshold for $J/\psi$ production on a single nucleon at rest, we assume that the production mechanism is the same as for production on a free nucleon.  The Fermi motion of the nucleon in the deuteron is what allows the production to occur, i.e. if the nucleon is moving towards the photon with a large enough momentum then the value of $s_1=(q+p)^2$, where $p$ is the 4-momentum of the nucleon in the deuteron, will be above the threshold value.  In the calculation of the amplitudes for $\nu=6.5\; GeV$ this condition was imposed on the internal nucleon momentum in the integrals involved.

\subsection{Calculated On-shell and Off-shell amplitudes}

\begin{figure}[hbp]
     \begin{center}
        \subfigure[]{%
            \label{fig:31}
            \includegraphics[width=0.4\textwidth]{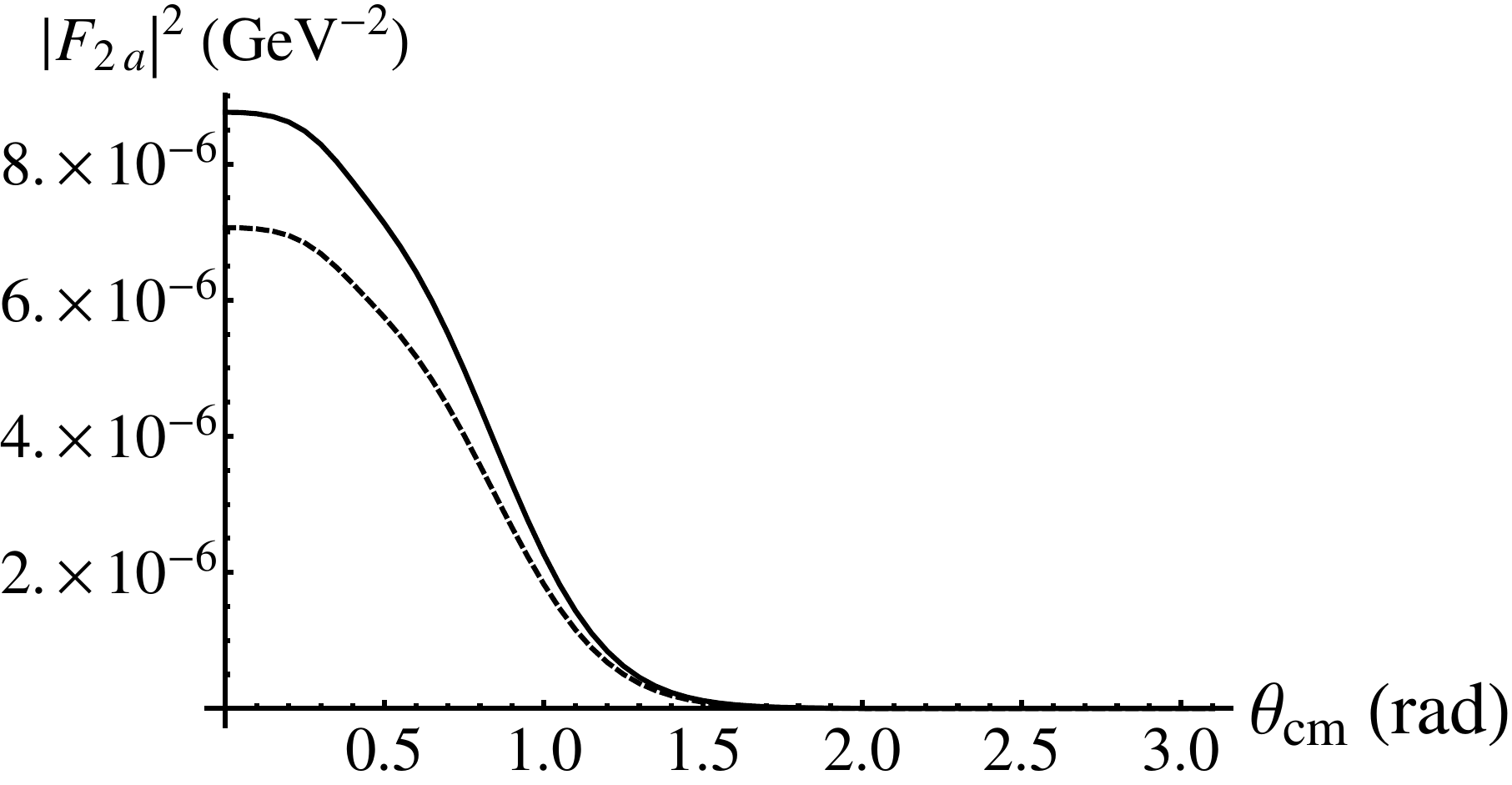}
        }%
        \hspace{0.5in}
         \subfigure[]{%
           \label{fig:32}
           \includegraphics[width=0.4\textwidth]{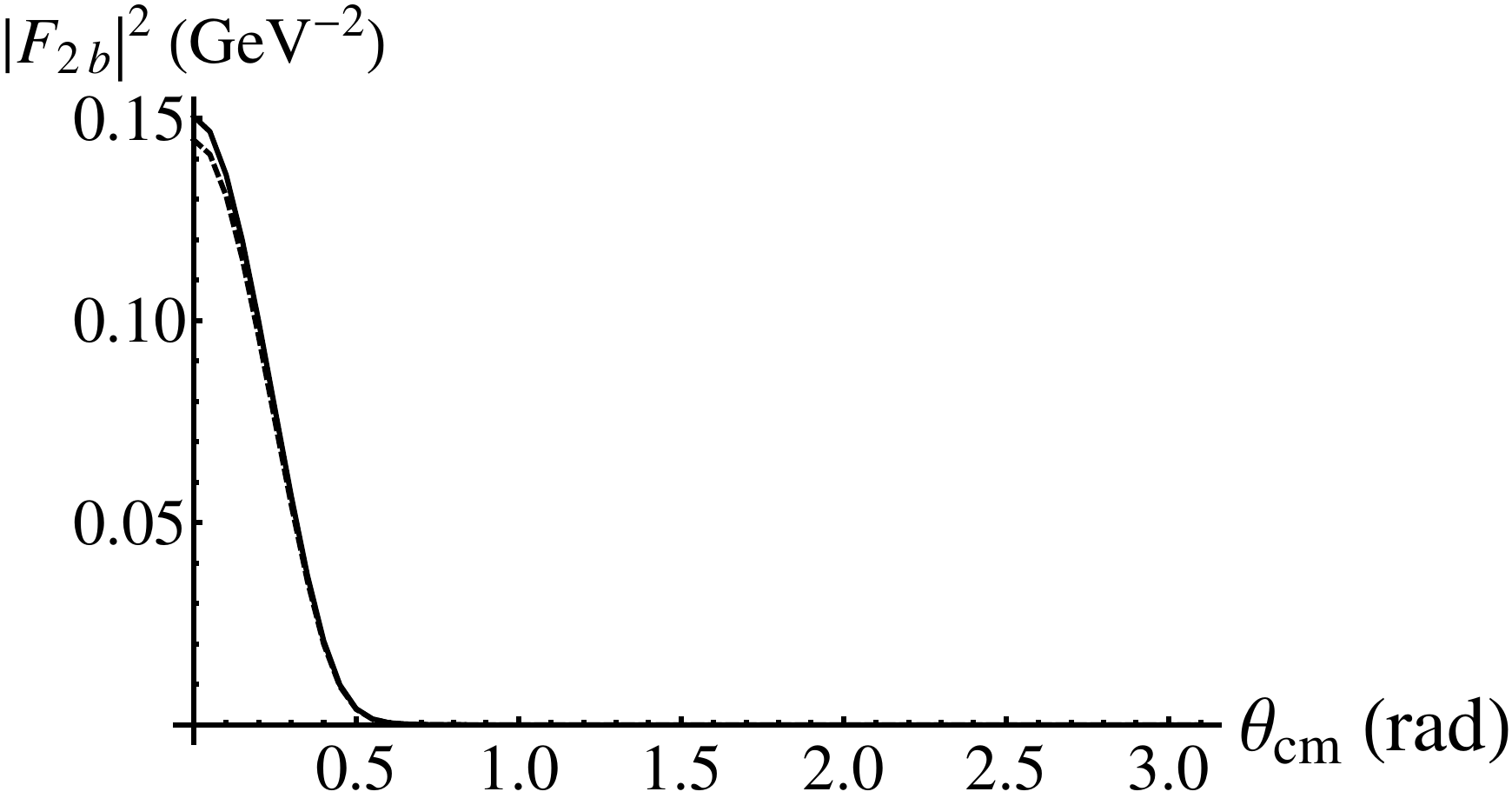}
        }\\ 

        \subfigure[]{%
            \label{fig:33}
            \includegraphics[width=0.4\textwidth]{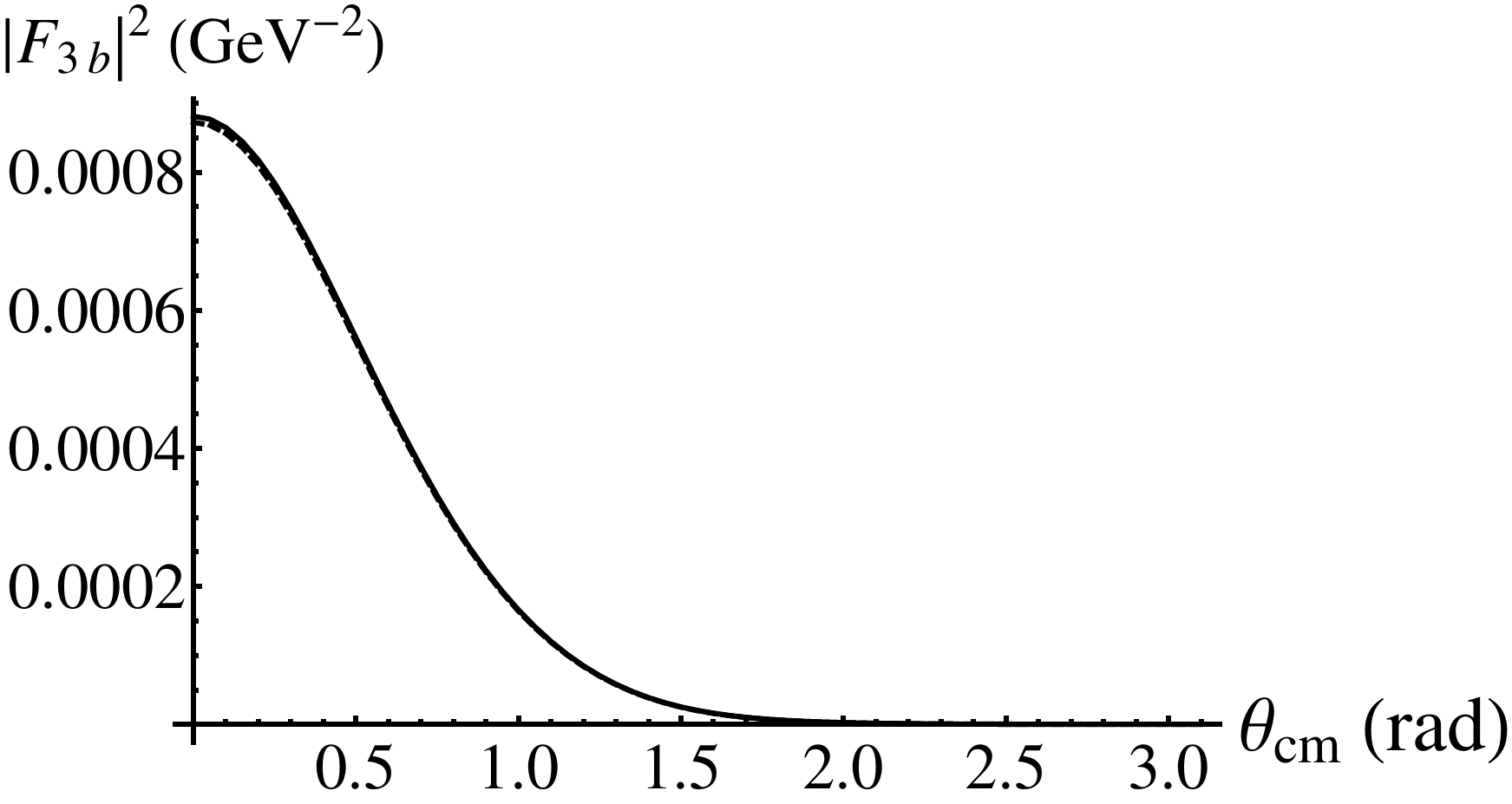}
        }%

    \end{center}
    \caption{%
       Squares of amplitudes for $\nu=9.0\;GeV$, $T^*_{Vn}=30$ MeV, $p_n=p_{n,min}$.  The solid curve is the total (on-shell plus off-shell parts), while the dashed curve is only including the on-shell part of the amplitude.
     }%
   \label{fig:ampstotandon}
\end{figure}

\begin{figure}[!hbp]
     \begin{center}
        \subfigure[]{%
            \label{fig:34}
            \includegraphics[width=0.4\textwidth]{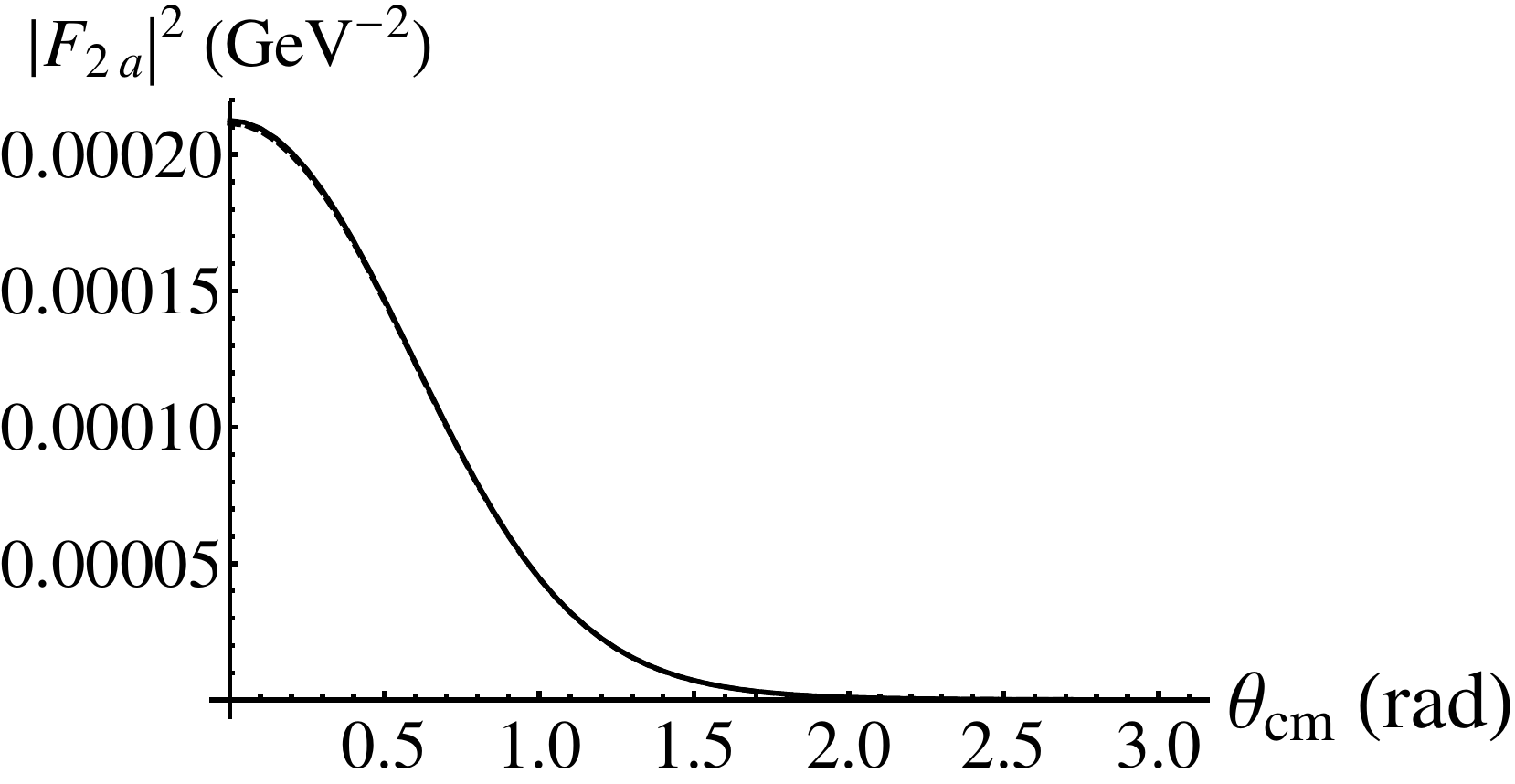}
        }%
        \hspace{0.5in}
         \subfigure[]{%
           \label{fig:35}
           \includegraphics[width=0.4\textwidth]{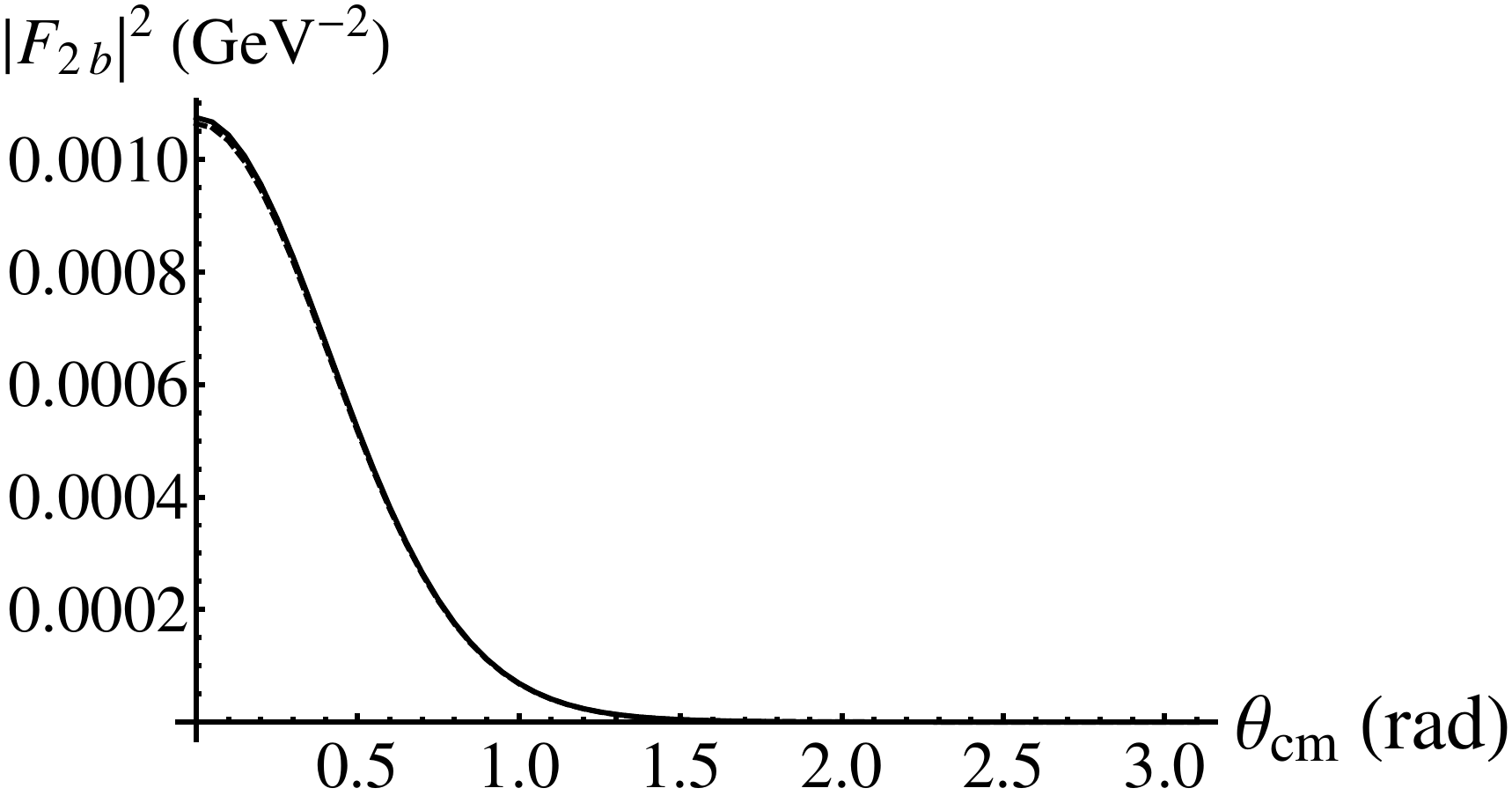}
        }\\ 

        \subfigure[]{%
            \label{fig:36}
            \includegraphics[width=0.4\textwidth]{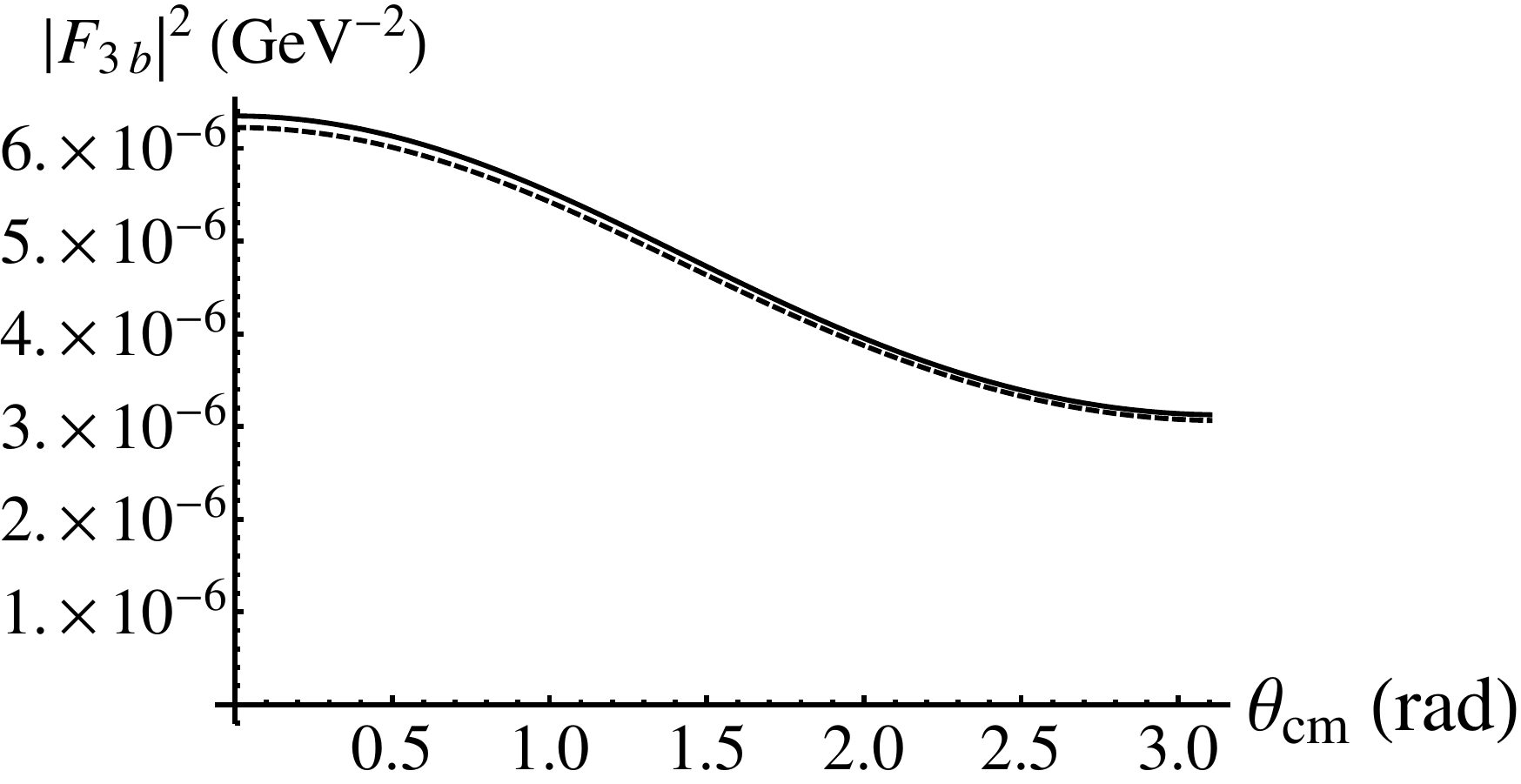}
        }%

    \end{center}
    \caption{%
       Squares of amplitudes for $\nu=6.5\;GeV$, $T^*_{Vn}=30$ MeV, $p_n=p_{n,min}$.  The solid curve is the total (on-shell plus off-shell parts), while the dashed curve is only including the on-shell part of the amplitude.
     }%
   \label{fig:ampstotandon65}
\end{figure}

Using the parameters in Table \ref{table:parameters}, the on-shell and off-shell parts of the amplitudes were calculated.  The squares of the individual amplitudes $F_{2a}$, $F_{2b}$, and $F_{3b}$ are shown in Figs. \ref{fig:ampstotandon} and \ref{fig:ampstotandon65}; shown in the graphs is a curve which includes only the (square of the) on-shell part of the amplitude, and also a curve which is the square of the total amplitudes including both the on-shell and off-shell parts.  For the off-shell parts, the same parametrizations of the elementary amplitudes ${\cal M}^{\gamma V}$ and ${\cal M}$ were used as for the on-shell parts.  As stated previously, the off-shell parts are very small compared to the on-shell parts, which means that knowledge of the exact forms of the off-shell elementary amplitudes ${\cal M}^{\gamma V}$ and ${\cal M}$ are not needed.

Since the $J/\psi$-nucleon scattering length is expected to be small (much smaller than e.g. the proton-neutron scattering length), the $J/\psi$-neutron rescattering diagram $F_{3a}$ should be a small contribution to the total amplitude.  This is born out in the next subsection, where $F_{3a}$ is calculated using a model potential and wavefunction, for a value of the scattering length on the order of that predicted by theoretical models.

\subsection{$J/\psi$-neutron Rescattering Diagrams and the Scattering Length}

Diagram $F_{3a}$ (Fig. \ref{fig:F3a}) is the $J/\psi$-neutron rescattering diagram.  This is the diagram where the $J/\psi$ and neutron scatter from each other with small relative momentum; hence this amplitude will involve the scattering length for the $J/\psi$-neutron interaction. 

For this diagram, we have $\mathbf{p}_{12}= \mathbf{p}_{Vn}\equiv \mathbf{p}_V +\mathbf{p}_n$ in Eqs. (\ref{Fon}) - (\ref{costheta}), and $\theta$ is the angle between $\mathbf{p}_{Vn}$ and $\mathbf{n}$ (see Fig. \ref{fig:coordsystem}).  For $T^*=0$, where $T^*=T_n^*+T_V^*$ ($^*$ indicates the c.m. frame of the $V-n$ system), we have $n_+=\vert n_-\vert$ (see \eq{nplusminus}), and so $F_{3a}^{on}=0$.  For small $T^*$ (below e.g. $100\;MeV$), $F_{3a}^{on}$ will be small because $n_-\gtrsim 0.45\;GeV$ for the possible JLab kinematics (see Fig. \ref{fig:nminusplus3a1}).  Thus the main contribution to $F_{3a}$ is from $F_{3a}^{off}$, for which the intermediate-state $J/\psi$ is always off-mass-shell.  However, because of the propagator denominator $f_{12}(n)+\cos{\theta}$, where 
\be
f_{12}(n)+\cos{\theta}=\frac{1}{2\vert\mathbf{p}_{Vn}\vert n}\Bigl(k^2-m_V^2\Bigr),
\ee
contributions to $F_{3a}^{off}$ from values of $\mathbf{n}$ for which $k$ is far off-mass-shell will be small.  To obtain estimates of $F_{3a}^{off}$, we will therefore evaluate it using on-mass-shell values of $ {\cal M}^{\gamma V}$ and ${\cal M}^{Vn}$.

The relation between the invariant amplitude ${\cal M}^{Vn}$ and the scattering amplitude $f(k,\theta)$ is~\cite{pdg}
\begin{equation}
{\cal M}=-8\pi\sqrt{s_{Vn}}\; f(k,\theta)  \end{equation}
for the on-energy-shell amplitudes.  The half-off-energy shell amplitudes are related in the same way.

Our normalization conventions are the following.  The scattering length $a$ is given by
\begin{equation}
\lim_{k\to 0} f(k,\theta)=-a.
\end{equation}
 The off-energy-shell amplitude is given by
\begin{equation}
\label{offshellamp}
f^{VN}(\mathbf{k}_1,\mathbf{k}_2)=-(2\pi)^2\mu \;\langle\mathbf{k}_2\vert V \vert \Psi_{\mathbf{k}_1}^{(+)} \rangle  = -(2\pi)^2\mu \;\langle\Psi_{\mathbf{k}_2}^{(-)}\vert V \vert {\mathbf{k}_1} \rangle = -(2\pi)^2\mu \;\langle\mathbf{k}_1\vert V \vert \Psi_{\mathbf{k}_2}^{(-)} \rangle^*  \end{equation}
where $V$ is the potential, $\mu$ is the reduced mass, $\mathbf{k}_1$ is the initial relative momentum (in terms of $\mathbf{n}$, $\mathbf{k}$ in Fig. \ref{fig:F3a}), $\mathbf{k}_2$ is the final relative momentum (in terms of $\mathbf{p}_V$, $\mathbf{p}_n$ in Fig. \ref{fig:F3a}), and $\Psi_{\mathbf{k}_2}$ is the exact scattering wavefunction for asymptotic relative momentum $\mathbf{k}_2$.  
Since this off-energy-shell scattering amplitude depends on the scattering wavefunction $\Psi_{\mathbf{k}_2}^{(-)}$ and the potential $V$, both of which are unknown for $J/\psi$-nucleon elastic scattering, we will resort to models in order to estimate the amplitude.

We normalize our $S$-wave scattering wavefunction $\Psi_{\mathbf{k}}$, and define the radial wavefunction $u(r)$, by:
\begin{equation}
\label{wfdef}
\Psi_{\mathbf{k}}(r)=\frac{1}{\sqrt{(2\pi)^3}}\;e^{i\delta(k)}\frac{u(r)}{r} 
 \end{equation}
where $\delta(k)$ is the $S$-wave phase shift.  
In order to calculate the matrix element, we will specify a model zero-energy wavefunction $u_0(r)$ which determines the potential $V(r)$ via the Schrodinger equation, and use that potential to solve for the wavefunction for $k\ne 0$ (the subcript zero on $u_0$ indicates it is for $k=0$).

We assume the $J/\psi$-nucleon potential is of finite range, and so is zero for $r$ larger than some distance $R$.   
The phase-shift $\delta(k)$ satisfies the following well-known properties as $k\to 0$:  
\begin{enumerate}
\item for a repulsive potential, or an attractive potential that doesn't admit a bound state:  $\delta\to -ak$ as $k\to 0$;
\item for an attractive potential which admits a single bound state:  $\delta\to \pi-ak$ as $k\to 0$
\end{enumerate}
The zero-energy wavefunction $\Psi_0(r)$ outside the range of the potential is then
\begin{equation}
\label{psiout}
\Psi_{0}^{out}(r)=\frac{1}{\sqrt{(2\pi)^3}}\;e^{i\delta(0)}\frac{u_{0}^{out}(r)}{r}=\frac{1}{\sqrt{(2\pi)^3}}\;\frac{r-a}{r}  
\end{equation}
for both cases, while the zero-energy radial wavefunction $u_{0}^{out}$ differs by a minus sign for the two cases; this is purely due to including the factor $e^{i\delta(k)}$ in the definition of $\Psi_k$ in \eq{wfdef}.  Our normalization conventions give $a>0$ for either a repulsive potential or an attractive potential with a bound state, and $a<0$ for an attractive potential that doesn't admit a bound state.  In all cases $a$ is the intercept on the $r$-axis of $u_{out}^0(r)$.

Theoretical calculations~\cite{brodsky97,kawanai2010} give values of $\vert a \vert\simeq 0.3\;fm$, with effective range $r_e\simeq 2.0\;fm$.  It is thought that the potential is attractive, but too weak to support a bound state.  Below, calculations of $F_{3a}$ are made for both cases of an attractive potential:  $a>0$ (bound state) and $a<0$ (no bound state).

\subsection{Positive scattering length $a$}

For the case of a positive scattering length and attractive potential (which possesses a bound state), we have $u_{0}^{out}(r)=-(r-a)$ .  We choose the simplest zero-energy wavefunction $u_0(r)$ consistent with the standard continuity requirements on $u(r)$ at $r=0$ and $r=R$ required by the Schrodinger equation.  This yields:

\begin{equation}
u_{0}^{in}(r)=\Bigl(-1+\frac{3a}{R}\Bigr) r + \Bigl(-\frac{3a}{R^2}\Bigr) r^2 +\Bigl (\frac{a}{R^3} \Bigr) r^3.
 \end{equation}
One further requirement on $u_{0}^{in}$ is that $u_{0}^{in}$ have no zeros on the interval $[0,R]$ (besides at $r=0$).  This ensures that the corresponding potential $V(r)$ is non-singular, since for $k=0$, $V(r)=\frac{1}{2\mu}\frac{u''}{u}$.  The zeros of $u_{0}^{in}$ are at $r=0$ and
\begin{equation}
\frac{r}{R}=\frac{3}{2}\pm\frac{1}{2}\sqrt{4\frac{R}{a}-3}  \end{equation}
and the right-hand-side must lie outside the range $[0,1]$.  This requires either
\begin{equation}
R<a \end{equation}
or
\begin{equation}
R>3a  \end{equation}
The potential is
\begin{equation}
\label{modelpot}
V(r)=\frac{1}{2\mu}\frac{1}{r}\frac{r-R}{R^2(3-\frac{R}{a})-3R r +r^2}  \end{equation}
and one can see that for $R>3a$ the potential is repulsive.  Therefore we require $R<a$. 
\newline

\subsubsection{Model wavefunction for $k\ne 0$}

Given this model potential we can proceed to calculate the off-shell scattering amplitude \eq{offshellamp} and the amplitude $F_{3a}^{off}$ once we calculate the wavefunction $u$ for non-zero $k$ for a given $a$ and $R$.  Taking $a=0.3\;fm$, $R=0.1\;fm$, and solving the Schrodinger equation for $k<100$ MeV numerically, we find that the wavefunction $\Psi_{\mathbf{k}}(r)$ and the off-shell amplitude $f(p,k)$ depend very weakly on $k$ over the entire momentum range from $k=0$ to $k=100$ MeV.

Since we are only interested in the $J/\psi$-neutron relative momentum up to around 100 MeV, it is legitimate to approximate the off-energy-shell amplitude $f^{Vn}(p,k)\simeq f^{Vn}(p,0)$ for the range of $k$ we are interested in.  For our model wavefunction $u_{0}^{in}$ we can evaluate $ f^{Vn}(p,0)$ analytically:
\be
\begin{split}
f^{Vn}(\mathbf{p},0)&=\frac{-2\pi^2}{\sqrt{(2\pi)^3}}\int d^3r e^{-i\mathbf{p}\cdot\mathbf{r}}U(r)\Psi_0(r) \\
&=\frac{6a}{p^2 R^2}\Bigl(-1+\frac{\sin{pR}}{pR}\Bigr).\\ 
\end{split} 
\ee
The momentum $p$ appearing in $f(p,0)$ is the relative momentum of the $J/\psi$-neutron pair in their center-of-mass frame, before they scatter in diagram $F_{3a}$; it is thus the magnitude of $\mathbf{n}$ (or -$\mathbf{k}$) in the outgoing $V-n$ center-of-mass frame, and so we must boost $n$ to that frame.

\subsection{Negative scattering length $a$}

If the potential is attractive but too weak to support a bound state, then $a<0$ and we have $u_{0}^{out}(r)=r-a$ .  In this case the zero-energy wavefunction is
\begin{equation}
\label{negau}
u_{0}^{in}=\Bigl(1-\frac{3a}{R}\Bigr) r + \Bigl(\frac{3a}{R^2}\Bigr) r^2 -\Bigl (\frac{a}{R^3} \Bigr) r^3.  
\end{equation}
The requirement that $u$ have no zeros on $[0,R]$ imposes no restriction on $a$ and $R$ in this case.  Theoretical calculations give $a$ around $-0.3\;fm$, and effective range $r_e\simeq 2.0\;fm$~\cite{brodsky97,kawanai2010}.  Using these values with our model wavefunction and potential implies $R=1.3\;fm$. Again there's very little variation with $k$ of the wavefunctions and off-shell amplitudes for $k$ from $0$ to $0.1\;GeV$, so to calculate $F_{3a}$ in this case we will again approximate $f^{Vn}(\mathbf{p},\mathbf{k})\simeq f^{Vn}(\mathbf{p},0)$. This yields
\be
f^{Vn}(\mathbf{p},0)=\frac{6a}{p^2 R^2}\Bigl(1-\frac{\sin{pR}}{pR}\Bigr). 
\ee

\subsection{Results}

\begin{figure}[tbp]
     \begin{center}
        \subfigure[Squares of individual amplitudes, for positive scattering length $a$.   Dashed curve is the square of the total amplitude.  $V-p$: $F_{3b}$; $V-n$: $F_{3a}$; $p-n$: $F_{2b}$.]{%
            \label{fig:squares9a}
            \includegraphics[width=0.4\textwidth]{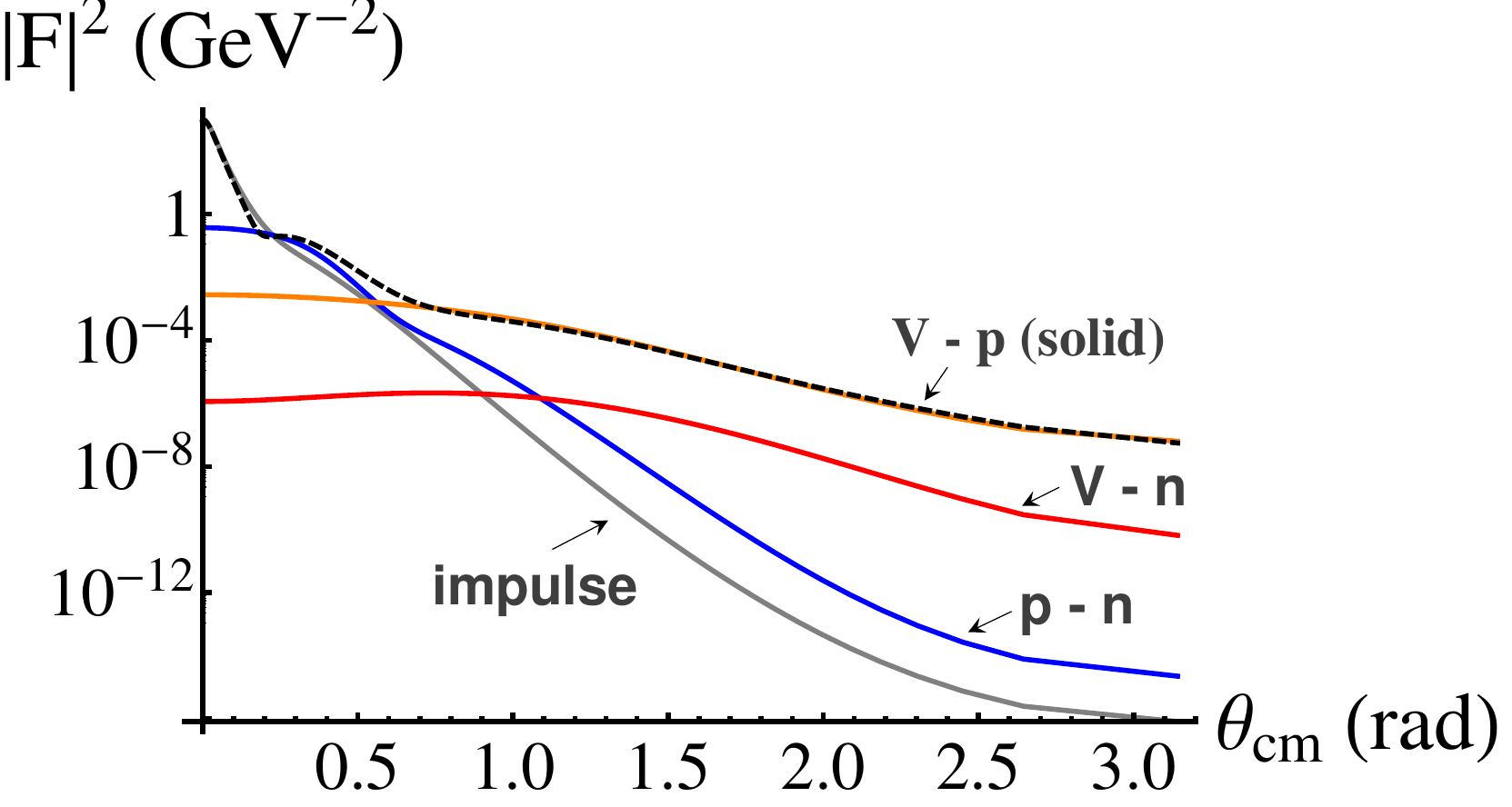}
        }%
        \hspace{0.5in}
         \subfigure[Electroproduction differential cross-section (logarithmic scale). ]{%
           \label{fig:squaretot9b}
           \includegraphics[width=0.4\textwidth]{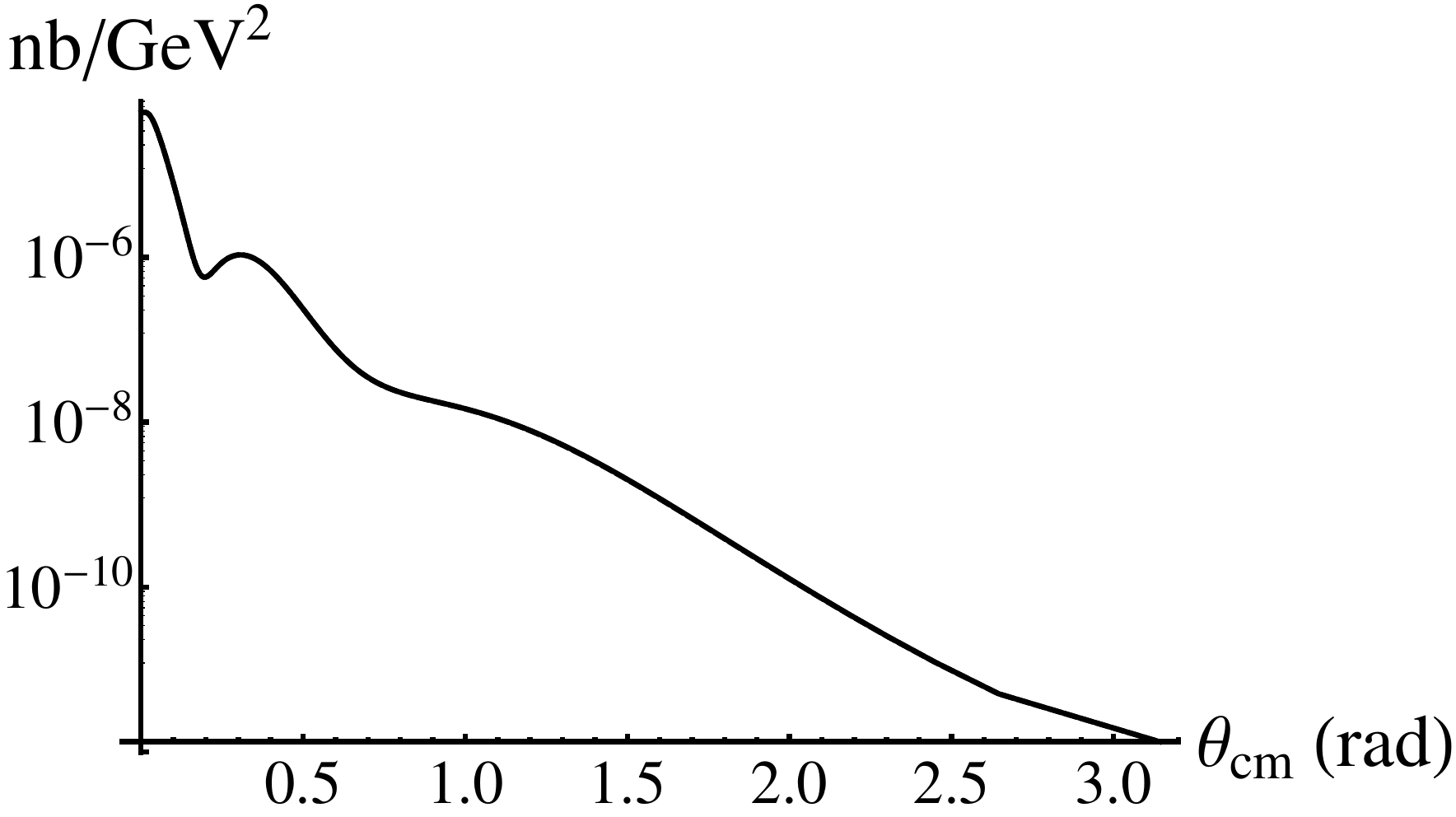}
        }\\ 

        \subfigure[Electroproduction differential cross-section (linear scale).  ]{%
            \label{fig:cross9linear}
            \includegraphics[width=0.4\textwidth]{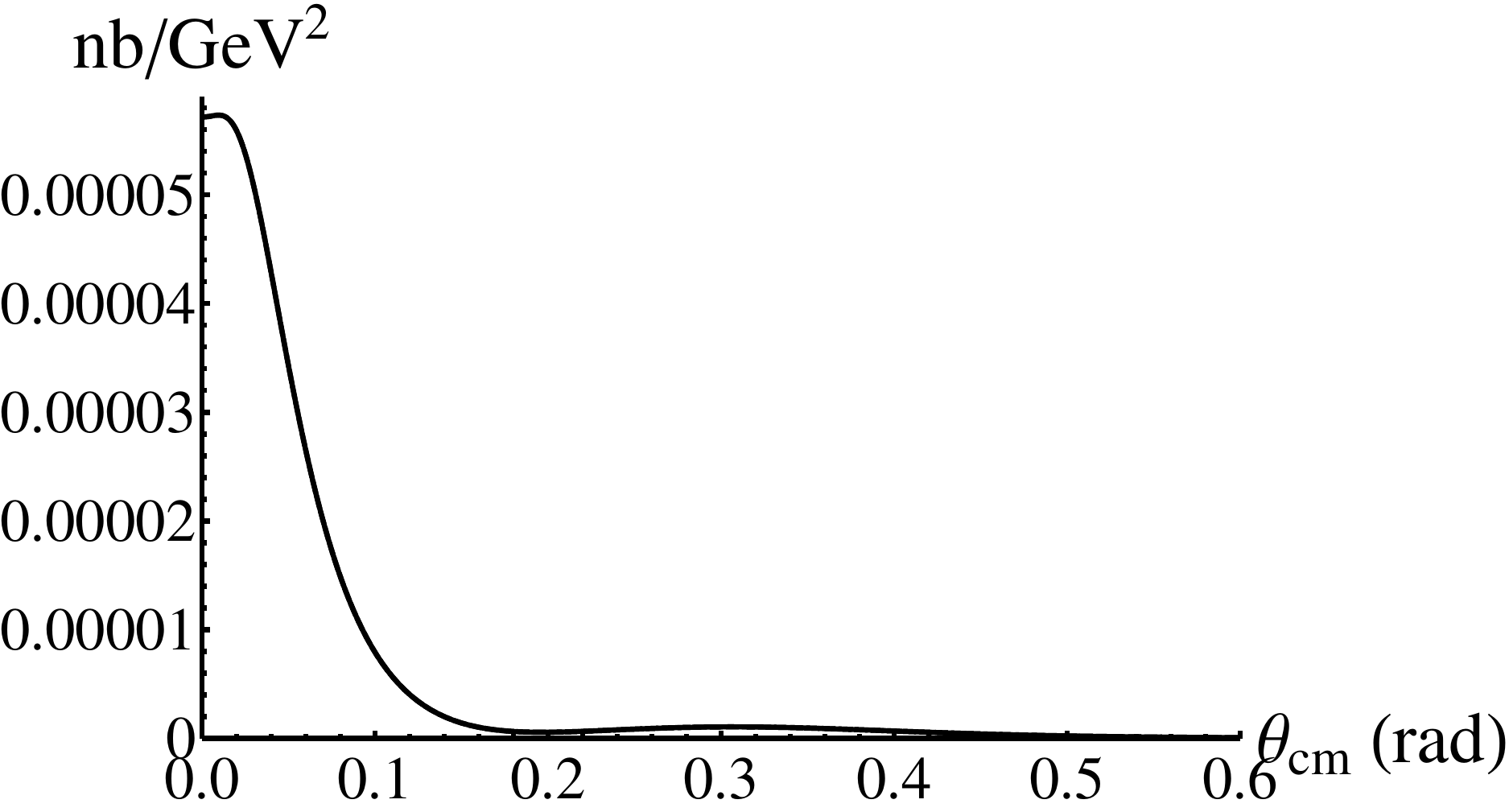}
        }%

    \end{center}
    \caption{%
     Squares of amplitudes, and electroproduction differential cross-section \eq{electrocross}, for $\nu=9\;GeV$,  $T^*_{Vn}=30$ MeV, $p_n=p_{n,min}$.  The solid curves in (b) and (c) includes all amplitudes, while the dashed curves (which are not distinguishable from the solid curve) omit $F_{3a}$ for both positive and negative scattering length $a$,  for the model potential of \eq{modelpot}.
     }%
   \label{fig:ampsSquared9}
\end{figure}

\begin{figure}[tbp]
     \begin{center}
        \subfigure[Positive scattering length $a$.  Solid curve includes all amplitudes, dashed is omitting $F_{3a}$. ]{%
            \label{fig:squaretot65a}
            \includegraphics[width=0.4\textwidth]{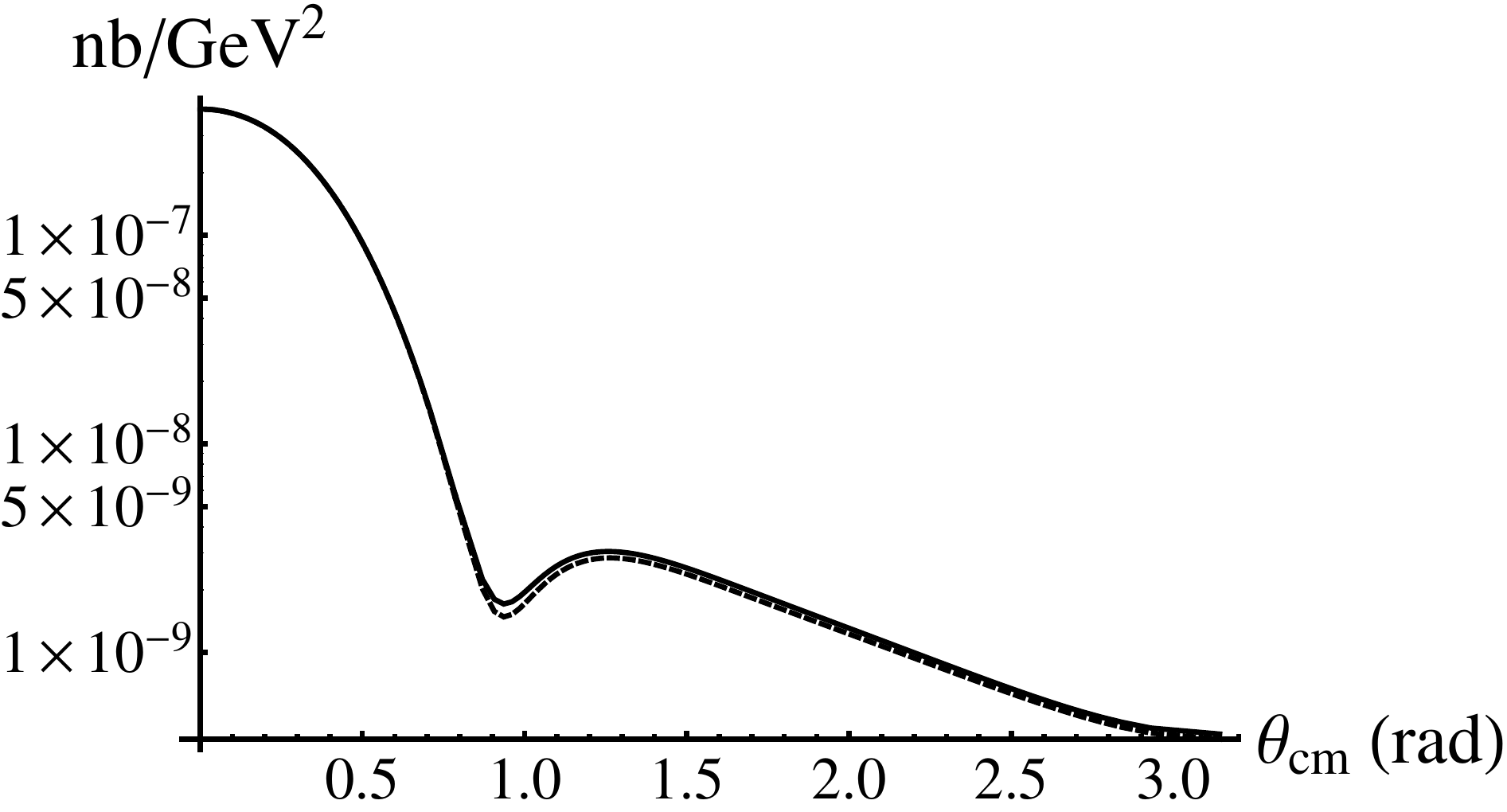}
        }%
        \hspace{0.5in}
         \subfigure[Negative scattering length $a$.  Solid curve includes all amplitudes, dashed is omitting $F_{3a}$. ]{%
           \label{fig:squaretot65b}
           \includegraphics[width=0.4\textwidth]{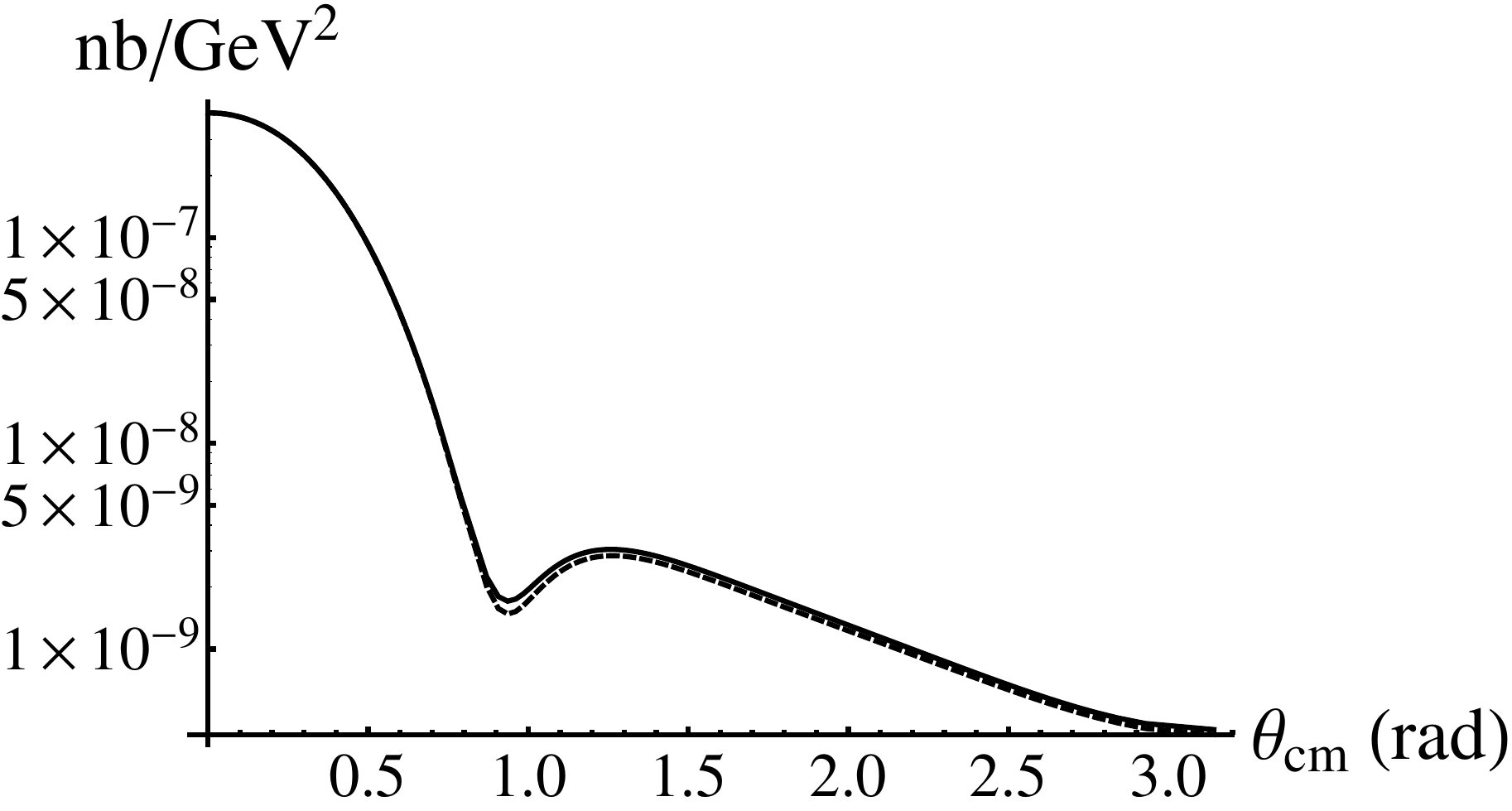}
        }\\ 

        \subfigure[Linear scale]{%
            \label{fig:linear65}
            \includegraphics[width=0.4\textwidth]{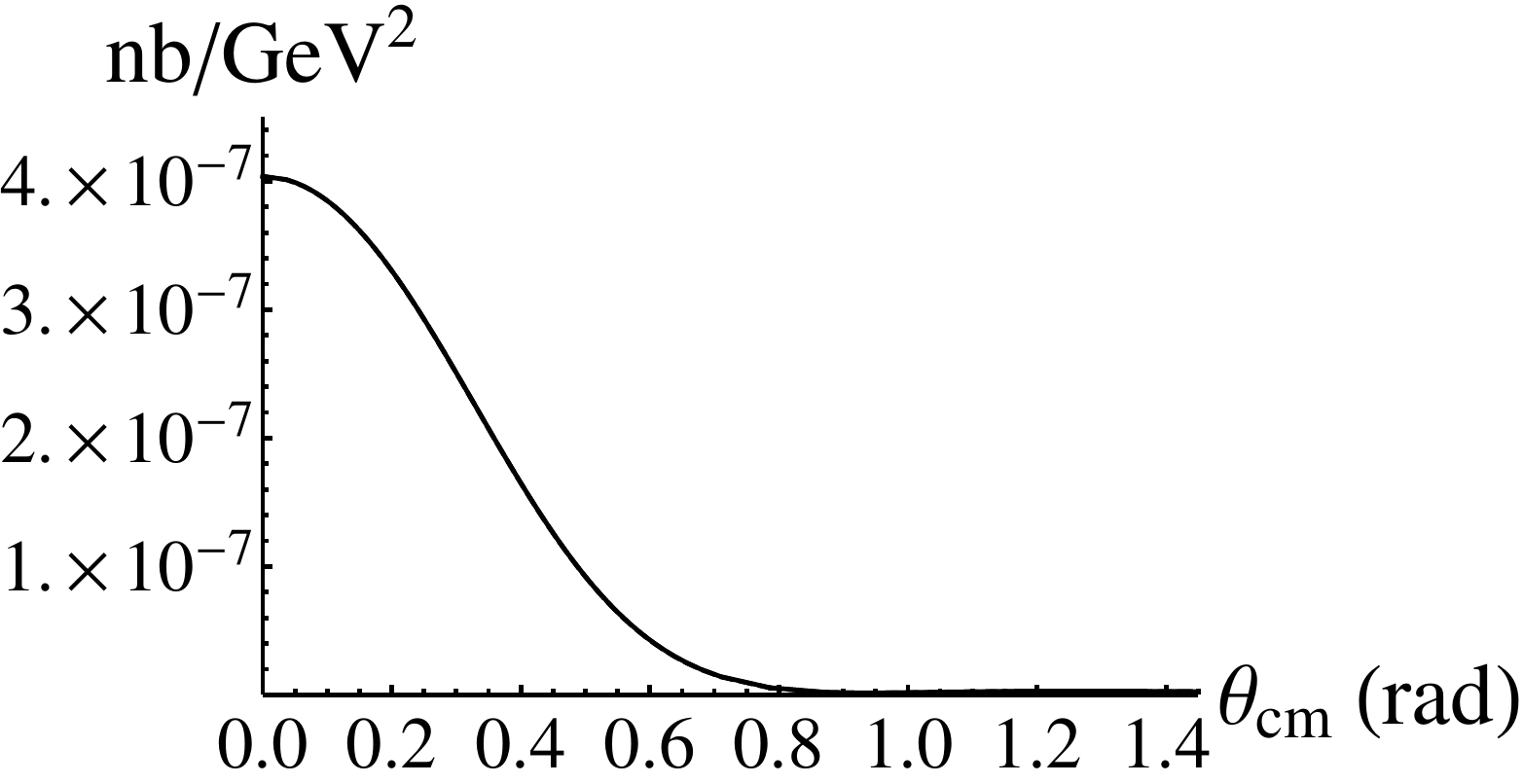}
        }%

    \end{center}
    \caption{%
      Electroproduction differential cross-section \eq{electrocross} for $\nu=6.5\;GeV$, $T^*_{Vn}=30$ MeV, $p_n=p_{n,min}$, and $a=\pm0.3\;fm$.  In (c), the dashed curves (not distinguishable from solid curve) are omitting $F_{3a}$ for both positive and negative $a$,  for the model potential of \eq{modelpot}.
     }%
   \label{fig:ampsSquared65}
\end{figure}

The electroproduction differential cross-section in the LAB frame, shown in Figs. \ref{fig:ampsSquared9} and \ref{fig:ampsSquared65}, is given by
\be
\label{electrocross}
\frac{d^8\sigma}{dE^{\prime} d\Omega^{\prime} dp_p d\Omega_p d\Omega_n}=\frac{ v_0 V_T\;E^{\prime}}{8 (2\pi)^3 M_d\; E}\;\times\frac{1}{8(2\pi)^5}\frac{p_p^2}{E_p}\frac{p_n^3}{ \vert E_V\;p_n^2 - E_n \mathbf{p}_n \cdot \mathbf{p}_V \vert}\;\vert F \vert^2
\ee 
where $F=F_{1a}+F_{1b}+F_{2a}+F_{2b}+F_{3a}+F_{3b}$ is the total amplitude for $J/\psi$ production from a virtual photon, and
\be
v_0=\sqrt{16 E^2 E^{\prime 2} - Q^4}=4 E E^{\prime} \cos^2(\theta^{\prime}/2),
\ee
\be
V_T = \frac{1}{2}\frac{Q^2}{\mathbf{q}^2}+\frac{Q^2}{v_0},
\ee
and
\be
E^{\prime}=E-\nu=12\;GeV - \nu.
\ee
In the above, $E$ is the initial electron energy (taken to be $12$ GeV), $E^{\prime}$ is the final electron energy, and $\theta^{\prime}$ is the scattering angle of the electron relative to the initial electron momentum (all quantities in the LAB frame).

The results  are shown in Figs. \ref{fig:ampsSquared9} and \ref{fig:ampsSquared65}.  For $\nu=9$ GeV, the squares of the individual amplitudes $F_{1b}$, $F_{2b}$, $F_{3b}$, and $F_{3a}$ are also shown in Fig. \ref{fig:squares9a}, along with the square of the total amplitude including all diagrams.   As can be seen in that graph, the amplitude $F_{3a}$ makes a negligible contribution to the total amplitude.  There are intervals of $\theta_{cm}$ where the amplitudes  $F_{1b}$, $F_{2b}$, and $F_{3b}$ individually dominate the  total amplitude.  However, by comparing Fig. \ref{fig:squares9a} with Fig. \ref{fig:cross9linear}, which shows the electroproduction differential cross-section on a linear scale, over the range of $\theta_{cm}$ for which the cross-section is non-negligible (for $0<\theta_{cm}<0.2$ rad) the cross-section is due exclusively to the impulse diagram $F_{1b}$.  The very small ``bump" visible in Fig. \ref{fig:cross9linear} at $\theta_{cm}\simeq 0.3$ rad is due to the proton-neutron rescattering amplitude $F_{2b}$. 

For $\nu=6.5$ GeV, the difference between the cross-section including $F_{3a}$ and omitting $F_{3a}$ is visible in the logarithmic-scale graphs (Figs. \ref{fig:squaretot65a} and \ref{fig:squaretot65b}), but not in the linear-scale graph, Fig. \ref{fig:linear65}.

$F_{3a}$ was also calculated using 3 other potentials:  a square-well potential yielding $a=0.3\;fm$ and $R=0.1\;fm$, and also the potential of \eq{modelpot} but with $a=0.3\;fm$, $R=0.29\;fm$, and a square-well potential yielding $a=0.3\;fm$ but with $R=0.29\;fm$.  There was a negligible difference between the values of $F_{3a}$ calculated with these potentials.

\subsection{Conclusion}
\label{conclusionFirst}

It does not appear to be possible to measure the $J/\psi$-nucleon scattering length via production on the deuteron, under the kinematic conditions available at JLab.  For small values of the relative momentum of the outgoing $J/\psi$-neutron pair, the initial momentum of the neutron inside the deuteron that is required for on-mass-shell rescattering of the $J/\psi$-neutron pair is larger than $\sim 0.6\;GeV$ (see Fig. \ref{fig:nminusplus3a1}), where the deuteron wavefunction is negligible.  The off-mass-shell part of the rescattering amplitude was calculated using model $J/\psi$-nucleon potentials and was found to make a negligible contribution to the total amplitude.   The vast majority of $J/\psi$ production events, for $T_{Vn}^*\le 0.03\;GeV$, will be at small values of $\theta_{cm}$, where the impulse diagram $F_{1b}$ dominates, and therefore information on $J/\psi$-nucleon elastic scattering at small relative energy cannot be obtained.

\section{Intermediate energy $J/\psi$ production on the deuteron}
\label{sec:intermedenergy}

It may be possible to extract the $J/\psi$-nucleon elastic scattering amplitude from the $\gamma^*+D\to J/\psi +p+n$ experiment, at higher relative energy of the $J/\psi$-nucleon pair, under different kinematic conditions for the final-state particles than was considered in the previous sections of this chapter.
Under certain kinematic conditions, the dominant contributions to the amplitude will come from rescattering diagrams (p-n rescattering and $J/\psi-n$ rescattering).  If we fix the magnitude of the outgoing neutron's momentum at a moderately large value (here taken to be 0.5 GeV) the contribution of the impulse diagram will be negligible, since the impulse diagram is proportional to the value of the deuteron wavefunction at that momentum (see Fig. \ref{fig:diagrams2} for the impulse and rescattering diagrams). For the analysis presented here, we:
 \begin{itemize}
   \item use coplanar kinematics
   \item fix the magnitude of the outgoing neutron momentum at $p_n=0.5\;GeV$
   \item fix the 4-momentum-transfer-squared $t=(q-p_V)^2$ at a particular value
   \item plot amplitudes or differential cross-sections vs. $\theta_n$ (the angle that the outgoing neutron momentum $\mathbf{p}_n$ makes with direction of the incoming photon momentum) for fixed $p_n$ and $t$	
\end{itemize}
(see Fig. \ref{fig:plusminpic}).  
For some range of $\nu$ and $t$, these graphs will display peaks due to $p-n$ and $J/\psi-n$ on-mass-shell rescattering.  For $\nu=10\;GeV$, the peak due to $J/\psi-n$ rescattering is evident (see Fig. \ref{fig:2a3atot}), but for $\nu=9\;GeV$ it is not evident (see Fig.  \ref{fig:nu9}).  This analysis is similar to what has been done in~\cite{laget81} for the reaction $\gamma+D\to \pi+N+N$.

The kinematics here are very different than in the previous sections.  There it was the relative energy of the $J/\psi$-neutron system which was kept fixed, at a small value, while the parameter which was varied was the production angle of the $J/\psi$ in the overall center-of-mass system.

\subsection{Intermediate-energy $J/\psi$ production}

We are interested here in kinematics available at JLab after the 12 GeV upgrade.  The maximum (virtual) photon energy is then around 11 GeV.  Here we evaluate the amplitude for virtual photon 4-momentum $q=(\nu,\mathbf{q})$ with $\nu=10\;GeV$, $Q^2=-q^2=0.5\;GeV$, keeping $p_n=0.5\;GeV$ and $t$ fixed.  The results presented here are for $t=-2\;GeV^2$; calculations were done for larger values of $\vert t\vert$, with similar results (although the total amplitude decreases with increasing $\vert t\vert$).  We consider the same set of diagrams as before, shown in Figs. \ref{fig:diagrams} and \ref{fig:diagrams2}.  The impulse diagrams, $F_{1a}$ and $F_{1b}$, are negligible for these kinematics.

\begin{figure}[tbp]
     \begin{center}
        \subfigure[]{%
            \label{fig:pluskin}
            \includegraphics[width=0.4\textwidth]{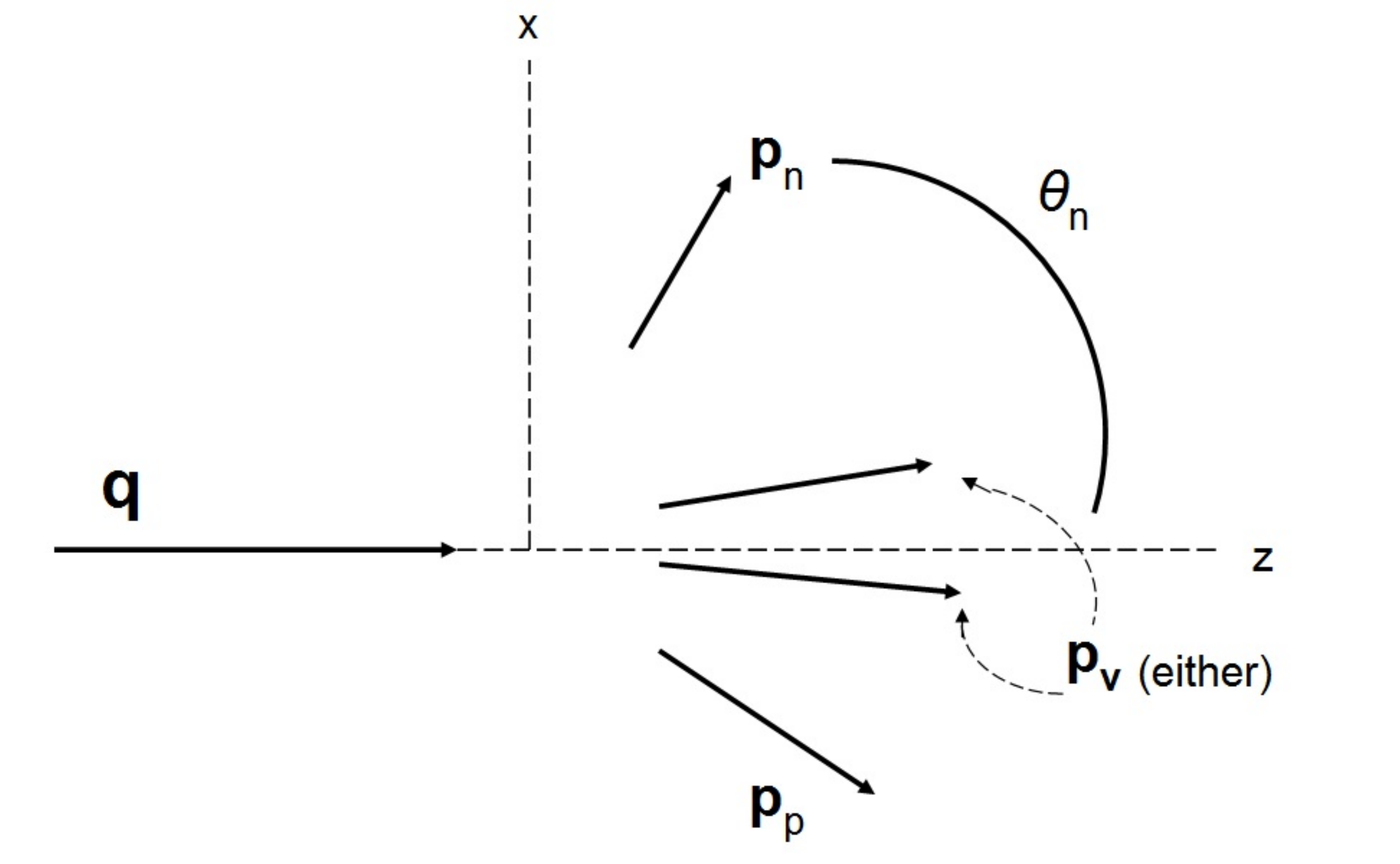}
        }%
        \hspace{0.5in}
         \subfigure[]{%
           \label{fig:minuskin}
           \includegraphics[width=0.4\textwidth]{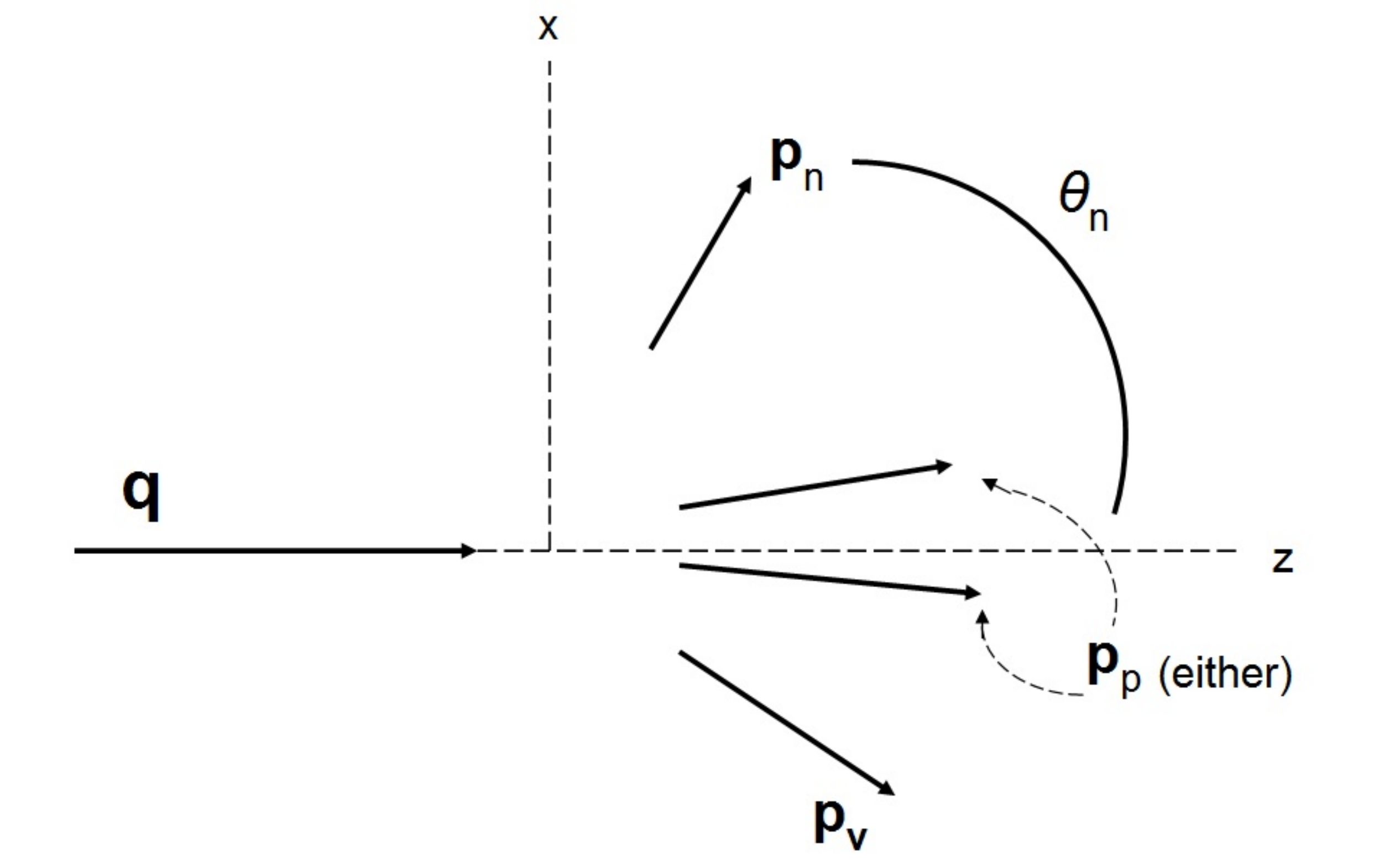}
        }\\ 


    \end{center}
    \caption{%
       (a) ``plus" kinematics and (b) ``minus" kinematics.  For ``plus", $\mathbf{p}_p$ is always on the opposite side of the photon momentum $\mathbf{q}$ direction as the neutron momentum.  For ``minus", $\mathbf{p}_V$ is always on the opposite side of the photon momentum $\mathbf{q}$ direction as the neutron momentum.
     }%
   \label{fig:plusminpic}
\end{figure}

Note that for a given $t$, $p_n$, and $\theta_n$, there are 2 sets of allowed values of the proton and $J/\psi$ momentum $\lbrace \mathbf{p}_p,\mathbf{p}_V \rbrace$; we've called the two sets the ``plus" set and the ``minus" set.  If we define $x$ and $z$ axes as in Fig. \ref{fig:plusminpic}, with the $x$-component of the neutron momentum always positive, then the ``plus" kinematics is as shown in Fig. \ref{fig:pluskin} and the ``minus" kinematics is as shown in Fig. \ref{fig:minuskin}.  For the ``plus" kinematics, $p_{px}$ is negative for all $\theta_n$ (while $p_{Vx}$ takes both positive and negative values over the range of $\theta_n$), while for the ``minus" kinematics, $p_{Vx}$ is negative for all $\theta_n$ (while $p_{px}$ takes both positive and negative values over the range of $\theta_n$).

\begin{figure}[tbp]
     \begin{center}
        \subfigure[p-n rescattering]{%
            \label{fig:nminpnmin}
            \includegraphics[width=0.4\textwidth]{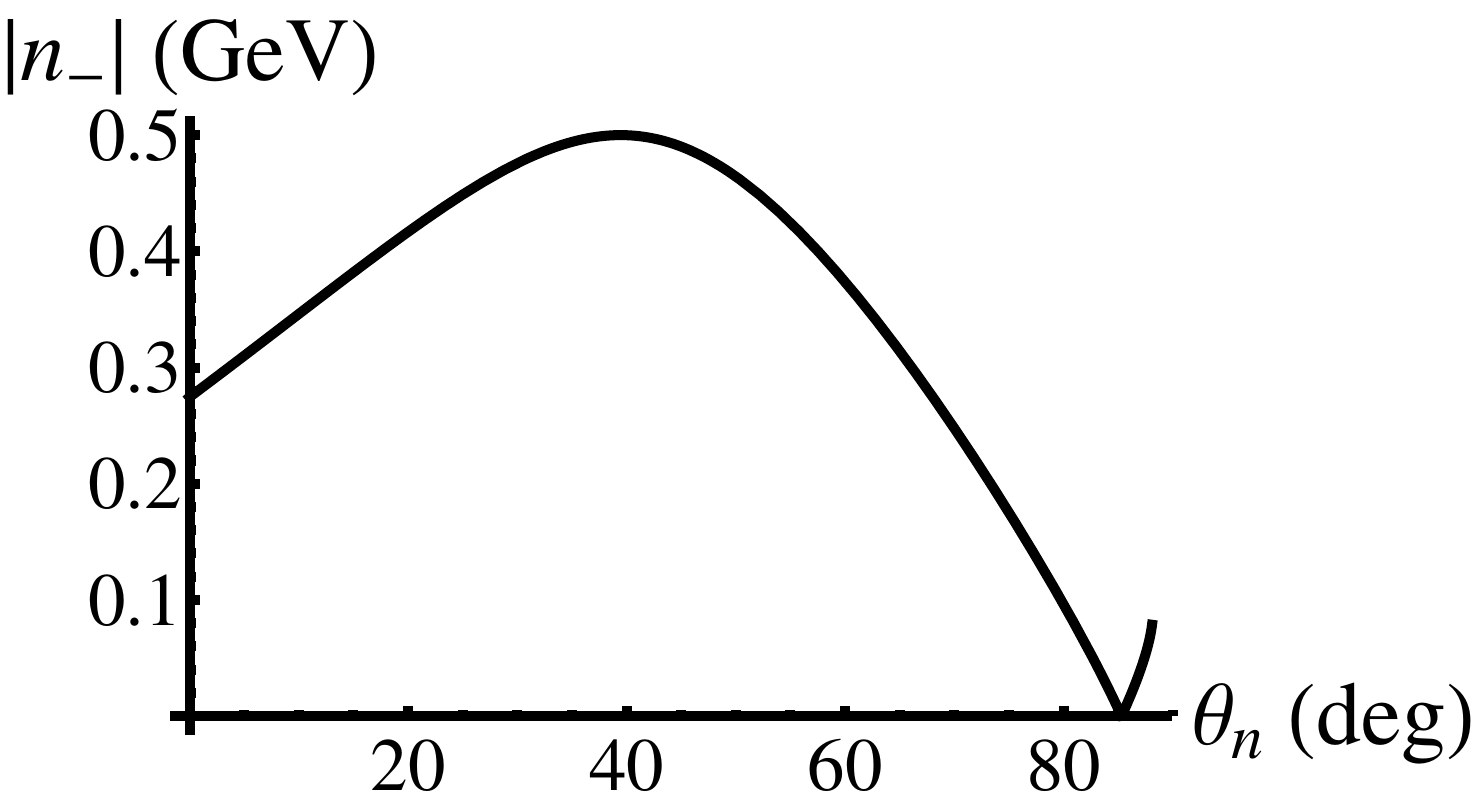}
        }%
        \hspace{0.5in}
         \subfigure[$J/\psi$-n rescattering]{%
           \label{fig:nminvnmin}
           \includegraphics[width=0.4\textwidth]{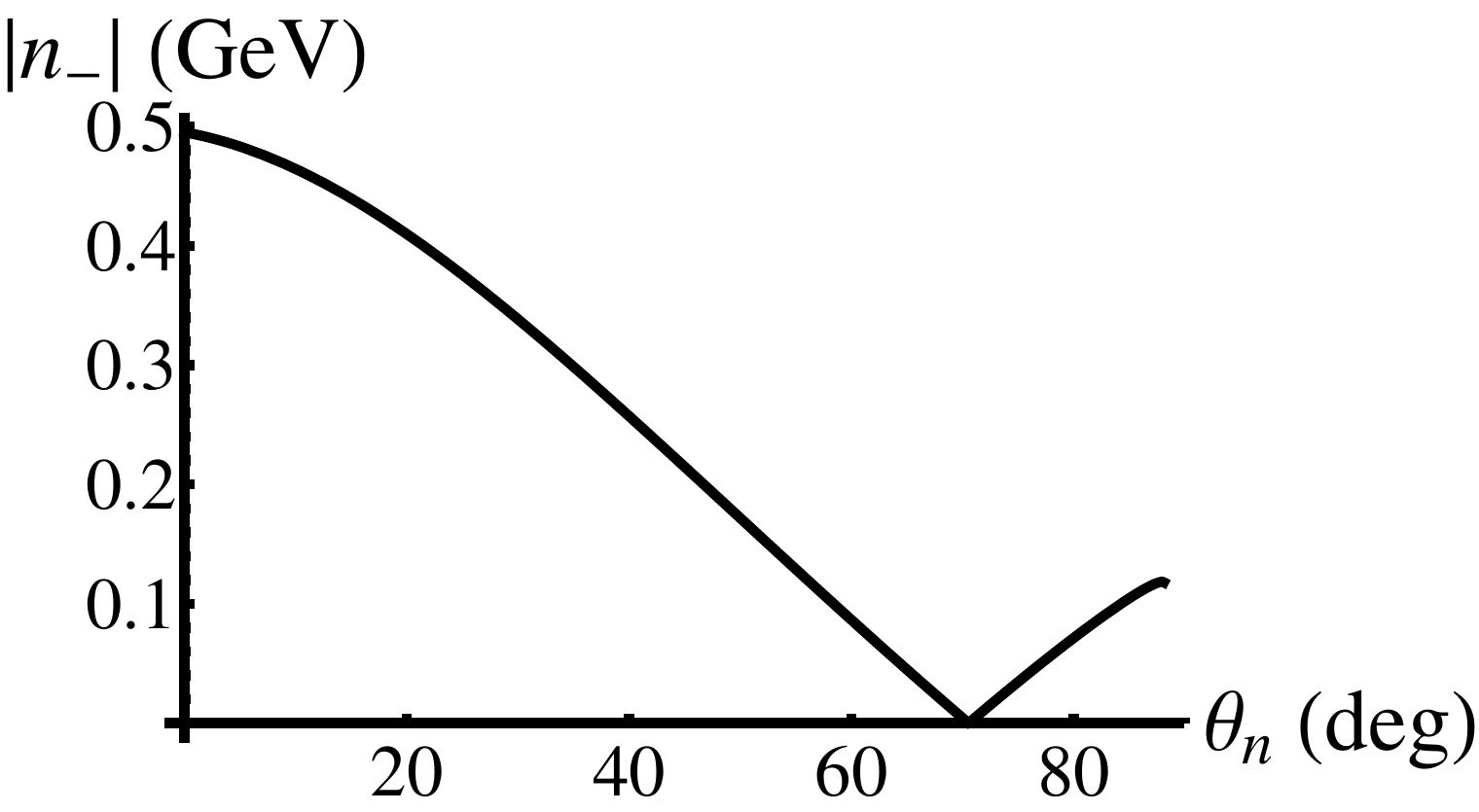}
        }\\ 

        \subfigure[p-n rescattering]{%
            \label{fig:nminpnpl}
            \includegraphics[width=0.4\textwidth]{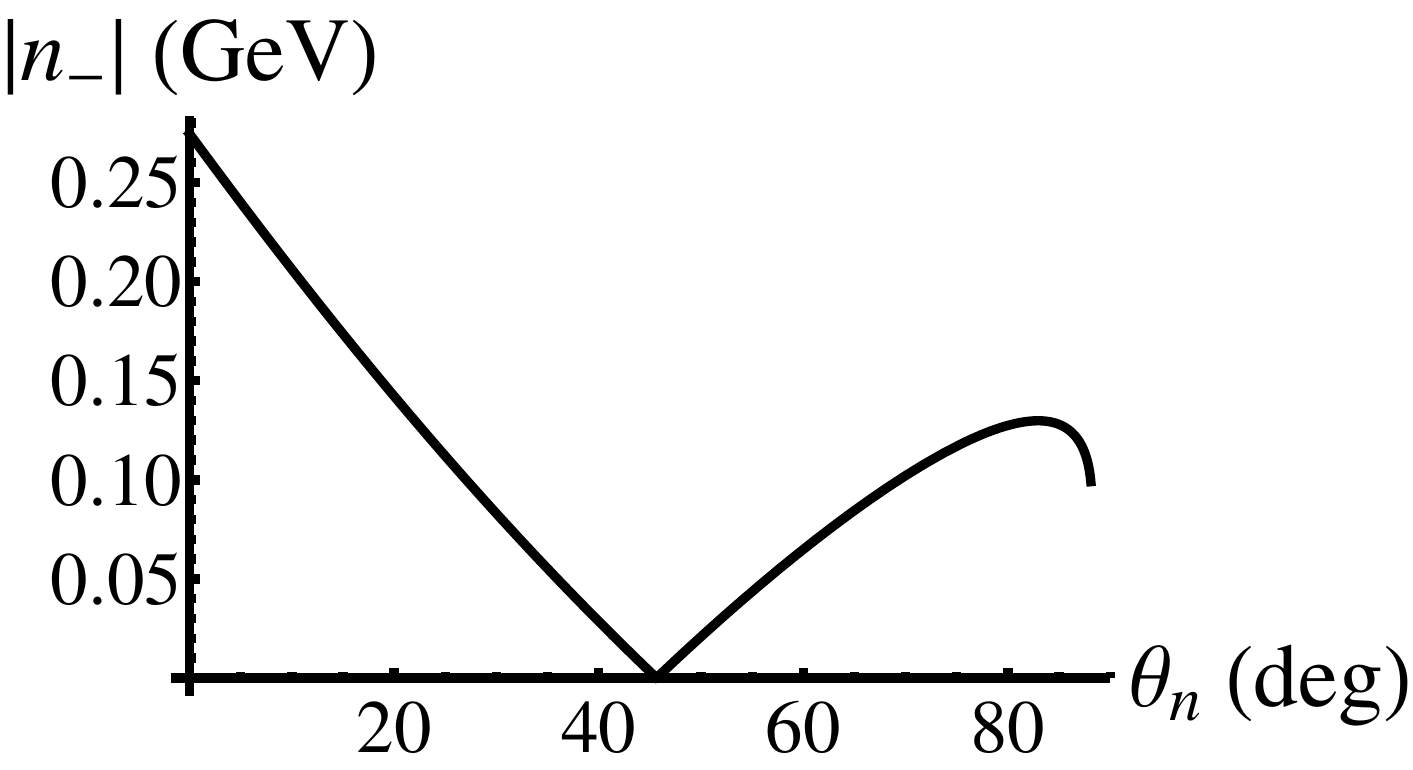}
        }%
         \hspace{0.5in} 
        \subfigure[$J/\psi$-n rescattering]{%
            \label{fig:nminvnpl}
            \includegraphics[width=0.4\textwidth]{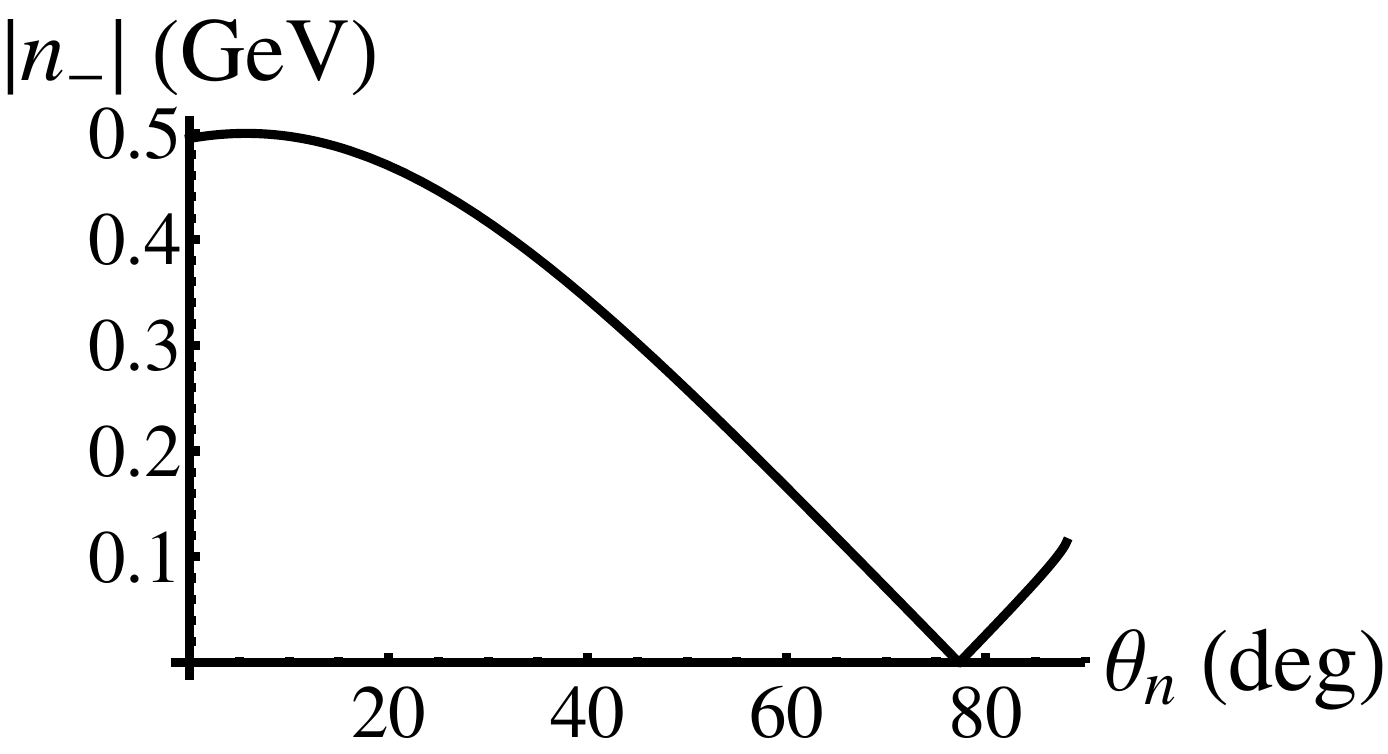}
        }%

    \end{center}
    \caption{%
       $\vert n_-\vert$ vs. $\theta_n$, for photon energy $\nu=10\;GeV$, and $t=-2\;GeV^2$.  (a) and (b) are for the ``minus" kinematics, (c) and (d) are for the ``plus" kinematics.
     }%
   \label{fig:nmin}
\end{figure}

\begin{figure}[tbp]
     \begin{center}
        \subfigure[$J/\psi$-p rescattering]{%
            \label{fig:nminpnmin}
            \includegraphics[width=0.4\textwidth]{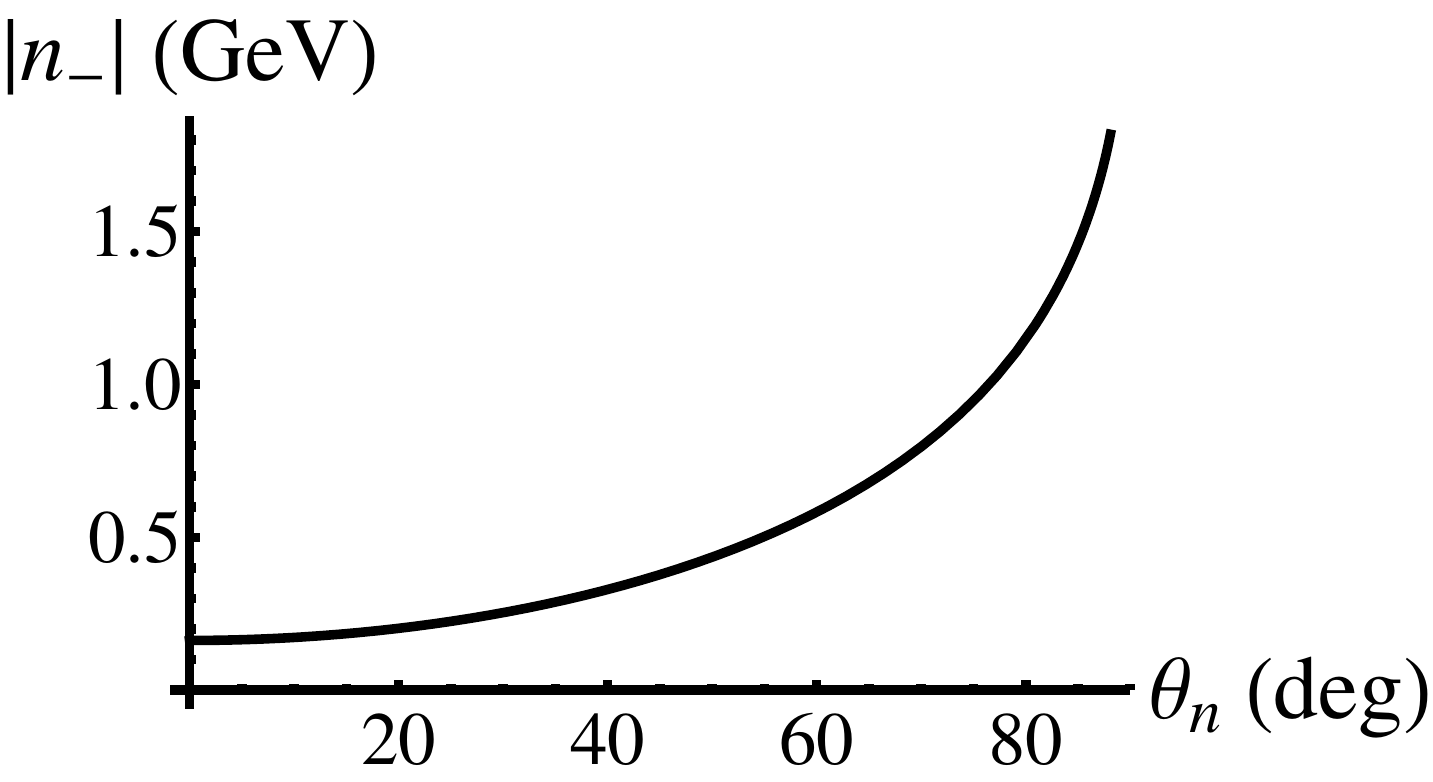}
        }%
        \hspace{0.5in}
         \subfigure[$J/\psi$-p rescattering]{%
           \label{fig:nminvnmin}
           \includegraphics[width=0.4\textwidth]{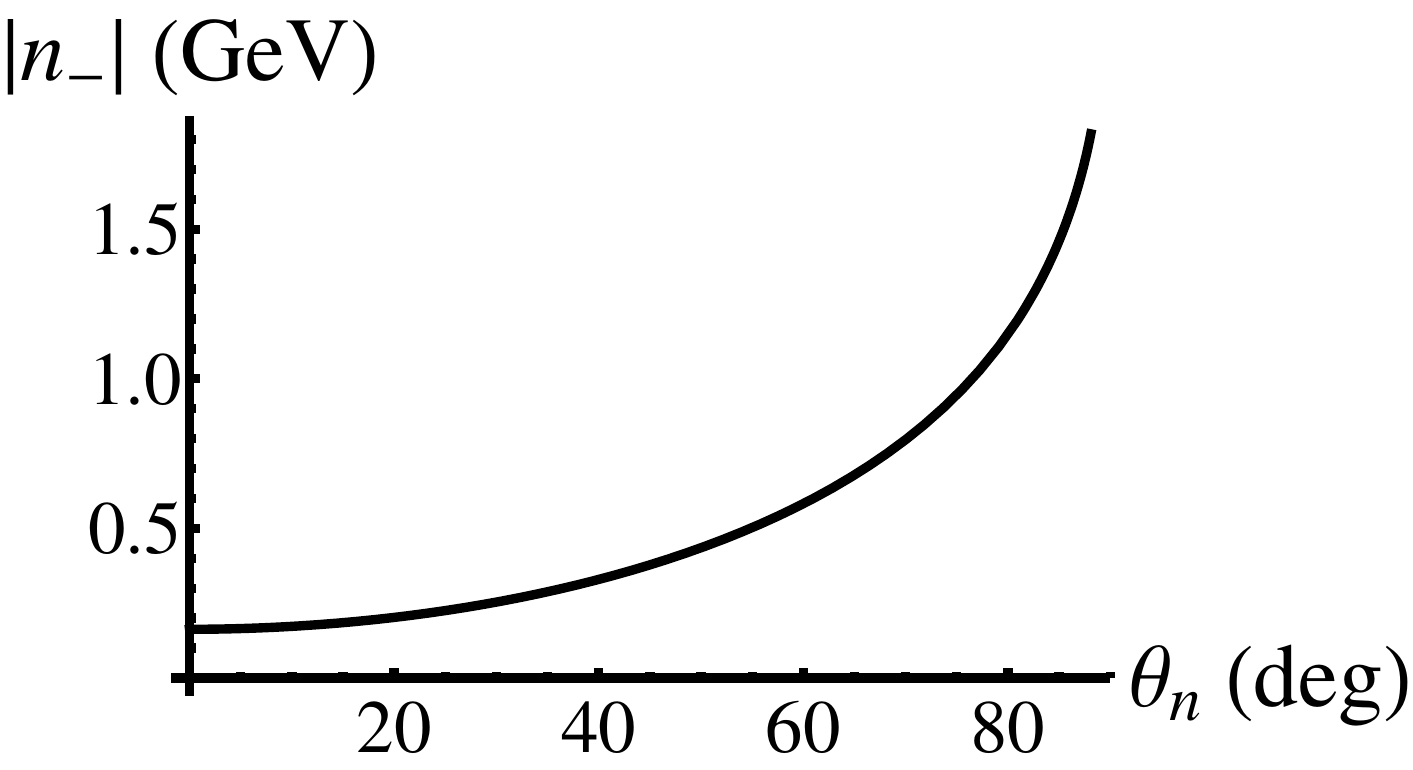}
        }\\ 


    \end{center}
    \caption{%
       $\vert n_-\vert$ vs. $\theta_n$ for $J/\psi$-p rescattering, for photon energy $\nu=10\;GeV$, and $t=-2\;GeV^2$.  (a) is for the ``minus" kinematics, (b) is for the ``plus" kinematics.
     }%
   \label{fig:nminvp}
\end{figure}

Figs. \ref{fig:nmin} and  Fig. \ref{fig:nminvp} show graphs of $\vert n_-\vert$ vs. $\theta_n$ for the 3 different pairs of outgoing particles.  Since the value of $\vert n_-\vert$ varies greatly with $\theta_n$, the value of the amplitude of the corresponding rescattering diagram varies greatly also, and has a prominent peak at the value of $\theta_n$ for which $\vert n_-\vert =0$.

\subsubsection{Calculation of amplitudes}
For the calculation of the amplitudes, the elementary 2-body amplitudes $ {\cal M}^{\gamma V}$ and ${\cal M}$ are taken to be of the diffractive form $Ae^{\frac{1}{2}B t}$ with parameters determined from existing experimental data.  For the $J/\psi$-nucleon rescattering diagrams, the only available data is from the experiment at SLAC~\cite{psidata77} discussed in Sec. \ref{sec:parameterssubsec}.  They determined the total $J/\psi$-nucleon cross-section to be $\sigma_{tot}^{J/\psi\; N}=3.5\pm0.8\;mb$, which gives via the optical theorem $A_{Vn}=1.61\pm0.4\;GeV^{-4}$.  The energy of the $J/\psi$ in this experiment was $\sim20$ GeV in the Lab frame (nucleon at rest).  However, for our kinematics the rescattering of the $J/\psi$ on the nucleon takes place at an energy in the outgoing neutron's rest frame of from 6 to 10 GeV, which is significantly smaller than in the SLAC experiment; thus the value of $A_{Vn}$ at our energy may be significantly different.  Since the entire reason for measuring the cross-section for this process is to extract the $J/\psi$-nucleon scattering amplitude in an energy region where it has not been measured before, we've used several different values of the parameter $A_{Vn}$ in the calculations, from 1 times the SLAC value up to 10 times the SLAC value.  Since the total cross-section $\sigma_{tot}$ for $J/\psi$-nucleon scattering goes like $\sqrt{A_{Vn}}$ (\eq{Aandsigma}), this corresponds to a range of $\sigma_{tot}$ (which is what was actually measured in the SLAC experiment) of from 1 to $\sim3$ times the SLAC value.  In~\cite{brodsky97}, theoretical calculation of the $J/\psi$-nucleon elastic scattering cross-section at threshold yielded $7\;mb$, which is twice the value measured at the higher energy at SLAC.

The full calculation of the amplitudes must of course include the off-shell parts.  If we use the same parametrization of the elementary amplitudes as in the on-shell part, then the off-shell parts were found to be very small compared to the on-shell parts.

\begin{figure}[tbp]
     \begin{center}
        \subfigure[]{%
            \label{fig:2a3a}
            \includegraphics[width=0.4\textwidth]{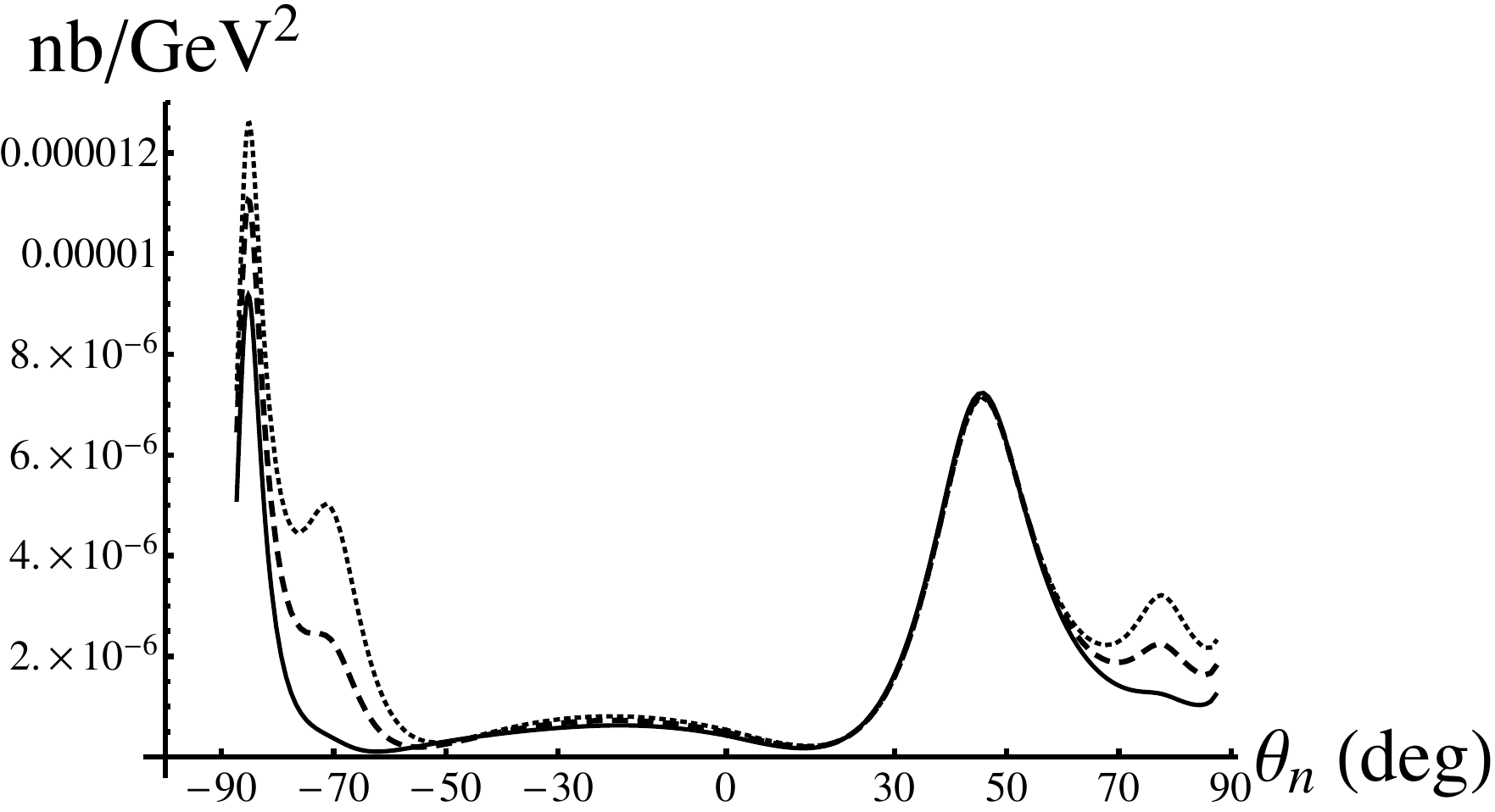}
        }%
        \hspace{0.5in}
         \subfigure[]{%
           \label{fig:2a3a5A}
           \includegraphics[width=0.4\textwidth]{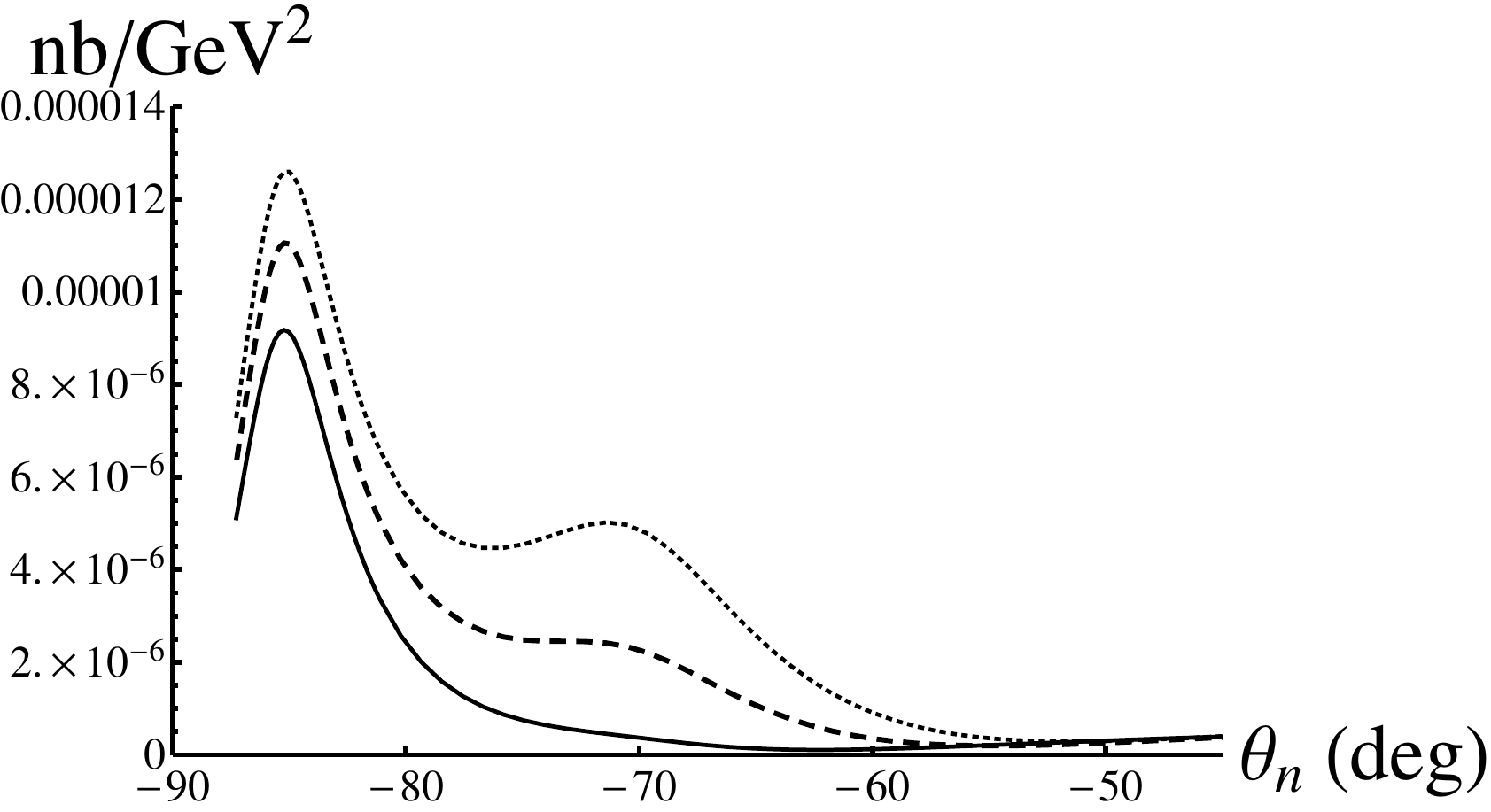}
        }\\ 
%

    \end{center}
    \caption{%
       Electroproduction differential cross-section vs. $\theta_n$, including all diagrams, for photon energy $\nu=10\;GeV$, $Q^2=0.5\;GeV^2$ and $t=-2\;GeV^2$, for 3 values of $A_{Vn}$:  Solid curve:  $A_{Vn}=1.6\;GeV^{-4}$.  Dashed curve:  $A_{Vn}=8.0\;GeV^{-4}$.  Dotted curve:  $A_{Vn}=16\;GeV^{-4}$.  In (a), the large peaks at $\theta_n\simeq 45^{\circ}$ and $\theta_n\simeq-85^{\circ}$ are due to $p-n$ rescattering, while the  small peaks (or bumps) at $\theta_n\simeq 80^{\circ}$ and $\theta_n\simeq-70^{\circ}$ are due to $J/\psi$-neutron rescattering.  (b) shows detail of left half of (a).  
     }%
   \label{fig:2a3atot}
\end{figure}

The amplitudes $F_{2b}$ and $F_{3b}$, where the $J/\psi$ is produced on the neutron and then rescattering (of the neutron or $J/\psi$, respectively) occurs on the proton, are much smaller than $F_{2a}$ and $F_{3a}$, and do not exhibit the well-defined peaks that $F_{2a}$ and $F_{3a}$ do.  Fig. \ref{fig:2a3atot} shows the 8-fold electroproduction differential cross-section, \eq{electrocross}, versus $\theta_n$.  In that figure,  the negative values of $\theta_n$ are for the ``minus" kinematics, and the positive values are for the ``plus" kinematics.  Graphs are shown for 3 different values of the $J/\psi$-neutron elastic scattering parameter $A_{Vn}$:  $A_{Vn}=1.6\;GeV^{-4}$ (which is the value determined in the experiment at SLAC),  $A_{Vn}=8.0\;GeV^{-4}$, and $A_{Vn}=16\;GeV^{-4}$.  It is seen that only if $A_{Vn}$ is of the order of 10 times as large as the previously measured value is there a noticeable peak due to the $J/\psi$-neutron rescattering, for the ``plus" kinematics.  The $p-n$ rescattering peak is much larger than, and close enough to, the $J/\psi$-neutron rescattering peak that it obscures the $J/\psi$ peak.  For the ``minus" kinematics, the same statement holds; in addition, however, the size of the $p-n$ rescattering peak varies (by $\sim 40\%$) as the value of $A_{Vn}$ is varied.  Note also that the position of each of the peaks is simply given by the value of $\theta_n$ where the value of the corresponding $\vert n_-\vert$ is zero (see Fig. \ref{fig:nmin}). 

\begin{figure}[tbp]
     \begin{center}

            \includegraphics[width=0.4\textwidth]{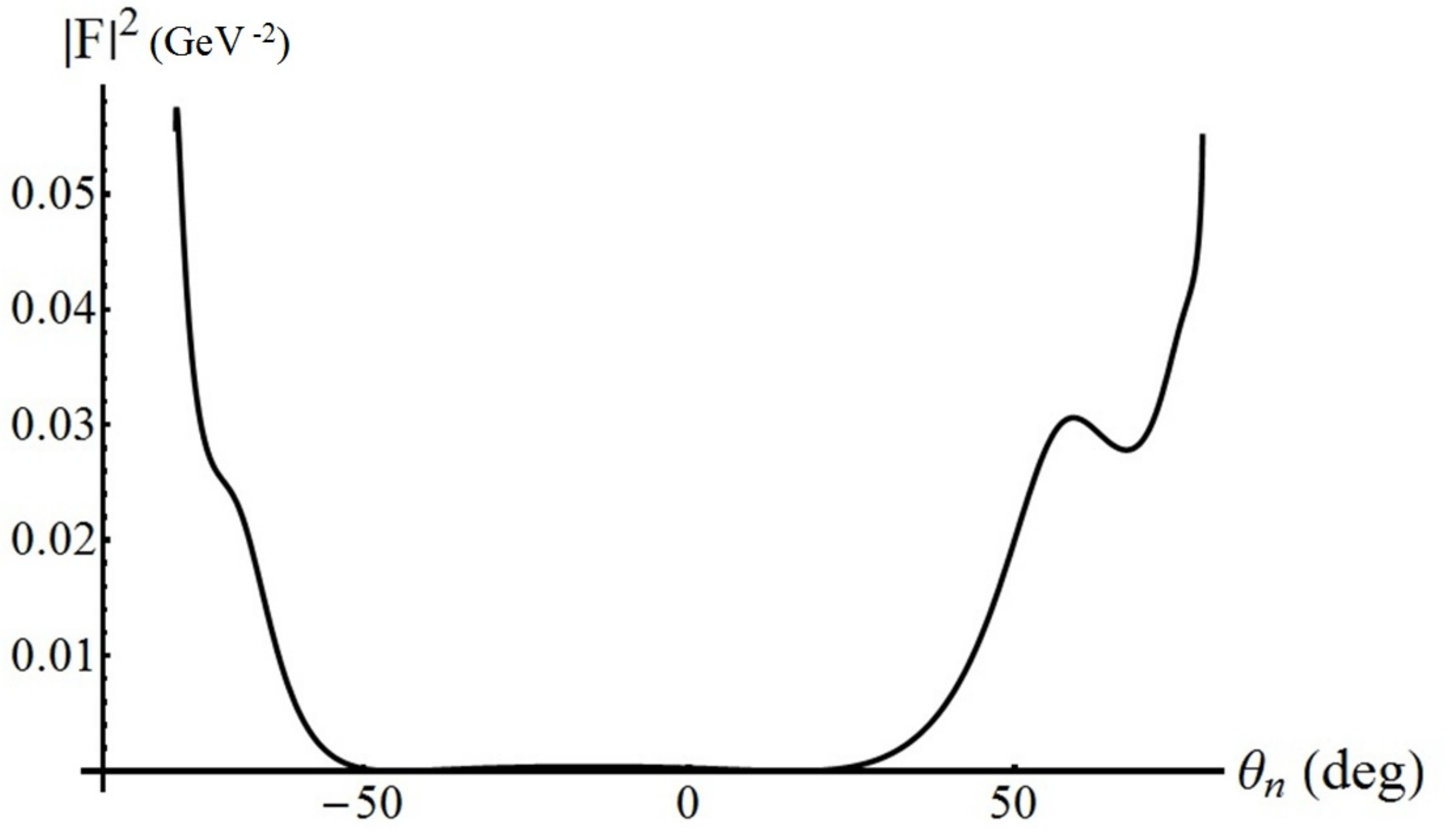}

    \end{center}
    \caption{%
         Amplitude squared vs. $\theta_n$, including all diagrams, for  $A_{Vn}=16\;GeV^{-4}$, for photon energy $\nu=9\;GeV$ and $t=-3\;GeV^2$.  The peak or bump on either side is due to $p-n$ rescattering, while the peak due to $J/\psi$-nucleon rescattering is not visible. 
     }%
   \label{fig:nu9}
\end{figure}

It's important to note that the peak due to the $J/\psi$-neutron rescattering isn't observable at lower energies.  Fig. \ref{fig:nu9} shows the square of the total amplitude for photon energy of $\nu=9\;GeV$ and $t=-3$, for $A_{Vn}=10\times 1.6\;GeV^{-4}$.  On this graph the peak due to $p-n$ rescattering is visible, but there's no visible peak due to $J/\psi$-neutron rescattering.

\subsection{Conclusion}
\label{conclusionSecond}

We have shown here the possibility of measuring the elastic $J/\psi$-nucleon scattering amplitude for energies significantly smaller than the energy of the only existing data.  If the total $J/\psi$-nucleon cross-section $\sigma_{tot}^{J/\psi N}$ at these energies is of the order of $2-3$ times the previously measured value, then the differential cross-section as a function of $\theta_n$ should exhibit well-defined peaks corresponding to on-mass-shell $p-n$ and $J/\psi - n$ rescattering, for virtual photon energy of $\nu=10\;GeV$ and 4-momentum-transfer-squared $t=(q-p_V)^2=-2\;GeV^2$.  However, at lower photon energy (below $9\;GeV$) the $J/\psi - n$ rescattering peak would not be distinguishable.  As it is expected~\cite{brodsky97} that $\sigma_{tot}^{J/\psi N}$ should increase as the energy decreases, it is not impossible that the lower energy cross-section could be larger than the measured value by a factor of $\sim 2$.

\section{Conclusion}
\label{sec:conclusion}

This paper is concerned with determining information regarding $J/\psi$-nucleon elastic scattering.  As discussed in Sec. \ref{conclusionFirst}, it does not appear possible to measure the $J/\psi$-nucleon scattering length via production on the deuteron under the kinematic conditions available at JLab.  As discussed in Sec. \ref{conclusionSecond}, however, it will be possible to extract the   $J/\psi$-nucleon scattering amplitude at higher relative momentum if $\sigma_{tot}^{J/\psi N}$ is large enough.

\section*{Acknowledgements} This work was partially supported by the DOE grant No. DE-FG02-97ER-41014.

 \end{document}